\documentclass[prd,aps, floatfix,preprintnumbers,showpacs,nofootinbib,notitlepage,superscriptaddress]{revtex4-1}
\usepackage{amsthm}   
\usepackage{amsmath}  
\usepackage{amssymb}  
\usepackage{mathrsfs}
\usepackage{color}
\usepackage[all]{xy}
\usepackage[pdftex]{graphicx}
\usepackage{indentfirst}
\usepackage[pdftex]{hyperref}
\usepackage{subfigure}
\usepackage{longtable}

\newcommand{\las}{\mathscr{L}}

\newcommand{\be}{\begin{equation}}
\newcommand{\ee}{\end{equation}}
\newcommand{\ba}{\begin{eqnarray}}
\newcommand{\ea}{\end{eqnarray}}

\newcommand{\en}{\nonumber\\}

\newcommand{\de}{\delta}

\newcommand{\kk}{\mathbf{k}}
\newcommand{\xx}{\mathbf{x}}

\newcommand{\pp}[1]{\mathbf{p}_{#1}}
\newcommand{\qq}{\mathbf{q}}

\newcommand{\cH}{\mathcal{H}}

\definecolor{darkred}{RGB}{175,0,0}
\definecolor{darkblue}{RGB}{0,0,175}

\newcommand{\tl}{{\bf t}_{\rm L}}
\newcommand{\ttt}{{\bf t}}
\newcommand{\qgal}{{\bf q}_{\rm gal} }
\newcommand{\qsub}{{\bf q}_{\rm sub} }
\newcommand{\qenv}{{\bf q}_{\rm env} }
\newcommand{\qlos}{{\bf q}_{\rm los } }
\newcommand{\qcos}{{\bf q}_{\rm cos } }
\newcommand{\asub}{\vec{\alpha}_{\rm sub } }

\newcommand{\psub}{\phi_{\rm sub } }
\newcommand{\ksub}{\kappa_{\rm sub}}

\newcommand{\qdm}{{\bf q}_{\rm DM } }

\newcommand{\mnras}{Mon.\ Not.\ R.\ Astron.\ Soc.}

\newcommand{\apjl}{Astrophys.\ J.\ L.} 
\newcommand{\aap}{Astronon.\ Astrophys.}

\begin{document}
\title{A Dark census: Statistically detecting the satellite populations of distant galaxies }
\author{Francis-Yan Cyr-Racine}
\affiliation{Department of Physics, Harvard University, Cambridge, Massachusetts 02138, USA}
\affiliation{NASA Jet Propulsion Laboratory, California Institute of Technology, Pasadena, California 91109, USA}
\affiliation{California Institute of Technology, Pasadena, California 91125, USA}

\author{Leonidas A. Moustakas}
\affiliation{NASA Jet Propulsion Laboratory, California Institute of Technology, Pasadena, California 91109, USA}
\affiliation{California Institute of Technology, Pasadena, California 91125, USA}

\author{Charles R. Keeton}
\affiliation{Department of Physics and Astronomy, Rutgers, The State University of New Jersey, Piscataway, New Jersey 08854, USA }

\author{Kris Sigurdson}
\affiliation{Department of Physics and Astronomy, University of British Columbia, Vancouver, British Columbia V6T 1Z1, Canada}

\author{Daniel A. Gilman}
\affiliation{Department of Physics and Astronomy, University of California, Los Angeles, California 90095, USA}

\date{\today}
\begin{abstract}
In the standard structure formation scenario based on the cold dark matter paradigm, galactic halos are predicted to contain a large population of dark matter subhalos. While the most massive members of the subhalo population can appear as luminous satellites and be detected in optical surveys, establishing the existence of the low mass and mostly dark subhalos has proven to be a daunting task. Galaxy-scale strong gravitational lenses have been successfully used to study mass substructures lying close to lensed images of bright background sources. However, in typical galaxy-scale lenses, the strong lensing region only covers a small projected area of the lens's dark matter halo, implying that the vast majority of subhalos cannot be directly detected in lensing observations. In this paper, we point out that this large population of dark satellites can collectively affect gravitational lensing observables, hence possibly allowing their statistical detection. Focusing on the region of the galactic halo outside the strong lensing area, we compute from first principles the statistical properties of perturbations to the gravitational time delay and position of lensed images in the presence of a mass substructure population. We find that in the standard cosmological scenario, the statistics of these lensing observables are well approximated by Gaussian distributions. The formalism developed as part of this calculation is very general and can be applied to any halo geometry and choice of subhalo mass function. Our results significantly reduce the computational cost of including a large substructure population in lens models and enable the use of Bayesian inference techniques to detect and characterize the distributed satellite population of distant lens galaxies.
\end{abstract}
\maketitle

\section{Introduction}
Dark matter forms the gravitational backbone of most of the observed structures in the Universe, from the largest galaxy clusters to the faintest dwarf galaxies. Despite this ubiquity, the nature of dark matter remains a mystery. On the one hand, the cold dark matter (CDM) paradigm has been extremely successful at describing observations on large cosmological scales such as the cosmic microwave background \cite{2014A&A...571A..16P}, the clustering of galaxies \cite{Sanchez:2013aa}, and cosmic shear measurements \cite{Kilbinger:2012aa}. On the other hand, this success constitutes a mixed blessing since there is a vast array of particle candidates that naturally fall under the CDM umbrella on large cosmological scales. One possible avenue to distinguish between this plethora of models is to look on much smaller length scales where clues of the particle nature of dark matter are more evident. For instance, the initial free-streaming of warm dark matter particle would damp the growth of structure on small scales \cite{1981ApJ...250..423D, Blumenthal:1982mv,Bode:2000gq,Dalcanton:2000hn,Zentner:2003yd,Boyanovsky:2008ab,Boyanovsky:2011aa,Smith:2011ev} while the fluctuation power spectrum of dark matter that couples to relativistic species until late times would display both acoustic oscillations and collisional damping \cite{2005A&A...438..419B,Cyr-Racine:2013ab,Cyr-Racine:2014aa,Boehm:2014aa,Dvorkin:2014aa,Buckley:2014ab}. On the other hand, if these physical phenomena are absent in the dark matter sector, standard structure formation theory predicts that galaxies should harbor a very large number of dark satellites \cite{1999ApJ...522...82K,Moore:1999aa}. The statistical detection of these numerous dark subhalos would validate a key prediction of standard CDM theory.

Strong gravitational lensing provides a way to probe the distribution of dark matter on the smallest scales \cite{Nierenberg:2014aa,2014MNRAS.442.2017V,Nierenberg:2013aa,2015MNRAS.447.3189X,Hezaveh:2012ai,MacLeod:2012aa,Li:2012aa,Zackrisson:2012aa,2012Natur.481..341V,Liesenborgs:2012aa,Fadely:2009aa,Chen:2011aa,Xu:2011aa,DAloisio:2010aa,Metcalf:2010aa,Xu:2010aa,Vegetti:2009aa,Bacon:2009aa,Keeton:2009aa,Xu:2009aa,Petters:2008yn,Petters:2008ik,Keeton:2009ab,Vegetti:2008aa,Chen:2008aa,Williams:2008aa,Chiba:2008aa,Peirani:2008aa,Alard:2008aa,Shin:2008aa,Chen:2007aa,Dobler:2006aa,Amara:2006aa,Rozo:2006aa,Maccio:2006aa,Inoue:2005ab,2005MNRAS.364.1451O,2005tyad.confE..12D,Keeton:2005aa,Hagan:2005aa,Taylor:2005aa,Inoue:2005aa,Mao:2004aa,Metcalf:ab,Taylor:aa,Metcalf:aa,Metcalf:2004aa,Kochanek:2004aa,Cohn:2004aa,Bradac:2004aa,Keeton:2003aa,Kochanek:2003aa,Moustakas:2003aa,Zentner:2003aa,Keeton:2003ab,Chen:2003aa,Taylor:ab,Sheth:2003aa,Taylor:ac,Dalal:2002aa,Dalal:aa,Bradac:aa,Metcalf:ad,Keeton:aa,Metcalf:ac,Chiba:aa,Mao:1998aa}. For instance, the observations of flux-ratio anomalies in multiply imaged lensed quasars have been used to study of abundance of mass substructures within the lens galaxy \cite{Nierenberg:2014aa,Metcalf:2010aa,Keeton:2003ab,Dalal:2002aa,Dalal:aa,Metcalf:ad,Metcalf:ac,Chiba:aa,Mao:1998aa}. More recently, direct gravitational imaging \cite{Koopmans:aa,Vegetti:2008aa,Vegetti:2009aa,2012Natur.481..341V,2014MNRAS.442.2017V} has enabled the detection of massive substructures along magnified arcs and Einstein rings. Similarly, resolved spectroscopy of strongly lensed dusty star-forming galaxies has been proposed to study mass substructures within the lens galaxy \cite{Hezaveh:2012ai,Hezaveh:2014aa,2016ApJ...823...37H}. By construction, these techniques are most sensitive to substructures lying close to lensed images, that is, substructures that appear close in projection to the Einstein radius of the lens. Since the typical Einstein radius of a galaxy-scale lens is a small fraction of its virial radius, only a small number of mass substructures are on average projected close to the region probed by strong lensing \cite{Chen:2011aa}. One therefore naturally expects that an order unity number of mass substructures could be detectable in each individual lens. Meaningful constraints on mass substructure inside lens galaxies can then be obtained by considering a sample of galaxies as was recently done in Ref.~\cite{2014MNRAS.442.2017V}.

While the vast majority of mass substructures in lens galaxies cannot be directly detected in lensing observations, the collective effect of substructures far from lensed images can nevertheless cause small perturbations to lensing observables that can be statistically detected. For instance, Ref.~\cite{Chen:2007aa} studied how astrometric perturbations could be used to probe mass substructures, while Ref.~\cite{Fadely:2009aa} used both astrometric and magnification perturbations to constrain the presence of dark clumps within the lens HE0435-1223. In addition, time-delay fluctuations in multiply imaged lensed quasars have been proposed \cite{Keeton:2009ab} as a tool to characterize broad properties of mass substructures within lens galaxies. Certainly, the overall population of mass substructures will perturb all lensing observables in a coherent and correlated way. 

In this manuscript, we develop a formalism to study stochastic millilensing from a large population of unresolved mass substructures inside the halos of galaxies acting as strong gravitational lenses. The aim of this formalism is to compute the joint effect of mass substructures on all lensing observables (image positions, magnifications, time delays), taking into account all possible correlations among those. Our work builds on the theory of stochastic stellar microlensing \cite{1983ApJ...267..488V,1984JApA....5..235N,1986ApJ...306....2K,1987A&A...179...80S,1987ApJ...319....9S,1987PhRvL..59.2814D,1988ApJ...335...67D,1994A&A...288....1S,1994A&A...288...19S,2003A&A...404...83N,2006MNRAS.370...91T} and generalizes the results of Refs.~\cite{Petters:2008yn,Petters:2008ik,Keeton:2009aa}. We focus our analysis on multiply imaged point sources (e.g.~quasar lenses) since these are the most relevant objects where gravitational time delays can in principle be measured. As we discuss below, time-delay measurements are crucial in probing the satellites populating the outskirts of distant lens galaxies. 

As in some of the stellar microlensing works, we use Markov's method (see e.g.~Ref.~\cite{1999Uchaikinzolotarev}) to compute from first principles the probability distribution of lensing potential and deflection perturbations in the presence of a population of mass substructures inside the lens galaxy.  By performing an Edgeworth expansion \cite{Blinnikov:1997jq}, we show that for a realistic structure formation scenario the probability distributions are nearly Gaussian. We also compute the leading order deviations from pure Gaussianity. The advantage of our analysis is that it allows one to determine which physical quantities control the behavior of lensing observables in the presence of mass substructures. This dependence on physical parameters is often obscured in studies relying purely on numerical simulations. By removing the need to perform such simulations, our approach has the potential to significantly speed up the analysis of substructures inside lens galaxies, and provide a convenient way to explore degeneracies between the macrolens and the substructure population. Most importantly, it provides a practical way to statistically detect dark satellites inside lens galaxies, hence providing a key test of standard cold dark matter theory. 

In this paper, we focus on analyzing the effect of mass substructures that are spatially located beyond a few Einstein radii. We leave the analysis of local mass substructures that are spatially located close to lensed images to future work. This paper is organized as follows. In Sec.~\ref{sec:stoc_lensing_gen_case}, we describe the challenge of statistically detecting a population of unresolved mass substructures and explain our approach to tackle this problem via the use of the characteristic function. In Sec.~\ref{sec:Actual_calculation}, we justify the division of the overall substructure population into two subpopulations (distributed and local), and we perform the actual calculation of the characteristic function for a population of distributed substructures. We show that in the cases of interest its behavior is quasi-Gaussian, and we discuss in which situations non-Gaussianities can become important. We also compare our results to the output of Monte Carlo realizations.  We then show in Sec.~\ref{marginalization} how our approach can be used to marginalize over the distributed substructure population. We discuss which physical properties of the distributed substructure population are most relevant to the lensing observables in Sec.~\ref{sec:discussion}, and we finally conclude in Sec.~\ref{sec:conclusion}.

\section{Stochastic Lensing: General Case}\label{sec:stoc_lensing_gen_case}
In this section, we present the general ideas behind our approach to substructure lensing. After brief remarks about our setup and notation, we introduce the challenges of lens modeling in the presence of a stochastic population of mass substructures. We then present the basic ideas behind our analytical stochastic lensing framework and derive important results regarding the joint distribution of gravitational lensing observables. These results are used throughout the rest of this paper. 
\subsection{Setup and notation}
For definiteness, we consider a situation where a high-redshift point-like source (such as a quasar) is multiply imaged by a massive foreground galaxy whose properties are described by a set of parameters $\qgal$. For instance, $\qgal$ could contain information about the lens Einstein radius, the dark matter and stellar density profiles, their ellipticity, etc. The characteristics of the main lens can also depend on the fundamental properties of dark matter (denoted by the parameters $\qdm$) such as its free-streaming length ($\lambda_{\rm fs}$), its sound horizon ($r_{\rm DAO}$), its self-interaction cross section ($\sigma_{\rm SIDM}$), and its temperature of kinematic decoupling ($T_{\rm kd}$).  In addition, the main lens galaxy lives in a local environment characterized by parameters $\qenv$ which contain information, for instance,  about the external shear and the properties of nearby luminous galaxies. We parametrize the line-of-sight structures (that is, exterior to the main lens plane) between the source and the observer via an array $\qlos$. Of course, all of these different sets of parameters have a dependence on the background cosmology, which we denote as $\qcos = \{H_0, \Omega_{\rm m},\Omega_\Lambda, A_{\rm s}, n_{\rm s}\}$, where $H_0$ is the Hubble parameter, $\Omega_{\rm m}$ and $\Omega_\Lambda$ are the energy densities in matter and dark energy, respectively, in units of the critical energy density of the Universe, $A_{\rm s}$ is the amplitude of the primordial power spectrum of scalar fluctuations, and $n_{\rm s}$ is the scalar spectral index. Throughout this paper, we use the numerical values for the cosmological parameters from the Planck 2015 data release \cite{Planck:2015xua}.

In addition to the spatially smooth component described by $\qgal$, the lens galaxy also contains mass substructures, the most luminous of which can appear as satellites orbiting the main lens. We collect the individual properties of these substructures in an array ${\bf c}_{\rm sub}$, which could, for instance, contain information about the position, virial mass, and concentration of each substructure.  Finally, we assume that individual substructure properties are sampled from an underlying probability distribution parametrized by an array $\qsub$ which encodes information about the substructure mass function, their spatial distribution within the lens, and their mass-concentration relation, which has a strong dependence on the parameters contained in $\qdm$. We summarize our notation in Table \ref{param_table} and indicate the interdependency of these different sets of parameters. 
\begin{table}[t]
\begin{tabular}{|c|c|c|c|}
\hline
{\bf Parameters} & {\bf Description} & {\bf Dependency} & {\bf Example} \\
\hline
$\qcos$ & Cosmological parameters & - &$H_0, \Omega_{\rm m},\Omega_\Lambda, A_{\rm s}, n_{\rm s}$\\
\hline
$\qdm$ & Dark matter parameters & $\qcos$  & $\lambda_{\rm fs}$, $r_{\rm DAO}$, $\sigma_{\rm SIDM}$, $T_{\rm kd}$ \\
\hline
$\qenv$ & Lens environment parameters & $\qcos$ & External shear, positions and mass of nearby galaxies \\
\hline
$\qgal$ & Macrolens parameters & $\qcos$, $\qdm$, $\qenv$ & Lens Einstein radius, ellipticity, density profile \\
\hline
$\qlos$ &Line-of-sight structure parameters& $\qcos$, $\qdm$ & Nonlinear matter power spectrum parameters\\
\hline
$\qsub$ & Substructure population parameters & $\qcos$, $\qdm$, $\qgal$ & Substructure mass function and their spatial distribution \\
\hline
${\bf c}_{\rm sub}$ & Individual substructure parameters & $\qsub$ & Positions, masses and concentrations of each subhalo\\
\hline
\end{tabular}
\caption{Summary of our notation for the different sets of parameters relevant to our gravitational lensing analysis. The third column indicates the relative dependency of the different sets of parameters while the last column gives examples of the different types of parameters. See main text for more details.}\label{param_table}
\end{table}

We reiterate that our goal is to use gravitational lenses to constrain the substructure population parameters $\qsub$ and then use this information to learn new insights about the microphysics of dark matter encoded in $\qdm$. Of course, determining the impact of a given choice of $\qcos$, $\qdm$, and $\qgal$ on the population parameters $\qsub$ is highly nontrivial and requires detailed numerical simulations. This is a very active area of research and tremendous progress has been made in the last decade \cite{2007ApJ...657..262D,2008ApJ...679.1260M,Springel:2008cc,Zavala:2009,Zolotov:2012xd,Brooks:2012ah,Rocha:2012jg,2013MNRAS.430..105P,2014ApJ...786...87B,2012MNRAS.423.3740V,2013MNRAS.434.3337A,2013MNRAS.431L..20Z,2014MNRAS.444..222G,2014MNRAS.445L..31B,2014PhRvD..90d3524B,2014MNRAS.444.3684V,2015MNRAS.448..792G}. In this work, we are interested in developing a formal understanding of the impact of a given choice of $\qsub$ on lensing observations. We defer to future work the interpretation of given $\qsub$ constraints in terms of dark matter microphysics. 
\subsection{Stochastic lensing by unresolved substructures: Main challenge}
In this subsection, we review the challenges of lens modeling in the presence of unresolved mass substructures. Let us imagine that we have a data vector ${\bf d}$. In general, ${\bf d}$ could include the position and surface brightness of the multiple images of the source, the time delays between the images, the velocity dispersion of the lens, and other data about the lens environment (external shear and convergence). Using these data, we would like to jointly constrain the arrays of parameters $\qq\equiv\{\qgal,\qenv, \qlos\}$ and ${\bf c}_{\rm sub}$. Given a choice of these parameters, one can compute the theory observables ${\bf t}(\qq,{\bf c}_{\rm sub})$ (encompassing, for instance, image magnifications and positions, as well as relative time delay between lensed images) and use them to compute the likelihood of measuring ${\bf d}$, $\las({\bf d}|\qq,{\bf c}_{\rm sub})$. The problem with the above procedure is that a given dark matter theory does not predict the positions and masses of individual subhalos within the lens galaxy. The fundamental dark matter physics rather makes predictions about the statistical properties of subhalos (described here by $\qsub$) such as their mass distribution, their concentration and their spatial distribution within the lens. Therefore, the elements of the array ${\bf c}_{\rm sub}$ are nuisance parameters that need to be integrated out. 

 One could however sidestep this issue by directly specifying the \emph{statistics} of the substructure population via the set of parameters $\qsub$, without having to explicitly draw random realizations ${\bf c}_{\rm sub}$. The immediate problem with this approach is that the theory observables are no longer unambiguously specified. Instead, for each choice of substructure population parameters $\qsub$, one obtains a probability density function for the theory observables $P({\bf t}|{\bf q},\qsub)$. Formally, this probability density  can be written as
\be\label{P_of_t_given_qw}
P({\bf t}|{\bf q},\qsub) = \int P_{\rm sub}({\bf c}_{\rm sub}|\qsub)\de_{\rm D}^k\left({\bf t}-{\bf t}({\bf q},{\bf c}_{\rm sub})\right) d{\bf c}_{\rm sub},
\ee
where $P_{\rm sub}({\bf c}_{\rm sub}|\qsub)$ is the probability of having a mass substructure population specified by ${\bf c}_{\rm sub}$, given a choice of population parameters $\qsub$, and where $\de_{\rm  D}^k$ is the $k$-dimensional Dirac delta function ($k$ is the length of the ${\bf t}$ vector).  Once $P({\bf t}|{\bf q},\qsub)$ is specified, the likelihood of the data ${\bf d}$ now takes the form,
\be\label{likelihood_qsub}
\las({\bf d}|{\bf q},\qsub)= \int d{\bf t} P({\bf t}|{\bf q},\qsub) \las({\bf d}|{\bf t}),
\ee
where $ \las({\bf d}|{\bf t})$ is the likelihood of the data given the theory observables.
Note that if we substitute Eq.~(\ref{P_of_t_given_qw}) into Eq.~(\ref{likelihood_qsub}), we obtain
 \be\label{standard_clump_marg}
 \las({\bf d}|{\bf q},\qsub)= \int P_{\rm sub}({\bf c}_{\rm sub}|\qsub)\las({\bf d}|{\bf q},{\bf c}_{\rm sub})d{\bf c}_{\rm sub},
 \ee
 which is just the standard marginalization over the substructure population. Once $\las({\bf d}|{\bf q},\qsub)$ is known, it is straightforward to construct the desired posterior distribution
$P({\bf q},\qsub|{\bf d})  \propto \las({\bf d}|{\bf q},\qsub)\Pi({\bf q},\qsub)$, where $\Pi({\bf q},\qsub)$ is the prior probability  distribution for ${\bf q}$ and $\qsub$.  

The above calculation of $P({\bf q},\qsub|{\bf d}) $ hinges on the accurate determination of the likelihood $\las({\bf d}|{\bf q},\qsub)$, either through Eq.~(\ref{likelihood_qsub}), or directly through Eq.~(\ref{standard_clump_marg}). Let us for now focus on the latter approach which has been used in various works on mass substructure inside gravitational lenses \cite{Nierenberg:2014aa,Metcalf:2010aa,Keeton:2003ab,Dalal:2002aa,Metcalf:ad,Metcalf:ac,Chiba:aa,Mao:1998aa}. If one could rapidly draw a large number of substructure realizations ${\bf c}_{\rm sub}$ from the distribution $P_{\rm sub}({\bf c}_{\rm sub}|\qsub)$ and compute the theory observables ${\bf t}({\bf q},{\bf c}_{\rm sub})$ for each of those, one could then replace the integral in Eq.~\eqref{standard_clump_marg} by a sum of all the realizations
\be
\las({\bf d}|{\bf q},\qsub)\simeq\sum_{{\bf c}_{\rm sub}\sim P_{\rm sub}(\qsub)} \las({\bf d},{\bf t}({\bf q},{\bf c}_{\rm sub})),
\ee
where the notation ${\bf c}_{\rm sub}\sim P_{\rm sub}(\qsub)$ means that ${\bf c}_{\rm sub}$ is drawn from the distribution $P_{\rm sub}(\qsub)$. This approach has several drawbacks. First, it is difficult to assess how many realizations are needed  to properly estimate the likelihood. A related issue is how to identify the substructure realizations that contribute most to the sum and make sure that these realizations are included in it. Second, for dark matter models that predict an abundance of subhalos, randomly drawing a realization can be a numerically costly process since thousands or millions of subhalos need to be included in the lensing calculation. Most importantly, a purely numerical approach obscures which key physical quantities control the impact of substructures on lensing observables. While this approach is viable if we are interested in accurately knowing the likelihood for a few points in parameter space, it is impractical if we are using a Markov Chain Monte Carlo approach to estimate the final posterior distribution of ${\bf q}$ and $\qsub$. To remedy these problems, we describe in the following section an approach that allows efficient computation of stochastic lensing observables while leaving the physics of substructure lensing transparent.
\subsection{Stochastic substructure lensing: Characteristic function approach}
We now turn our attention to an analytical approach to the computation of lensing observables in the presence of a population of unresolved mass substructures. The calculation presented here draws from the theory of stochastic microlensing in the presence of a uniform star field  \cite{1983ApJ...267..488V,1984JApA....5..235N,1986ApJ...306....2K,1987A&A...179...80S,1987ApJ...319....9S,1987PhRvL..59.2814D,1988ApJ...335...67D,1994A&A...288....1S,1994A&A...288...19S,2003A&A...404...83N,2006MNRAS.370...91T}. As a starting point, our technique takes full advantage of two simplifying facts about the impact of mass substructures on the lensing observables:
\begin{itemize}
\item For deflection, shear, convergence, and projected gravitational potential, the overall impact of the subhalo population is the sum of the contributions from each mass substructure. 
\item Each subhalo is independent of all other subhalos in the lens. 
\end{itemize}
The first assumption is always true and is a direct consequence of the linearity of Poisson's equation. The second assumption is not strictly true since mass substructures may be themselves clustered within galactic halos. However, the relative importance of substructure clustering will be diminished by projection effects since lensing in only sensitive to the mass distribution integrated along the line of sight. Moreover, tidal interactions between subhalos and the tidal field of the host galaxy will tend to erase correlations among substructures within a few dynamical times \cite{2014arXiv1407.2648C}. Thus, to a good approximations, we can use the above simplifying facts to make progress in evaluating $P({\bf t}|{\bf q},\qsub)$. For the moment, let us focus on the lensing quantities that receive purely additive corrections from the substructures. These include the projected gravitational potential $\phi$, the deflections $\vec{\alpha}=\vec{\nabla}\phi$, the convergence $\kappa = (\phi_{xx}+\phi_{yy})/2$, as well as the shears $\gamma_{\rm c}=(\phi_{xx}-\phi_{yy})/2$ and $\gamma_{\rm s}=\phi_{xy}$. We denote these ``linear'' lensing quantities\footnote{Note that we use the nomenclature ``quantities" and not ``observables" since $\phi$, $\vec{\alpha}$, $\kappa$ ,$\gamma_{\rm c}$ and $\gamma_{\rm s}$ are not directly observable.} by $\tl = \big\{\ldots ,\{\phi^{(j)},\vec{\alpha}^{(j)},\kappa^{(j)},\gamma_{\rm c}^{(j)}, \gamma_{\rm s}^{(j)}\},\ldots\big\}$, where the index $j$ denotes that these quantities are evaluated at the position of the $j$th image. Our goal is to take advantage of the linearity to first compute $P(\tl|{\bf q},\qsub)$ and then reconstruct $P({\bf t}|{\bf q},\qsub)$ using the relation
\be\label{eq:linear_to_actual}
P({\bf t}|{\bf q},\qsub)=\int d\tl P(\tl|{\bf q},\qsub) \de_{\rm D}(\ttt-\ttt(\tl)).
\ee
The linear quantities $\tl$ receives contribution from both the smooth mass model and its environment (described by ${\bf q}$) and the mass substructures themselves
\be\label{eq:smooth+subs}
\tl({\bf q},{\bf c}_{\rm sub}) =\bar{\bf t}_{\rm L}({\bf q}) + \sum_{i=1}^N\de\tl^{(i)},
\ee
where $\bar{\bf t}_{\rm L}({\bf q})$ describes the contribution from the smooth model and its environment, while $\de\tl^{(i)}\equiv\de\tl({\bf q},{\bf c}_i)$ is the contribution from mass substructure $i$. Here, $N$ is the total number of subhalos within the lensing galaxy. Since $\bar{\bf t}_{\rm L}({\bf q})$ is completely fixed by a given choice of ${\bf q}$, the inherent stochasticity of $\tl$ is entirely caused by the random mass substructures inside the lens galaxy. Defining $\Delta\tl\equiv\sum_{i=1}^N \de\tl^{(i)}$, all relevant statistical information about the mass substructures is contained in the probability density function $\Phi_N(\Delta\tl|{\bf q},\qsub)$ for the collective effect $\Delta\tl$ of $N$ substructures on the linear lensing quantities. Once $\Phi_N$ is known, the probability density $P(\tl|{\bf q},\qsub)$ appearing in Eq.~(\ref{eq:linear_to_actual}) is simply given by
\be\label{eq:deltatL_2_tL}
P(\tl|{\bf q},\qsub) = \Phi_N(\tl -\bar{\bf t}_{\rm L}({\bf q})|{\bf q},\qsub).
\ee
Effectively, $\Phi_N(\Delta\tl|{\bf q},\qsub)$ is an $l$-dimensional joint probability distribution for $l$ sums of $N$ independent random variables, where $l$ refers to the number of linear lensing quantities included in the analysis. Take $\Phi_1(\de\tl^{(i)}|{\bf q},\qsub)$ to be the joint probability distribution for the linear lensing quantities in the presence of a \emph{single} substructure. For now, we assume that we know the functional form of $\Phi_1(\de\tl^{(i)}|{\bf q},\qsub)$; its formal derivation is given in the next subsection. Since the subhalos are assumed to be independent of each other,  $\Phi_N(\Delta \tl|{\bf q},\qsub)$ is given by the $N$-fold convolution of $\Phi_1(\de\tl^{(i)}|{\bf q},\qsub)$ \cite{1999Uchaikinzolotarev}. We then take advantage of the convolution theorem to write the characteristic function\footnote{In this work, the characteristic function is simply the Fourier transform of the probability density function.} of $\Phi_N(\Delta\tl|{\bf q},\qsub)$ in terms of that of $\Phi_1(\de\tl^{(i)}|{\bf q},\qsub)$,
\be
Q_N(\kk_{\rm L}|{\bf q},\qsub) = q_1(\kk_{\rm L}|{\bf q},\qsub)^N,
\ee
where $\kk_{\rm L}$ is the Fourier conjugate variable to $\Delta \tl$, $Q_N(\kk_{\rm L}|{\bf q},\qsub)$ is the characteristic function of $\Phi_N(\Delta\tl|{\bf q},\qsub)$, and $q_1(\kk_{\rm L}|{\bf q},\qsub)$ is the characteristic function of $\Phi_1(\de\tl^{(i)}|{\bf q},\qsub)$. 

Now, in a typical galactic dark matter halo the number of mass substructure $N$ is large but unknown. Given a total convergence in dark matter substructures and a subhalo mass function, we can compute the average expected total number of mass substructures $\langle N\rangle$ [see, e.g.~, Eq.~\eqref{eq:average_N} below]. Since the evolution of mass substructures within lens galaxies is determined by the complex interplay of accretion, dynamical friction, tidal striping, baryonic feedback, and mergers, the actual number of subhalos will typically differ from this average value. Detailed $N$-body simulations \cite{2011MNRAS.410.2309G} of massive galaxies show that the scatter about the mean is consistent with that of a Poisson distribution. Then, the resulting characteristic function for the whole substructure population is a sum over all possible values of $N$, weighted by their Poisson probability with mean $\langle N\rangle$,
\ba\label{eq:QN}
Q_{\langle N\rangle }(\kk_{\rm L}|{\bf q},\qsub) &=& e^{-\langle N \rangle}\sum_{N=0}^{\infty}\frac{\langle N\rangle^N}{N!}q_1(\kk_{\rm L}|{\bf q},\qsub)^N\en
&=&\exp{[\langle N\rangle(q_1(\kk_{\rm L}|{\bf q},\qsub) -1)]}.
\ea
This result states that if one could compute $q_1(\kk_{\rm L}|{\bf q},\qsub)$ for a \emph{single} mass substructure, then one could obtain the characteristic function for the \emph{whole population} of unresolved subhalos by taking the exponential of $\langle N\rangle (q_1(\kk_{\rm L}|{\bf q},\qsub)-1)$. Finally, $\Phi_{\langle N\rangle}(\Delta\tl|{\bf q},\qsub)$ can be obtained by Fourier transforming $Q_{\langle N\rangle}(\kk_{\rm L}|{\bf q},\qsub)$. Therefore, we have reduced the computation of $P(\tl|{\bf q},\qsub)$ for $\langle N\rangle$ subhalos to that of computing $q_1(\kk_{\rm L}|{\bf q},\qsub)$ for a single substructure which is a considerable simplification. 

\subsection{Characteristic function for a single substructure}
To complete our formalism, we need an expression for $q_1(\kk_{\rm L}|{\bf q},\qsub)$, the characteristic function of the linear lensing quantities in the presence of a single mass substructure. We begin by writing down an expression for $\Phi_1(\de\tl|{\bf q},\qsub)$,
\be\label{P_of_tL_given_qw}
\Phi_1(\de\tl|{\bf q},\qsub)= \int P_{\rm sub}({\bf c}_{\rm sub}^{(1)}|\qsub)\de_{\rm D}^l\left(\de\tl-\de\tl({\bf q},{\bf c}_{\rm sub}^{(1)})\right) d{\bf c}_{\rm sub}^{(1)},
\ee
where ${\bf c}_{\rm sub}^{(1)}$ are the parameters describing the properties of a single mass substructure. Here, $P_{\rm sub}({\bf c}_{\rm sub}^{(1)}|\qsub)$ is the probability density function describing the probability of finding a clump of dark matter with parameters ${\bf c}_{\rm sub}^{(1)}$, given a set of substructure population parameters $\qsub$.  The characteristic function of the above distribution, $q_1(\kk_{\rm L}|{\bf q},\qsub)$, is simply the Fourier transform of Eq.~(\ref{P_of_tL_given_qw}),
\be\label{eq:q1_start}
q_1(\kk_{\rm L}|{\bf q},\qsub) = \int d{\bf c}^{(1)}_{\rm sub}\, e^{i\de\tl({\bf q},{\bf c}_{\rm sub}^{(1)})\cdot\kk_{\rm L}} P_{\rm sub}({\bf c}_{\rm sub}^{(1)}|\qsub).
\ee
Computing this integral requires us to specify the spatial geometry over which the mass substructure is distributed as well as the subhalo mass function inside the lens galaxy. In the next section we describe our strategy to evaluate this characteristic function.
\section{Characteristic Function for Substructure Population}\label{sec:Actual_calculation}
Up to this point, we emphasize that our analysis has been very general and is purely based on the linearity and independence of mass substructures inside galactic halos. In this section, we consider how the geometry of the substructure distribution inside galactic halos can be used to simplify the calculation of $q_1(\kk_{\rm L}|{\bf q},\qsub)$. As we describe below, it is advantageous to divide the substructure population into a distributed subpopulation that contains the vast majority of subhalos and contributes small perturbations to lensing observables, and into a local subpopulation that contains a few strong perturbers to lensing observables. 
\subsection{Local versus distributed substructure populations}\label{sec:local_vs_distant}
\begin{figure}[t]
\begin{center}
\includegraphics[width=0.59\textwidth]{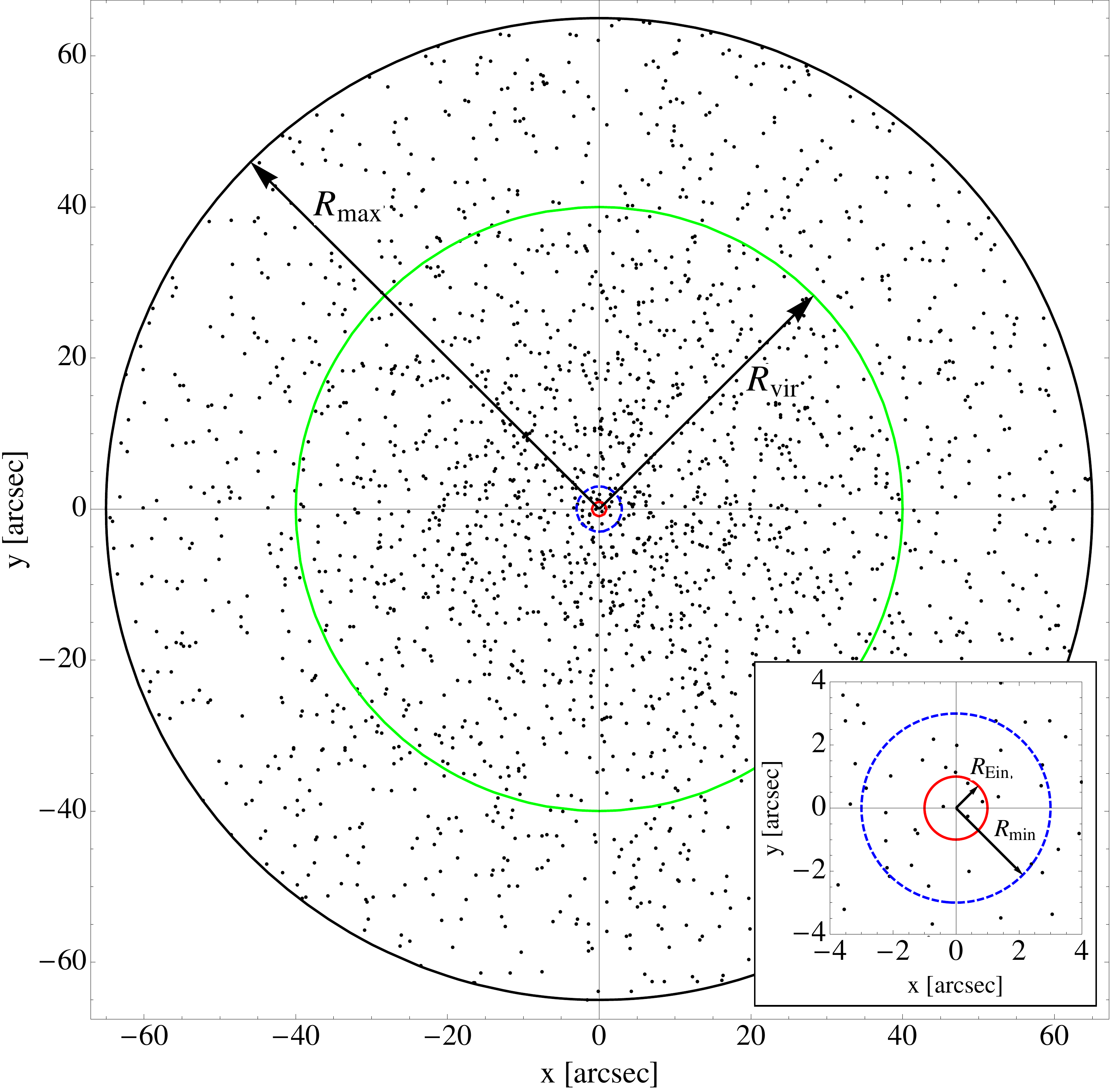}
\caption{Illustration of the various scales involved in galaxy-scale substructure lensing. The typical Einsteins radius $R_{\rm ein}$ of lens galaxy $(\sim 1$ arcsec) is indicated in red, while the typical virial radius of the galactic halo is indicated in green. The dashed blue circle and the outer black circle indicate our choice of $R_{\rm min}(=3 R_{\rm ein})$ and $R_{\rm max}(=65 R_{\rm ein})$ for the computation of the substructure characteristic function, respectively. The scattered dots represent a realization of a substructure population with the spatial distribution given in Eq.~(\ref{eq:sub_spatial_dist}) with a core radius given by $r_c = 30 R_{\rm ein}$. Here, we have assumed a power law mass function as given in Eq.~(\ref{eq:subhalo_mass_function}) with $\beta=-1.9$, $M_{\rm low} = 10^7 M_\odot$, and $M_{\rm high} = 10^{10} M_\odot$. The average convergence in mass substructure is taken to be $\langle\kappa_{\rm sub}(R_{\rm ein})\rangle =  6\times10^{-4}$. See main text for a description of the notation. The inset at the bottom right shows an enlargement of the halo's central region. In general, only an order unity number of substructures are projected close to the Einstein radius of the lens. }
\label{fig:scaling}
\end{center}
\end{figure}

We wish to compute the characteristic function for the linear lensing quantities in the presence of a single substructure at typical lensed image locations $\{\xx_i\}$ situated close to the Einstein radius $R_{\rm ein}$ of the lens. Similar to the analysis of Ref.~\cite{Keeton:2009aa}, our strategy is to divide the image plane into two regions: (i) an inner disk of radius $R_{\rm min}$ containing all the lensed images and a relatively small number of substructures (denoted ``local'' substructures), and (ii) an annulus with inner radius $R_{\rm min}$ and outer radius $R_{\rm max}$ containing the vast majority of the substructure population, which we shall refer to as the ``distributed'' population. This choice is illustrated in Fig.~\ref{fig:scaling}.
The first thing that is evident from Fig.~\ref{fig:scaling} is that the strong lensing region (red innermost circle) of typical galaxy-scale lenses probes the very inner part of the galactic halo. This is the region where flux ratio anomalies have been used to probe mass substructures within lens galaxies  \cite{Mao:1998aa,Chiba:aa,Metcalf:ac,Metcalf:ad,Dalal:2002aa,Keeton:2003ab,Metcalf:2010aa,Nierenberg:2014aa}. This is also the region where direct gravitational imaging \cite{Koopmans:aa,Vegetti:2008aa,Vegetti:2009aa,2012Natur.481..341V,2014MNRAS.442.2017V} and spatially resolved spectroscopy \cite{Moustakas:2003aa,Hezaveh:2012ai,Hezaveh:2014aa} can be used to detect individual mass substructures within galaxy-scale lenses.  In this area of the lens plane, it is possible for a mass substructure to cause a large perturbation to lensing observables which are known to cause non-Gaussian ``heavy tails" \cite{Keeton:2009aa} in stochastic lensing probability density functions. Furthermore, subhalos can have significant overlap with lensed images, implying that the internal properties of mass substructures such as their concentrations and tidal radii can be probed in this regime \cite{Nierenberg:2014aa}.  Due to its small size compared to the overall spatial extent of the dark matter halo, the inner region contains a relatively small fraction of the total number of mass substructures in the gravitational lens. 

On the other hand, the outer region of the lens halo (the area outside the dashed blue circle) contains the vast majority of the lens galaxy mass substructures. Since they are quite distant from any lensed image, these subhalos cannot significantly affects the lensing observables on an individual basis. However, their collective effect is not necessarily negligible. Furthermore, because of their relative position with respect to the strong lensing region, we do not expect their internal structure to play a significant role in their lensing signatures. Importantly, these properties of the distributed mass substructures considerably simplify the calculation of the characteristic function $q_1(\kk_{\rm L}|{\bf q},\qsub)$.
 
It is instructive to compare the relative contribution of the distributed and local subhalo populations to the linear lensing quantities. Writing the total contribution from substructures as 
\be\label{eq:local+distant}
\Delta \tl  = \Delta \tl^{\rm local} + \Delta\tl^{\rm dist},
\ee
let us compare the local and distributed pieces for potential fluctuations, deflections, convergence, and shears. To do so, we generate $10^4$ Monte Carlo realizations of mass substructure population. We assume the substructures to have smoothly truncated Navaro-Frenk-White (NFW) three-dimensional density profiles given by \cite{Baltz:2007vq}
\be\label{eq:truncated_NFW}
\rho(r_{\rm sub}) = \frac{M_{\rm NFW}}{4\pi r (r_{\rm sub}+r_{\rm s})^2}\left(\frac{r_{\rm t}^2}{r_{\rm sub}^2+r_{\rm t}^2}\right),
\ee
where $r_{\rm sub}$ is the three-dimensional distance from the center of the subhalo, $r_{\rm s}$ is the scale radius, and $r_{\rm t}$ is the tidal radius. We note that our choice of NFW profile is more conservative than the often used Pseudo-Jaffe profile since the latter has a steeper inner density slope and has thus a larger lensing efficiency. It is important to keep in mind that observations of low-mass galaxies show mild preference for even shallower density profiles, implying that the magnitude of the local substructure perturbations discussed in this section are likely to be conservative upper bounds. For the truncated NFW profile, the mass scale $M_{\rm NFW}$ is related to the total mass $M_{\rm sub}$ of a substructure via the relation \cite{Baltz:2007vq}
\be\label{eq:total_tNFW_mass}
M_{\rm sub} = M_{\rm NFW} \frac{\tau^2}{(\tau^2+1)^2}\left[(\tau^2-1)\ln{\tau}+\pi\tau-(\tau^2+1)\right],
\ee
where $\tau \equiv r_{\rm t}/r_{\rm s}$. We take the relation between the substructure mass and its scale radius to be (see Appendix \ref{app:scale_and_trunc})
\be\label{eq:r_s_vs_M_sub}
\left(\frac{r_{\rm s}}{1\,{\rm kpc}}\right) = (1.0\pm0.3) \left(\frac{M_{\rm sub}}{10^9 M_{\odot}}\right)^{0.735},
\ee
where we have taken into account the scatter in this relation as inferred by $N$-body simulations \cite{Garrison-Kimmel:2014vqa}. We also take the tidal truncation radius to obey the standard relation \cite{Binney2008,Tormen:1997ik} 
\be\label{eq:truncation1}
r_{\rm t} =  \left(\frac{M_{\rm sub}}{[2-{\rm d}\ln{M_{\rm main}}/{\rm d}\ln{r_{\rm 3D}}]M_{\rm main}(<r_{\rm 3D})}\right)^{1/3}r_{\rm 3D},
\ee
where $r_{\rm 3D}$ is the three-dimensional distance between the mass substructure and the center of the main lens halo and $M_{\rm main}(<r_{\rm 3D})$ is the fraction of the mass of the main halo contained in a sphere of radius $r_{\rm 3D}$. For a spherical main lens with a power-law convergence profile
\be
\kappa_{\rm main}(r) = \frac{1}{2}\left(\frac{b}{r}\right)^{2-\alpha_{\rm main}},\qquad (\alpha_{\rm main}\neq2)
\ee
where $b$ is a length scale closely related to the Einstein radius of the main lens, $r$ is the projected two-dimensional distance from the center of the lens, and $\alpha_{\rm main}$ is the power-law index of the density profile, the tidal truncation radius takes the form (see Appendix \ref{app:scale_and_trunc} for more details)
\be\label{eq:truncation2}
r_{\rm t} =  \left(\frac{\alpha_{\rm main}}{2-\alpha_{\rm main} }\frac{ \Gamma\left(\frac{2-\alpha_{\rm main}}{2}\right)}{ \Gamma\left(\frac{3-\alpha_{\rm main}}{2}\right)} \frac{ M_{\rm sub}}{2\sqrt{\pi}\Sigma_{\rm crit}b^2}\right)^{1/3}\left(\frac{b}{r_{\rm 3D}}\right)^\frac{\alpha_{\rm main}}{3}r_{\rm 3D}, \qquad (\alpha_{\rm main}\neq2),
\ee
where $\Gamma(x)$ is the gamma function and where $\Sigma_{\rm crit}$ is the critical density for lensing. The substructures are taken to be spatially distributed in two-dimensional projections according to a ``cored" profile for $0<r<R_{\rm max}$ given by
\be\label{eq:sub_spatial_dist}
 \mathcal{P}_r(r,\theta)  =\left( \frac{1}{2\pi r_{\rm c}^2}\frac{1}{W(R_{\rm max}/r_{\rm c})-1}\right)\frac{1}{(1+ (r/r_{\rm c}))^2},\quad\text{where}\quad W(x) = \frac{1}{1+x}+\ln{(1+x)},
 \ee
and where $r_{\rm c}$ is the core radius of the substructure distribution. This spatial distribution profile constitutes a good approximation to the radial substructure distribution found in $N$-body simulations \cite{Springel:2008cc}.  The core radius $r_{\rm c}$ is found to be a large fraction of the main halo virial radius. Here, we take $r_{\rm c} = 30 R_{\rm ein}$, where $R_{\rm ein}$ is the Einstein radius of the smooth lens. For a typical galaxy-scale gravitational lens with $R_{\rm ein}\sim 1''$, this gives $r_{\rm c}\sim 189$ kpc for a lens at redshift $z_{\rm lens}\sim0.5$. We define the boundary between the local and distributed population to lie at $R_{\rm min} = 3 R_{\rm ein}$ and also choose $R_{\rm max} = 65 R_{\rm ein}$. We note that $r_{\rm 3D}$ is related to $r$ via $r_{\rm 3D}=\sqrt{r^2+h^2}$, where $h$ is the projection of $r_{\rm 3D}$ along the line of sight, which must lie in the range $-\sqrt{R_{\rm max}^2-r^2}\leq h\leq \sqrt{R_{\rm max}^2-r^2}$ for a spherical halo. When we generate the Monte Carlo realization, we first choose $r$ from the probability distribution in Eq.~(\ref{eq:sub_spatial_dist}), and then randomly pick $h$ from the above range in order to generate the value of $r_{\rm 3D}$. We note though that in a realistic halo, the values of $h$ will not in general be \emph{uniformly} distributed within the above range. However, since $h$ only enters in the calculation of the truncation radius, the impact of this approximation on our results is very small.

We take the substructure to be distributed in mass according to a power-law mass probability distribution
\be\label{eq:subhalo_mass_function}
\mathcal{P}_M(M_{\rm sub}) \equiv \frac{1}{N}\frac{dN}{dM_{\rm sub}}= \frac{(\beta+1) M_{\rm sub}^\beta}{M_{\rm high}^{\beta+1} -M_{\rm low}^{\beta+1}},\qquad M_{\rm low} < M_{\rm sub} < M_{\rm high},\qquad (\beta\neq-1),
\ee
where $\beta$ is the power law index, and where $M_{\rm high}$ and $M_{\rm low}$ are the highest and lowest subhalo masses inside the lens galaxy, respectively. As was found numerically in Ref.~\cite{Springel:2008cc}, we take $\beta=-1.9$. We also choose $M_{\rm high} = 10^{10} M_\odot$ and $M_{\rm low}=10^7 M_\odot$. While $M_{\rm low}$ is typically much lower in standard cold dark matter models \cite{Hofmann:2001bi,Green:2003un}, this latter choice ensures that the number of mass substructures inside the lens galaxy is manageable within our Monte Carlo realizations.  The actual number of mass substructures indeed the lens galaxy is taken to be Poisson distributed around a mean value given by 
\be\label{eq:average_N}
\langle N \rangle= \frac{\langle\kappa_{\rm sub}(R_{\rm ein})\rangle}{ \int d M_{\rm sub} \int r dr d\theta\, \mathcal{P}_M(M_{\rm sub})  \mathcal{P}_r(r,\theta) \kappa_{\rm tNFW}(|{\bf r} - {\bf R}_{\rm ein}|)}
\ee  
where the angular bracket denotes ensemble averaging over many substructure realizations of the lens halo and $ \kappa_{\rm tNFW}({\bf r})$ is the convergence profile of a single smoothly truncated NFW subhalo as given in Ref.~\cite{Baltz:2007vq}. Equation~(\ref{eq:average_N}) follows from the independence of subhalos within the lens galaxy. We take the average convergence in mass substructures at the Einstein radius of the main lens to be $\langle\kappa_{\rm sub}(R_{\rm ein})\rangle =  0.001$. We note that setting $\langle\kappa_{\rm sub}\rangle$ as above is equivalent to choosing an overall normalization for the subhalo mass function [see Eq.~(\ref{eq:kappa_sub}) below for more details].
\begin{figure}[t]
\begin{center}
\includegraphics[width=0.49\textwidth]{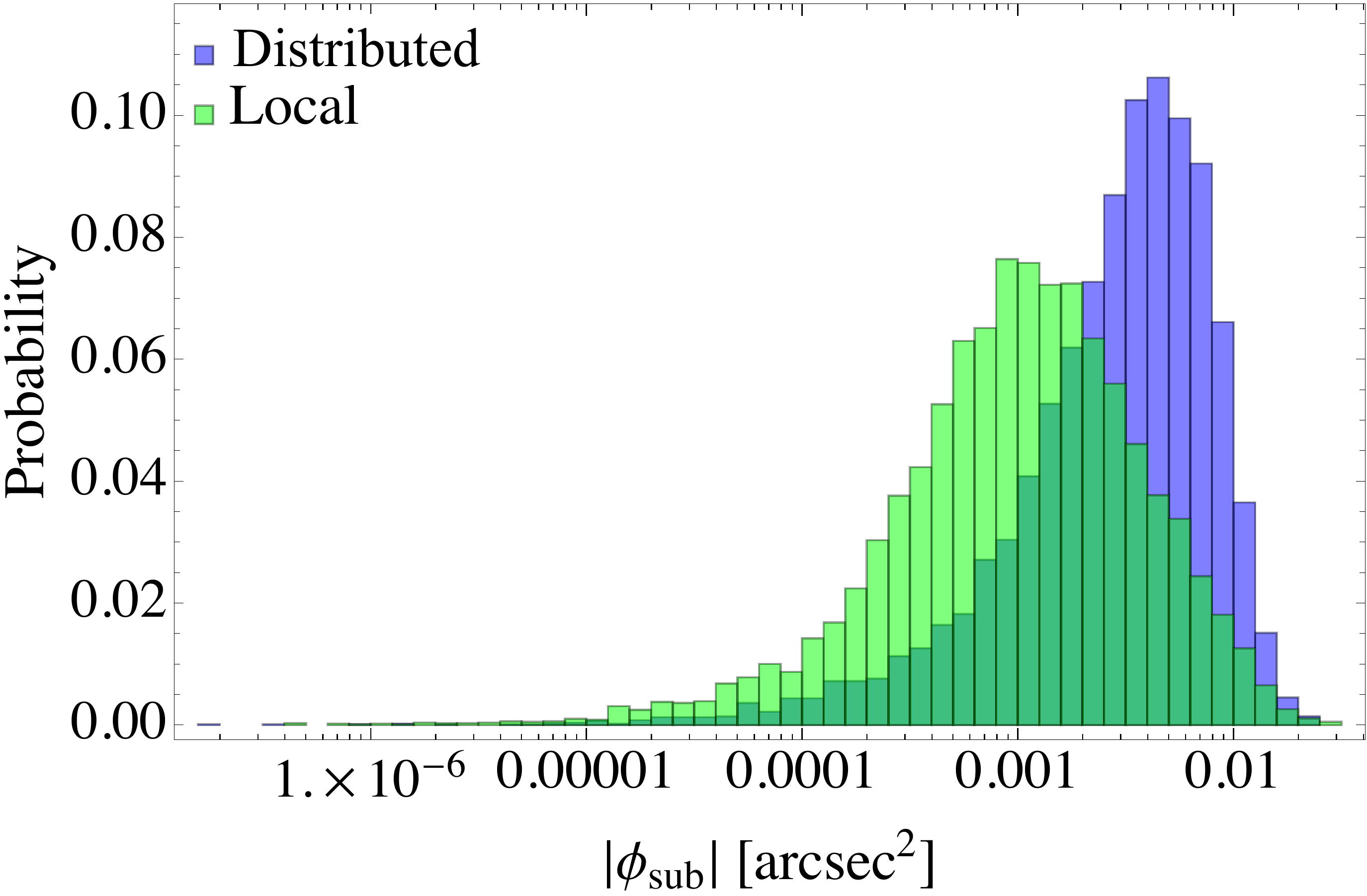}
\includegraphics[width=0.49\textwidth]{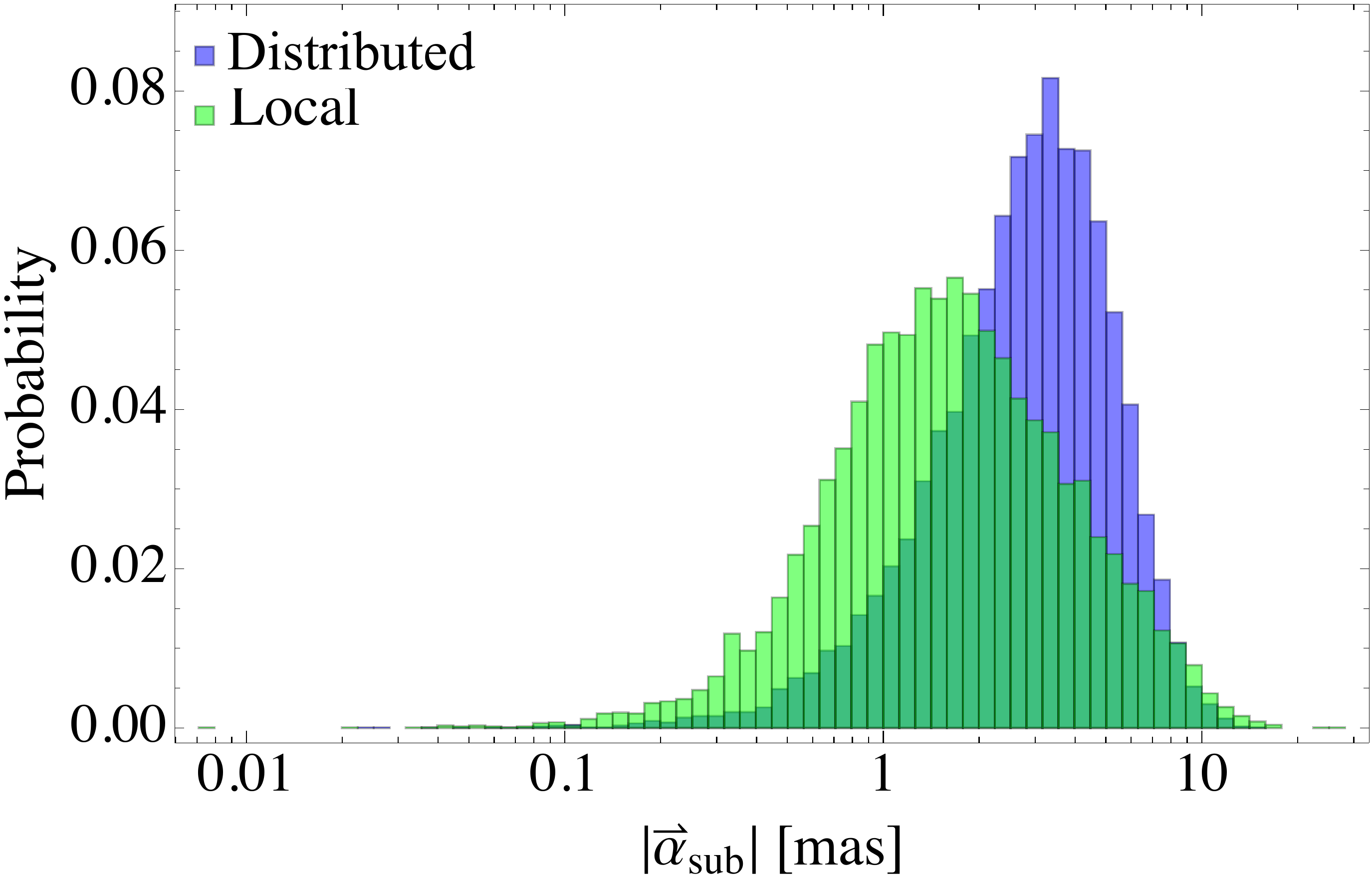}\\
\includegraphics[width=0.49\textwidth]{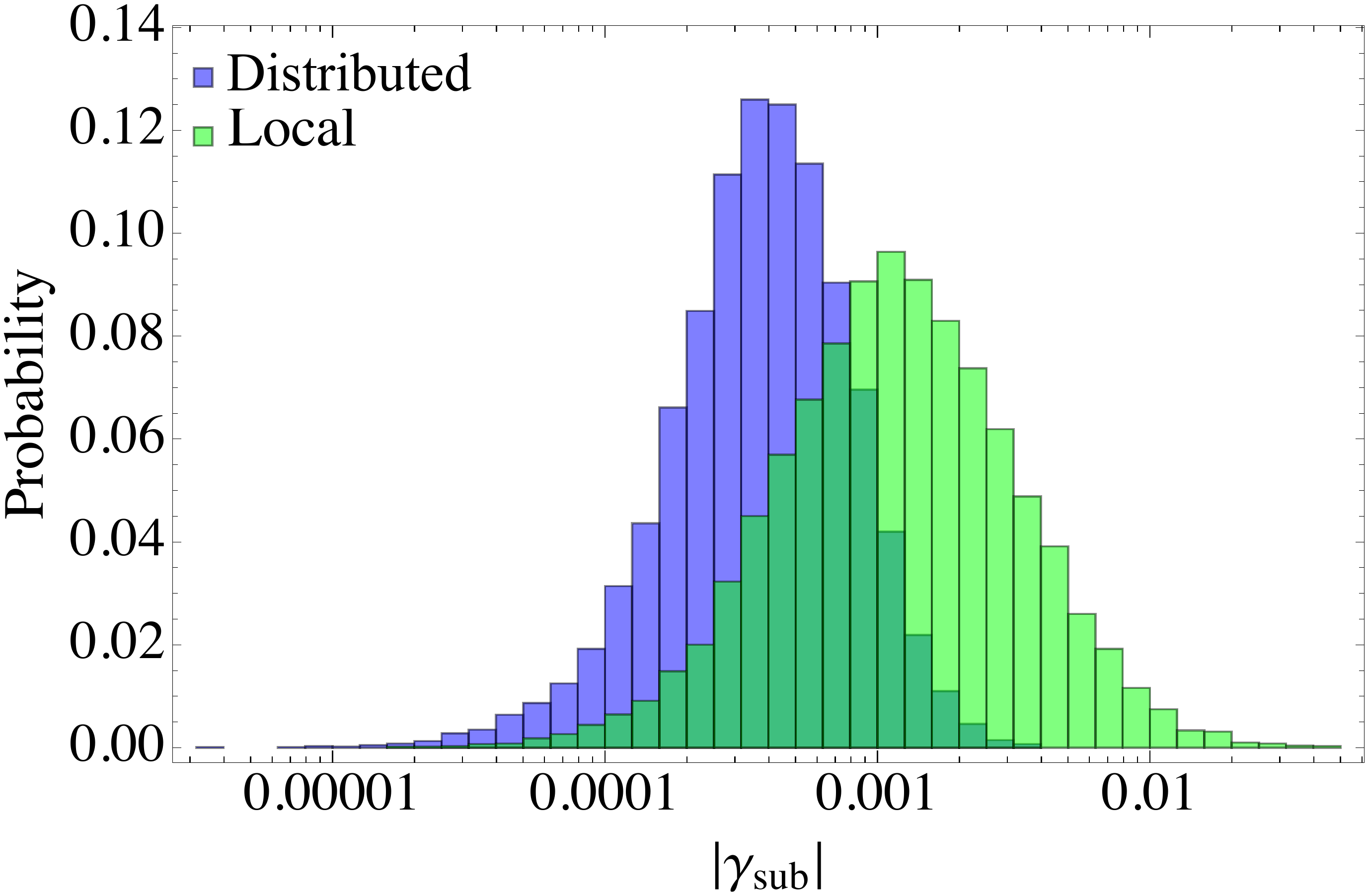}
\includegraphics[width=0.49\textwidth]{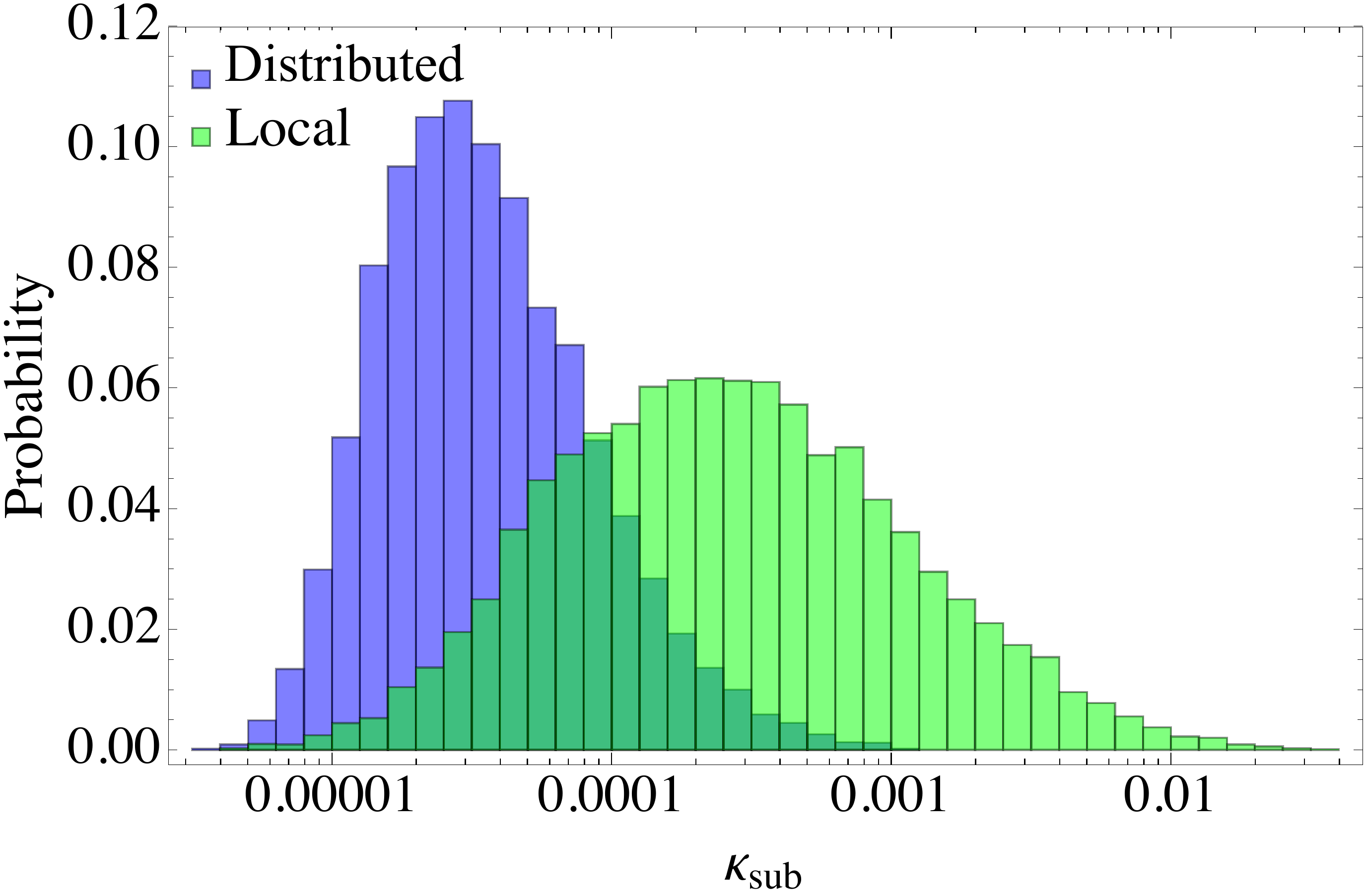}
\caption{Probability distributions for the local and distributed contributions to the linear lensing quantities $\phi_{\rm sub}$, $\alpha_{\rm sub}$, $\gamma_{\rm sub}$, and $\kappa_{\rm sub}$. These quantities are evaluated at the Einstein radius of the main lens, which we take to be $R_{\rm ein}=1''$.  We assume the lens to be at redshift $z_{\rm lens} =0.5$ with a source at redshift $z_{\rm source}=1$, yielding a critical density for lensing $\Sigma_{\rm crit} = 1.19\times 10^{11} M_\odot/{\rm arc sec}^2$. We define the divide between the distributed and local contributions to lie at $R_{\rm min} = 3R_{\rm ein}$ and include mass substructures up to $R_{\rm max} = 65 R_{\rm ein}$.  The substructures are spatially distributed according to a cored profile [Eq.~(\ref{eq:sub_spatial_dist})] with core radius $r_{\rm c} = 30 R_{\rm ein}$. Mass substructures are taken to have a smoothly truncated NFW profile with tidal truncation radius that depends on subhalo mass and halo-centric distance as given in Eqs.~(\ref{eq:truncation1}) and (\ref{eq:truncation2}). We use a  subhalo mass-concentration relation derived from $N$-body simulations \cite{Garrison-Kimmel:2014vqa} and also implement the scatter about this relation [see Eq.~(\ref{eq:r_s_vs_M_sub})]. We assume a power law subhalo mass function with slope $\beta =-1.9$ between $M_{\rm low} = 10^7M_\odot$ and $M_{\rm high} = 10^{10} M_\odot$.  We take the average lensing convergence in mass substructure at the Einstein radius to be $\langle\kappa_{\rm sub}(R_{\rm ein})\rangle =  0.001$.}
\label{fig:distant_vs_local}
\end{center}
\end{figure}

We illustrate in Fig.~\ref{fig:distant_vs_local} the probability distributions for both the distributed and local contributions to the total linear lensing quantities for our $10^4$ Monte Carlo realizations. For lensing potential and deflection fluctuations, we observe that the most \emph{probable} fluctuations are dominated by the distributed substructures. This result can be explained by a simple geometrical argument. Indeed, while the contribution to the net deflection from substructures inside a thin ring of radius $r$ decays as $1/r$ for increasing $r$, the number of mass substructures inside the thin ring grows as $r$ for $r<r_{\rm c}$. Thus, inside the core radius of the substructure distribution, mass substructures located in a distant ring can contribute just as much to the total deflection as substructures much closer to the lensed images. A similar argument applies to lensing potential fluctuations. This indicates that detailed analyses of time delays and astrometric fluctuations caused by mass substructures can yield important information about the distributed population of satellite surrounding lens galaxies.

On the other hand, the substructure contribution to the shear and convergence (which determine the magnification perturbations) at a typical image position is largely dominated by the local subhalos. This is due to the fact that shear perturbations decay as $r^{-2}$, while convergence fluctuations decay even faster ($r^{-4}$ for our choice of truncated NFW profile). It implies that the contribution from distant rings of substructures is always subdominant compared to the local contribution, although the collective shear perturbations from the distributed substructures is not entirely negligible. Furthermore, for deflections, shears, and convergence, the largest fluctuations are always dominated by the local contributions. These large local perturbations, often caused by a single substructure close to a lensed image, have been used to detect individual mass substructures \cite{Vegetti:2009aa,2014MNRAS.442.2017V,Nierenberg:2014aa}. What Fig.~\ref{fig:distant_vs_local} is showing however is that by combining magnification information (largely sensitive to $\gamma_{\rm sub}$ and $\kappa_{\rm sub}$) with astrometric (sensitive to $\alpha_{\rm sub}$) and time delay (sensitive to $\phi_{\rm sub}$) measurements, one could infer important properties about \emph{both} the local and distributed substructure populations inside lens galaxies. This highlights the importance of developing a unified framework to jointly handle the different lensing observables, which is a major goal of this work.

Splitting the mass substructures into two subpopulations allows us to factorize the characteristic function $Q_{\langle N\rangle}$ given in Eq.~(\ref{eq:QN}) as 
\be
Q_{\langle N\rangle }(\kk_{\rm L}|{\bf q},\qsub) = Q^{\rm local}_{\langle N_{\rm l}\rangle }(\kk_{\rm L}|{\bf q},\qsub)Q^{\rm dist}_{\langle N_{\rm d}\rangle }(\kk_{\rm L}|{\bf q},\qsub),
\ee
which follows from Eq.~(\ref{eq:local+distant}) and the independence of each substructure. Here, $\langle N\rangle = \langle N_{\rm l}\rangle + \langle N_{\rm d}\rangle $, where $ \langle N_{\rm l}\rangle$ is the average number of substructures in the local population, and where $\langle N_{\rm d}\rangle $ is the average number of substructures in the distributed population. We note that we generically have $\langle N_{\rm d}\rangle \gg \langle N_{\rm l}\rangle$. In terms of the characteristic function for a single subhalo, this implies
\be\label{eq:q1_loc+q1_dis}
q_1(\kk_{\rm L}|{\bf q},\qsub) = \frac{\langle N_{\rm l}\rangle q_1^{\rm local}(\kk_{\rm L}|{\bf q},\qsub) + \langle N_{\rm d}\rangle q_1^{\rm dist}(\kk_{\rm L}|{\bf q},\qsub) }{\langle N\rangle}.
\ee
We can therefore separately compute the characteristic function for the local and distributed subpopulation and then combine them according to Eq.~(\ref{eq:q1_loc+q1_dis}) to compute the overall characteristic function of linear lensing quantities. In this work, we focus on statistically characterizing the distributed population of mass substructures inside typical lens galaxies, which is the dominant contribution for the projected lensing potential and deflections. We leave to future work the characterization of the local substructure population, but we note that gravitational imaging techniques \cite{Koopmans:aa,Vegetti:2008aa,Vegetti:2009aa,2014MNRAS.442.2017V} and resolved spectroscopy \cite{Moustakas:2003aa,Hezaveh:2012ai,Hezaveh:2014aa} can provide information about certain regions of the local substructure population. 

\subsection{Distributed substructure analysis for potential and deflection perturbations}\label{sec:distant_analysis}
In this section, we outline our calculations of the characteristic function $q_1^{\rm dist}(\kk_{\rm L}|{\bf q},\qsub)$ for the distributed population of mass substructures. We focus exclusively on computing the characteristic function for the projected lensing potential and the deflection perturbations since the contribution to shear and convergence perturbations from the distributed population of mass substructures is subdominant. As described above, there are key simplifying facts for the distributed substructure population:
\begin{itemize}
\item Their overall impact on the lensing observables is small.
\item We can approximate them as a collection of point masses. 
\end{itemize}
In the point-mass approximation, a single subhalo can be described by three parameters: its total mass $M_{\rm sub}$ and its radial and angular position in the lens plane. In the notation from Sec.~\ref{sec:stoc_lensing_gen_case}, this implies ${\bf c}_{\rm sub}^{(1)} = \{M_{\rm sub},r,\theta\}$. In order to construct the characteristic function, we need to specify the probability density function $P_{\rm sub}(M_{\rm sub},r,\theta;\qsub)$ for these parameters. As in our Monte Carlo examples of Sec.~\ref{sec:local_vs_distant}, we assume that this density function is separable into the product of the subhalo mass function with a spatial density distribution
\be\label{eq:separability}
P_{\rm sub}(M_{\rm sub},r,\theta;\qsub) = \mathcal{P}_M(M_{\rm sub};\qsub) \mathcal{P}_r(r,\theta;\qsub).
\ee
For cold dark matter, this separability is supported by $N$-body simulations over a wide range of subhalo masses \cite{Ludlow:2008qf,Springel:2008cc,Diemand:2004kx}. It remains to be seen whether this separability holds for more general dark matter models or when baryonic feedback is taken into account. In the cases for which the mass and spatial distributions are not separable, one could split the subhalo population into several subpopulations that each have with their own spatial distribution.  For simplicity, we assume here that Eq.~(\ref{eq:separability}) is valid, but it is clear that our analysis could also be carried out without this assumption. As we see below, we do not need to specify an explicit form for the subhalo mass function and position distribution at this point since all important quantities can be expressed as statistical moments of these distributions.

Before going through the detailed derivation of $q_1^{\rm dist}(\kk_{\rm L}|{\bf q},\qsub)$, it is informative to heuristically derive what we expect the answer to be.  As we discussed in the previous subsection, potential and deflection perturbations obtain contributions from a broad spatial projected area surrounding the strong lensing region. The resulting large number of mass substructures contributing to the total potential and deflection perturbations implies that the central limit theorem is applicable, and we thus expect Gaussian statistics to be approximately valid for these linear lensing quantities. In this approximation, the statistical properties of the linear lensing quantities are entirely specified by a covariance matrix ${\bf C}_{\rm sub}$ with general scaling given by
\be
{\bf C}_{\rm sub}^{ij} \propto \langle N_{\rm d}\rangle \left[ \int dm\, \mathcal{P}_m(m;\qsub) m^2\right]\left[\int d^2{\bf r} \mathcal{P}_r(r,\theta;\qsub) \mathcal{O}_{\rm L}^i \mathcal{O}_{\rm L}^j \right],
\ee
where the leading factor arises since the variance of the sum of $\langle N_{\rm d}\rangle$ normal random variables is $\langle N_{\rm d}\rangle$ times the variance of a single normal random variable. The second factor arises because the linear lensing quantities are always proportional to the subhalo mass, and the third factor is the spatial two-point function of the linear lensing quantities. Here, $\mathcal{O}_{\rm L}^i$ stands for the spatial dependence of the $i$th linear lensing quantity. As we see below, this scaling comes out naturally of our analysis.

We now turn our attention to the detailed derivation of the above scaling as well as the leading order deviations from the Gaussian approximation. The lensing potential difference $\phi_{\rm sub}(\xx_i)$ between an image position $\xx_i$ and a reference point $\xx_{\rm ref}$ caused by a point mass $M$ at position $\xx$ is given by
\be
\phi_{\rm sub} (\xx_i)= m\ln{\left[\frac{|\xx_i-\xx|}{|\xx_{\rm ref}-\xx|}\right]},
\ee
where $m\equiv M_{\rm sub}/(\pi\Sigma_{\rm crit})$. Since $\Sigma_{\rm crit}$ is the critical mass density for lensing, $m$ has units of area.  Since $|\xx_i|\sim R_{\rm ein}\ll|\xx|$ for a typical distributed dark matter substructure, we can write down the lensing potential difference at an image location as a multipole expansion. Converting to polar coordinates with $\xx=(r\cos{\theta},r\sin{\theta})$, we obtain
\be\label{eq:key_pot_expansion}
\phi_{\rm sub} (\xx_i) = - m \sum_{p=1}^\infty \frac{1}{r^p}\left[A_p(\xx_i)\cos{(p\,\theta)} + B_p(\xx_i) \sin{(p\,\theta)}\right],
\ee
where the dimensionless series coefficients are
\be
A_p(\xx_i) =  \frac{1}{p}\left(r_i^p \cos{(p\,\theta_i)}-r_{\rm ref}^p \cos{(p\,\theta_{\rm ref})}\right), \qquad B_p(\xx_i) =  \frac{1}{p}\left(r_i^p \sin{(p\,\theta_i)}-r_{\rm ref}^p \sin{(p\,\theta_{\rm ref})}\right),
\ee
where we have used $\xx_i = (r_i\cos{\theta_i},r_i\sin{\theta_i})$ and $\xx_{\rm ref} = (r_{\rm ref}\cos{\theta_{\rm ref}},r_{\rm ref}\sin{\theta_{\rm ref}})$. Since the deflections are simply related to the lensing potential by derivatives, that is, $\vec{\alpha}_{\rm sub}(\xx_i)  = \vec{\nabla}_{\xx_i} \phi_{\rm sub}$, we can write expansions similar to Eq.~\eqref{eq:key_pot_expansion} for each of these quantities. The only difference is that the series coefficients for $\vec{\alpha}_{\rm sub}$ are derivatives of $A_p(\xx_i)$ and $B_p(\xx_i)$. Taking $\vec{\mathcal{O}}_{\rm L}\equiv\Delta\tl/m$ to denote the vector containing all the stochastic random variables describing the perturbations to the linear lensing quantities, we can thus write
\be\label{eq:def_O}
\vec{\mathcal{O}}_{\rm L} =  - \sum_{p=1}^\infty \frac{1}{r^p}\left[\vec{A}_p\cos{(p\,\theta)} + \vec{B}_p \sin{(p\,\theta)}\right] .
\ee
We note that we have divided out the leading factor of the subhalo mass in the above definition since it only leads to an overall rescaling of $\vec{\mathcal{O}}_{\rm L}$.  In general, $\vec{\mathcal{O}}_{\rm L}$ would contain the stochastic variables $\phi_{\rm sub}^{(i)}$ and $\vec{\alpha}_{\rm sub}^{(i)}$  for the lensing potentials and deflections, respectively, evaluated at all possible image positions $i\in N_{\rm img}$. For instance, in the case of a single image with label $i$, we have $\vec{\mathcal{O}}_{\rm L}^{(i)} = \frac{1}{m}\left\{\phi_{\rm sub}^{(i)},\alpha_{{\rm sub}, x}^{(i)},\alpha_{{\rm sub}, y}^{(i)}\right\}$ and
\ba
\vec{A}_p &=& \Bigg(\frac{r_i^p\cos{\{p \theta_i\}}-r_{\rm ref}^p\cos{\{p\theta_{\rm ref}\}}}{p},r_i^{p-1}\cos{\{(p-1) \theta_i}\},-r_i^{p-1}\sin{\{(p-1) \theta_i}\}\Bigg),
\ea
\ba
\vec{B}_p &=&   \Bigg(\frac{r_i^p\sin{\{p \theta_i\}}-r_{\rm ref}^p\sin{\{p \theta_{\rm ref}\}}}{p},r_i^{p-1}\sin{\{(p-1) \theta_i}\},r_i^{p-1}\cos{\{(p-1) \theta_i}\}\Bigg),
\ea
where $p \geq1$. We emphasize that $\vec{A}_p $ and $\vec{B}_p$ are constant vectors that only depend on the configuration of lensed images and are thus independent of the mass substructure population. Taking $\kk_{\rm L}$ to be the Fourier conjugate of the stochastic vector $\vec{\mathcal{O}}_{\rm L} $, the characteristic function for a single dark matter substructure can be written as
\ba
q_1^{\rm dist}(\kk_{\rm L}|\qq,\qsub) &=& \int_{\cH_{\rm d}} d^2r \, \int dm\, e^{i\, m\, \kk_{\rm L}\cdot \vec{\mathcal{O}}_{\rm L} } P_{\rm sub}(m,r,\theta;\qsub)\en
&=&1+\int_{\cH_{\rm d}} d^2r \, \int dm\, (e^{i\, m\, \kk_{\rm L}\cdot \vec{\mathcal{O}}_{\rm L} } -1)P_{\rm sub}(m,r,\theta;\qsub)
\ea
where $\cH_{\rm d}$ denotes the area of the distributed domain of the lens halo and where we pulled out the leading factor of unity since we are only interested in the difference $q_1(\kk_{\rm L}|\qq,\qsub)-1$. Evaluating the above integrals is the most difficult part of the calculation. Clearly, for $|\kk_{\rm L}|\ll 1 /(m |\vec{\mathcal{O}}_{\rm L}|)$, the phase factor is nearly equal to unity and $q_1 \rightarrow 1$, while $q_1\rightarrow 0$ for  $|\kk_{\rm L}|\gg 1 /(m |\vec{\mathcal{O}}_{\rm L}|)$ since the phase is rapidly oscillating in this regime.  We expand $q_1(\kk_{\rm L}|\qq,\qsub)$ in a power series of mass and spatial moments
\ba\label{eq:moment_exp}
q_1^{\rm dist}(\kk_{\rm L}|\qq,\qsub) 
&=&1+ \sum_{n=1}^\infty\frac{i^n\langle m^n\rangle}{n!}\int_{\cH_{\rm d}} d^2r \mathcal{P}_r(r,\theta;\qsub)  ( \kk_{\rm L}\cdot \vec{\mathcal{O}}_{\rm L} )^n,
\ea
where the mass moments are given by
\be
\langle m^n\rangle\equiv \int dm\, \mathcal{P}_m(m;\qsub) m^n.
\ee
For conciseness, the simplification of the spatial integral appearing in Eq.~\eqref{eq:moment_exp} is presented in Appendix \ref{app:spatial_momts}. After these simplifications, the characteristic function for the linear quantities $\vec{\mathcal{O}}_{\rm L} $ in the presence of a single mass substructure then takes the form
\be
q_1^{\rm dist}(\kk_{\rm L}|\qq,\qsub) =1+ \sum_{n=1}^\infty\frac{(-1)^n\langle m^n\rangle}{n!}\left(\sum_{\Vert p\Vert =n} \binom{n}{\pp{}} K_{ \pp{}} \prod_{t=1}^{N_{\rm mult}} (i\kk_{\rm L}\cdot\vec{A}_t)^{p_t}\prod_{s=1}^{N_{\rm mult}}(i\kk_{\rm L}\cdot\vec{B}_s)^{p_{N_{\rm mult}+s}}\right).
\ee
where $\pp{} = \{p_1,p_2,\ldots,p_{2 N_{\rm mult}}\}$ is a multi-index with $\Vert \pp{}\Vert = \sum_{j=1}^{2N_{\rm mult}} p_j$, and where the kernel $K_{\pp{}}$ is given by Eq.~\eqref{app:eq:spatial_kernel}. It is understood that if a given $\vec{A}_t$ or $\vec{B}_t$ vanishes, then the corresponding $p_t$ must also vanish. We emphasize that the kernel $K_{\bf p}$ encodes all the information about the spatial distribution of mass substructures within the lens halo. This kernel can be computed for any halo geometry and mass substructure distribution.  Finally, we can use Eq.~(\ref{eq:QN}) to compute the characteristic function in the presence of a whole population of mass substructures
\be\label{eq:nongaussian_terms}
Q_{\langle N_{\rm d}\rangle }^{\rm dist}(\kk_{\rm L}|{\bf q},\qsub) = \exp{\left[\langle N_{\rm d}\rangle\sum_{n=1}^\infty\frac{(-1)^n\langle m^n\rangle}{n!}\left(\sum_{\Vert p\Vert =n} \binom{n}{\pp{}} K_{ \pp{}} \prod_{t=1}^{N_{\rm mult}} (i\kk_{\rm L}\cdot\vec{A}_t)^{p_t}\prod_{s=1}^{N_{\rm mult}}(i\kk_{\rm L}\cdot\vec{B}_s)^{p_{N_{\rm mult}+s}}\right)\right]}.
\ee
At leading order, this characteristic function has a Gaussian behavior, 
\be\label{eq:gaussian_terms}
Q_{\langle N_{\rm d}\rangle }^{\rm dist}(\kk_{\rm L}) \propto e^{i {\bf u}\cdot\kk_{\rm L}-\frac{1}{2}\kk_{\rm L}^T {\bf C}_{\rm sub} \kk_{\rm L}}, 
\ee
where ${\bf u} \equiv \langle \Delta\tl\rangle$ is the mean vector and ${\bf C}_{\rm sub} \equiv \langle \Delta\tl  \Delta\tl \rangle$ is the covariance matrix.
We note that in the case of circular symmetry of the galactic halo, the mean vector ${\bf u}$ exactly vanishes. We give in Appendix \ref{app:cov_mat} some useful expressions for the covariance matrix in the case of circular symmetry for two different spatial distributions. 

The non-Gaussian terms in Eq.~\eqref{eq:nongaussian_terms} essentially forms a multivariate Edgeworth expansion (see, e.g.~, Refs.~\cite{Blinnikov:1997jq,SJOS:SJOS091}) with successive term decaying as $\langle N_{\rm d}\rangle^{1-n/2}$. We show the details of this expansion in Appendix \ref{app:edgeworth_exp}, but it is instructive to consider the magnitude of the non-Gaussian contributions to $Q_{\langle N_{\rm d}\rangle }^{\rm dist}(\kk_{\rm L})$ in order to assess the validity of the leading Gaussian approximation. At each order $n$ in the $1/\langle N_{\rm d} \rangle^{n/2-1}$ expansion, the leading order non-Gaussian contribution takes the general form
\be\label{eq:non_gaussian_gen_term}
\frac{1}{n! \langle N_{\rm d} \rangle^{n/2-1}}\frac{\langle m^n\rangle}{\langle m^2\rangle^{n/2}}\frac{\langle\mathcal{O}_{\rm L}^n\rangle}{\langle\mathcal{O}_{\rm L}^2\rangle^{n/2}}, \qquad (n\geq3).
\ee
Here, we use the compact notation $\langle\mathcal{O}_{\rm L}^n\rangle$ to represent all possible spatial n-point functions of the different linear lensing quantities. In order to evaluate the above expression, we need to specify the mass function and spatial distribution of mass substructures. For illustration, we take the spatial distribution given in Eq.~\eqref{eq:sub_spatial_dist}, and write the mass function as
\be\label{eq:actual_mass_function}
\frac{dN}{dM_{\rm sub}} = a_0 \left(\frac{M_{\rm sub}}{M_0}\right)^\beta,
\ee
where $a_0$ is the mass function normalization and $M_0$ is a reference mass scale. Using Eq.~(\ref{eq:average_N}), the expected number of mass substructures in the distributed region is then
\be\label{eq:how_to_compute_N}
\langle N_{\rm d}\rangle = \frac{a_0}{M_0^\beta}\frac{M_{\rm high}^{\beta+1}-M_{\rm low}^{\beta+1}}{\beta+1}\left(1-\mathcal{P}_r(<R_{\rm min})\right),\qquad (\beta\neq-1)
\ee
where $\mathcal{P}_r(<R_{\rm min})$ is the cumulative probability of finding a mass substructure within a disk of radius $R_{\rm min}$. We take $M_0 = M_{\rm high}$ throughout this work. We note that in the point-mass limit, the convergence in mass substructures is related to the mass function given in Eq.~(\ref{eq:actual_mass_function}) via
\be\label{eq:kappa_sub}
 \langle\kappa_{\rm sub}({r}_{\rm ref})\rangle =\frac{a_0}{M_0^\beta}\frac{1}{\Sigma_{\rm crit}}\frac{M_{\rm high}^{\beta+2}-M_{\rm low}^{\beta+2}}{\beta+2}\mathcal{P}_r(r_{\rm ref}), \qquad (\beta\neq-2),
\ee  
where $r_{\rm ref}$ is a reference radius (e.g. $R_{\rm ein}$) where the convergence is evaluated.
\begin{figure}[t!]
\includegraphics[width=0.49\textwidth]{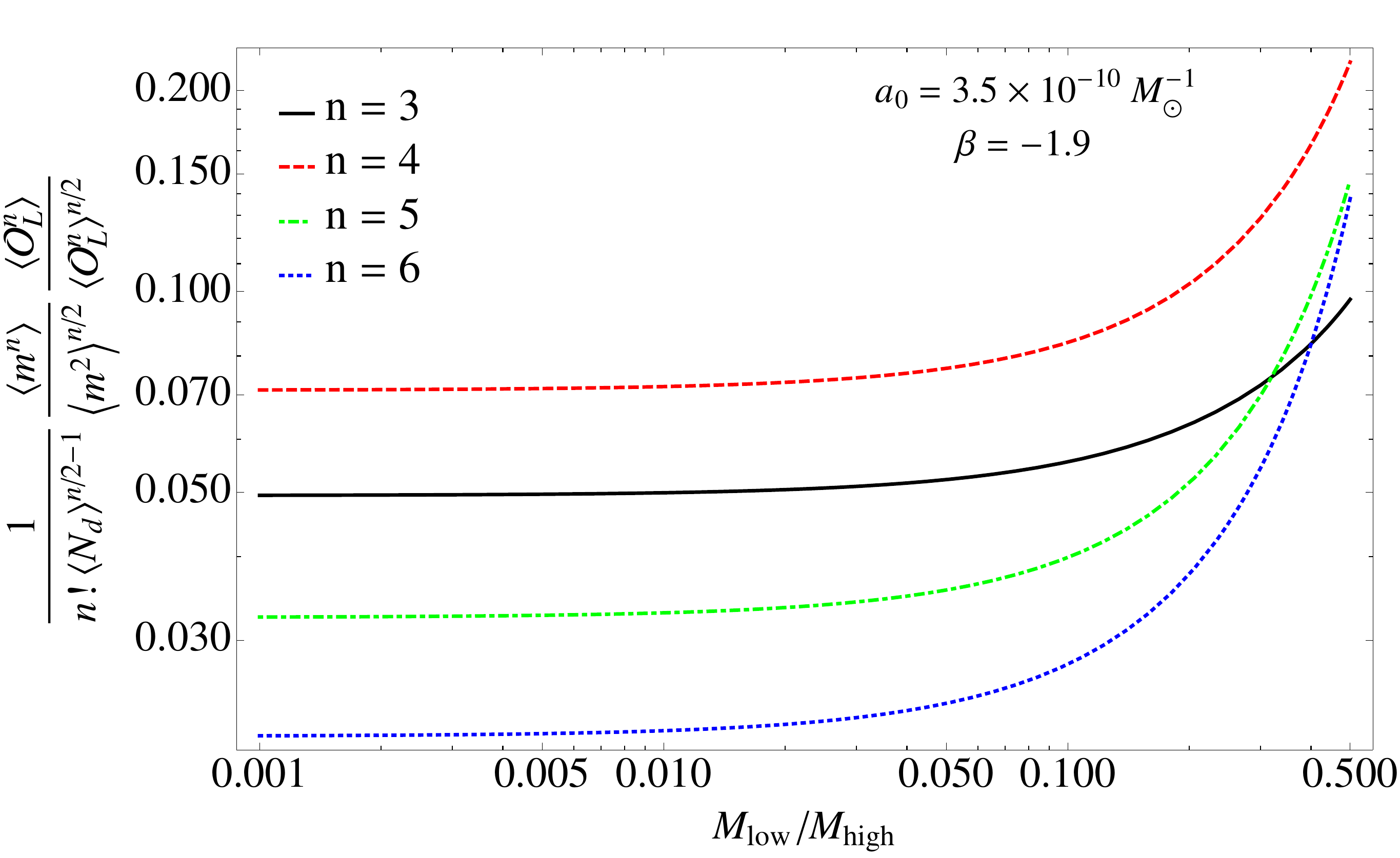}
\includegraphics[width=0.49\textwidth]{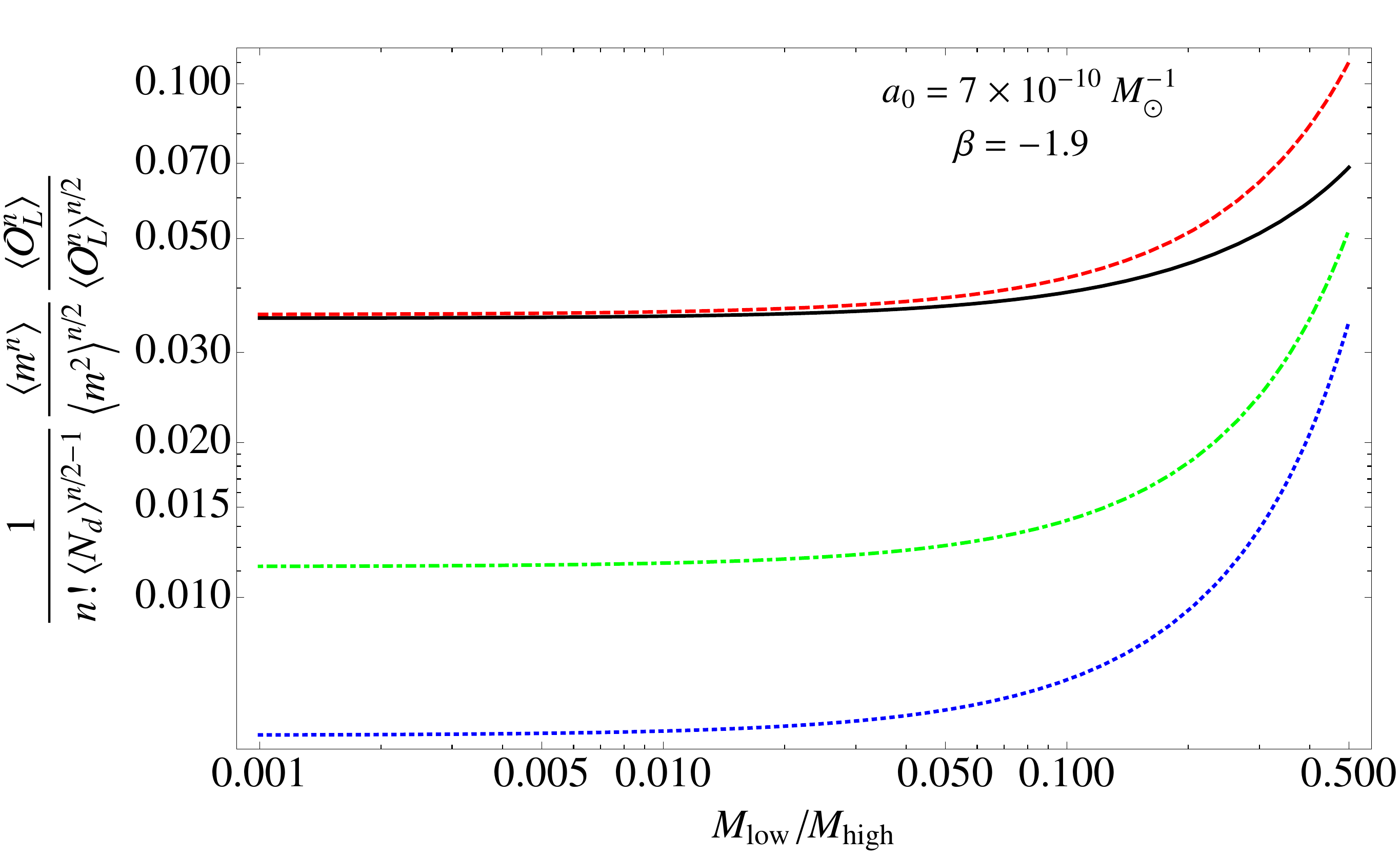}\\
\includegraphics[width=0.49\textwidth]{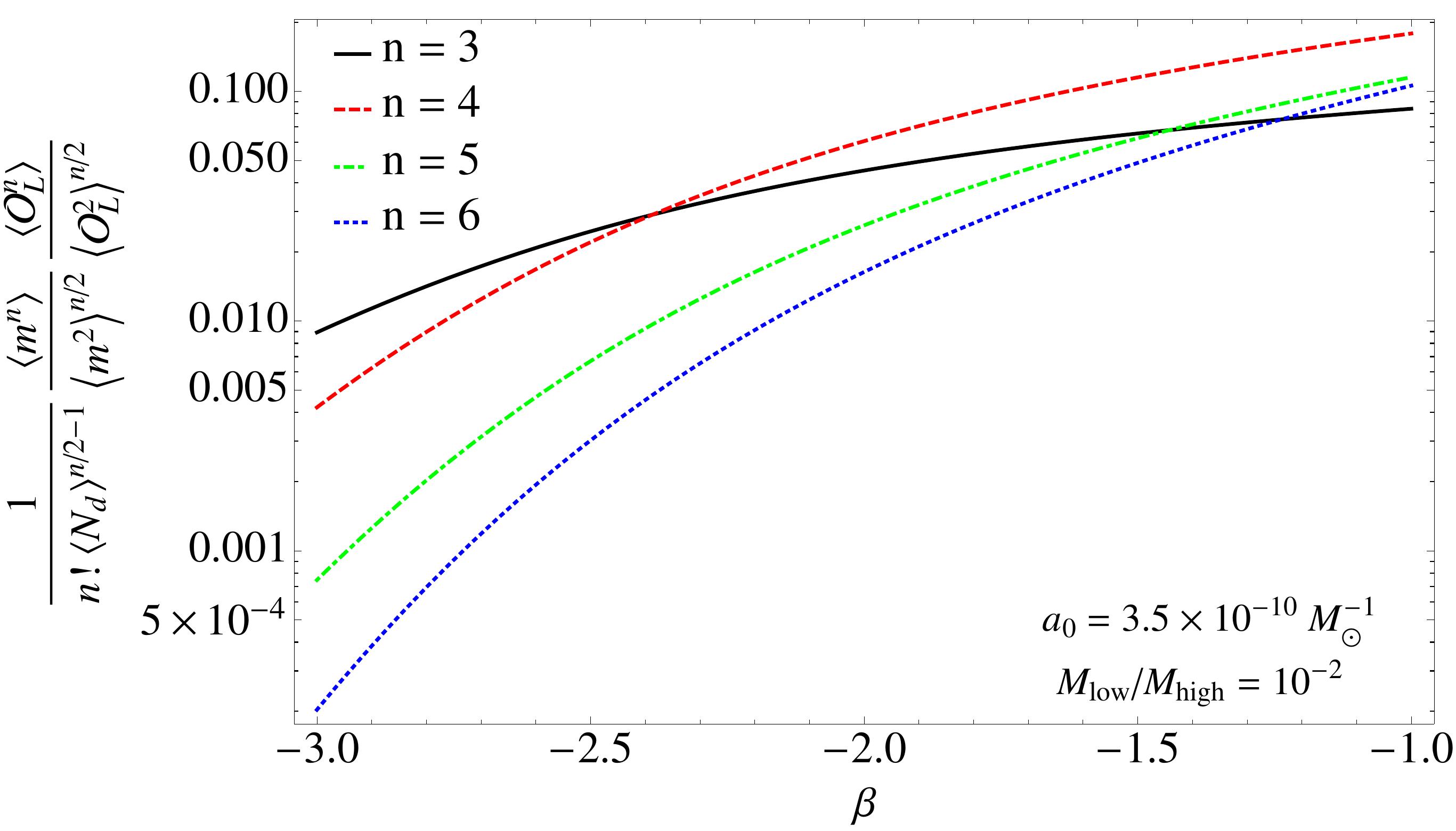}
\includegraphics[width=0.49\textwidth]{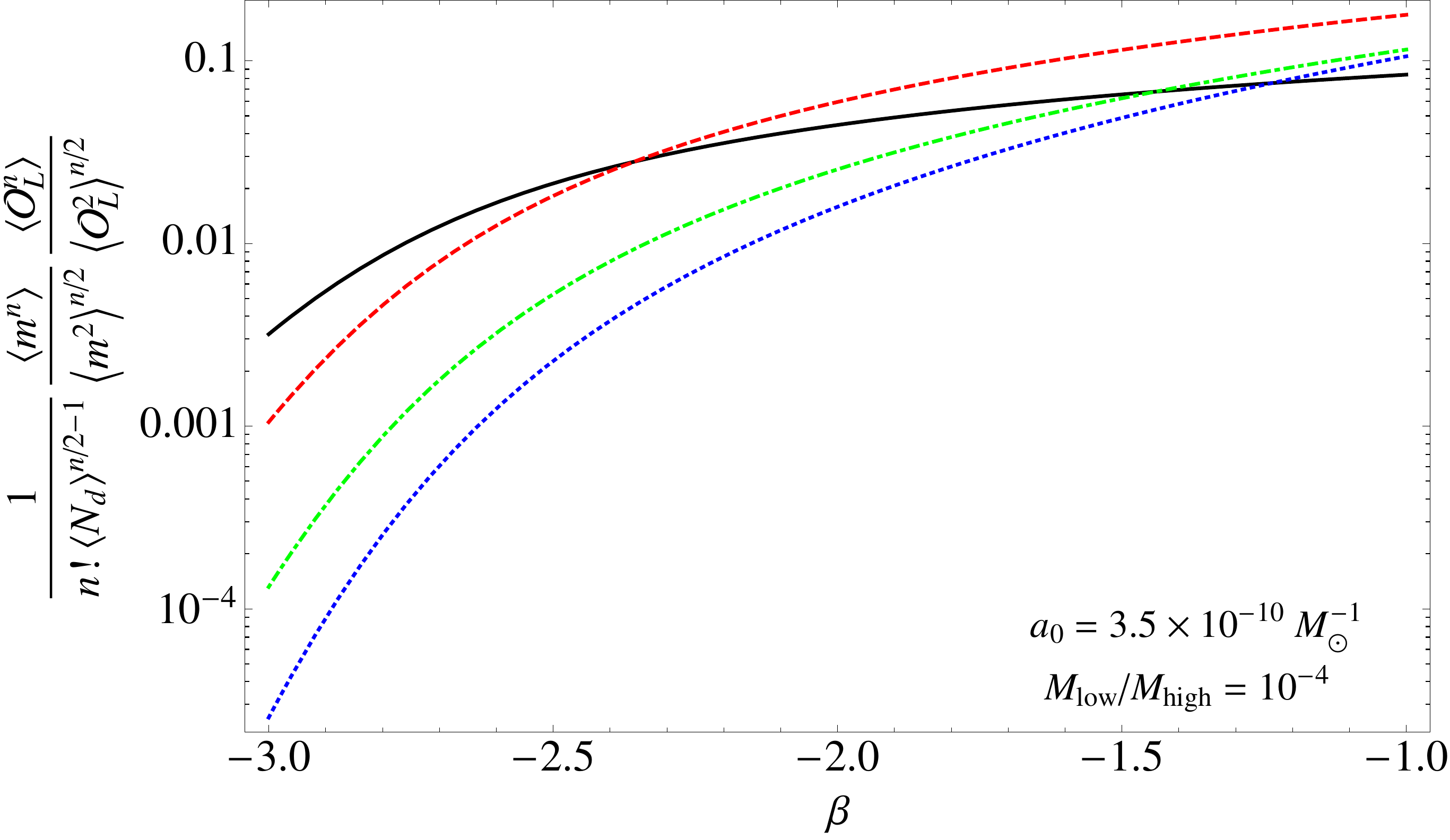}
\caption{Non-Gaussian contributions to the Edgeworth expansion of the characteristic function $Q_{\langle N_{\rm d}\rangle }^{\rm dist}(\kk_{\rm L})$ for different values of the mass function parameters. These curves characterize the degree of non-Gaussianity of the probability distribution of the linear lensing quantities. A value of unity on these plots indicate an $O(1)$ deviation from Gaussianity. We assume a power law mass function as given in Eq.~(\ref{eq:actual_mass_function}) with $M_{\rm high}=10^{10} M_\odot$ and also take the distributed mass substructures to be spatially located between $R_{\rm min} = 3R_{\rm ein}$ and $R_{\rm max} = 65 R_{\rm ein}$ according to Eq.~(\ref{eq:sub_spatial_dist}) with $r_{\rm c} = 30 R_{\rm ein}$. We illustrate the leading contribution at each order $n$ for $n = 3$ to $n=6$. Here, $\langle N_{\rm d} \rangle$ is computed as in Eq.~(\ref{eq:how_to_compute_N}). The spatial moments $\langle \mathcal{O}_{\rm L}^n\rangle$ are computed assuming that $ \mathcal{O}_{\rm L}$ is a deflection at a single image position, but similar results would be obtained for the lensing potential. Each panel illustrates different mass function parameters as indicated. The top panels fix $\beta = -1.9$ and display the dependence of the non-Gaussian corrections on the ratio $M_{\rm low}/M_{\rm high}$ for two different values of $a_0$. In the bottom panels, we fix $a_0$ and display the dependence on the mass function slope for two values of $M_{\rm low}/M_{\rm high}$ .}
\label{fig:nongaussian_moments}
\end{figure}

We illustrate in Fig.~\ref{fig:nongaussian_moments} the non-Gaussian contributions given in Eq.~(\ref{eq:non_gaussian_gen_term}) evaluated from $n=3$ to $n=6$ for different mass function parameters. Here, we take $\mathcal{O}_{\rm L}$ to represent a lensing deflection, but similar results would be obtained for the lensing potential. The upper panels illustrate the dependence of the non-Gaussian contributions on the lowest subhalo mass for two different values of the mass function normalization $a_0$ with $\beta=-1.9$. For comparison, pure cold dark matter simulations  yield $a_0\sim 3.8\times10^{-10} M_\odot^{-1}$ at the pivot point $M_0 = 10^{10} M_\odot$ with $\beta=-1.9$ \cite{Springel:2008cc}. We observe that for $M_{\rm low}/M_{\rm high} \lesssim 0.1$ the non-Gaussian contributions are subdominant for the fiducial values of $a_0 = 3.5\times10^{-10} M_\odot^{-1}$ and $\beta = -1.9$. Interestingly, the largest non-Gaussian contribution comes from the $n=4$ term, which implies that the probability density function of linear lensing quantities will primary pick up a nonzero excess kurtosis in this case. Further increasing the normalization of the subhalo mass function suppresses non-Gaussianities even more (upper right panel) since $\langle N_{\rm d}\rangle \propto a_0$. However, as $M_{\rm low}/M_{\rm high}\rightarrow 1$, the non-Gaussian contributions rapidly rise since the mass substructure population becomes dominated by a very limited number of massive subhalos and the applicability of the central limit theorem wanes.

The lower panels of Fig.~\ref{fig:nongaussian_moments} display the dependence of the non-Gaussian corrections on the slope of the substructure mass function. Here, we fix the ratio $M_{\rm low}/M_{\rm high}$ and the amplitude of the mass function at $M_0=M_{\rm high}$. We observe that as $\beta$ is made steeper (more negative) the non-Gaussian corrections rapidly decay since the number of mass substructures quickly rises with a steepening slope. Decreasing the ratio $M_{\rm low}/M_{\rm high}$ has little effect for $\beta>-2$ but does lead a faster decay of non-Gaussianities for $\beta<-2$. Again, we observe that the $n=4$ term dominates the non-Gaussianities when the mass function slope $\beta \gtrsim -2.3$ for the realistic normalization of the mass function shown. We confirm this observation by comparing our analytical results to Monte Carlo realizations in the next subsection.

\begin{figure}[b!]
\includegraphics[width=0.69\textwidth]{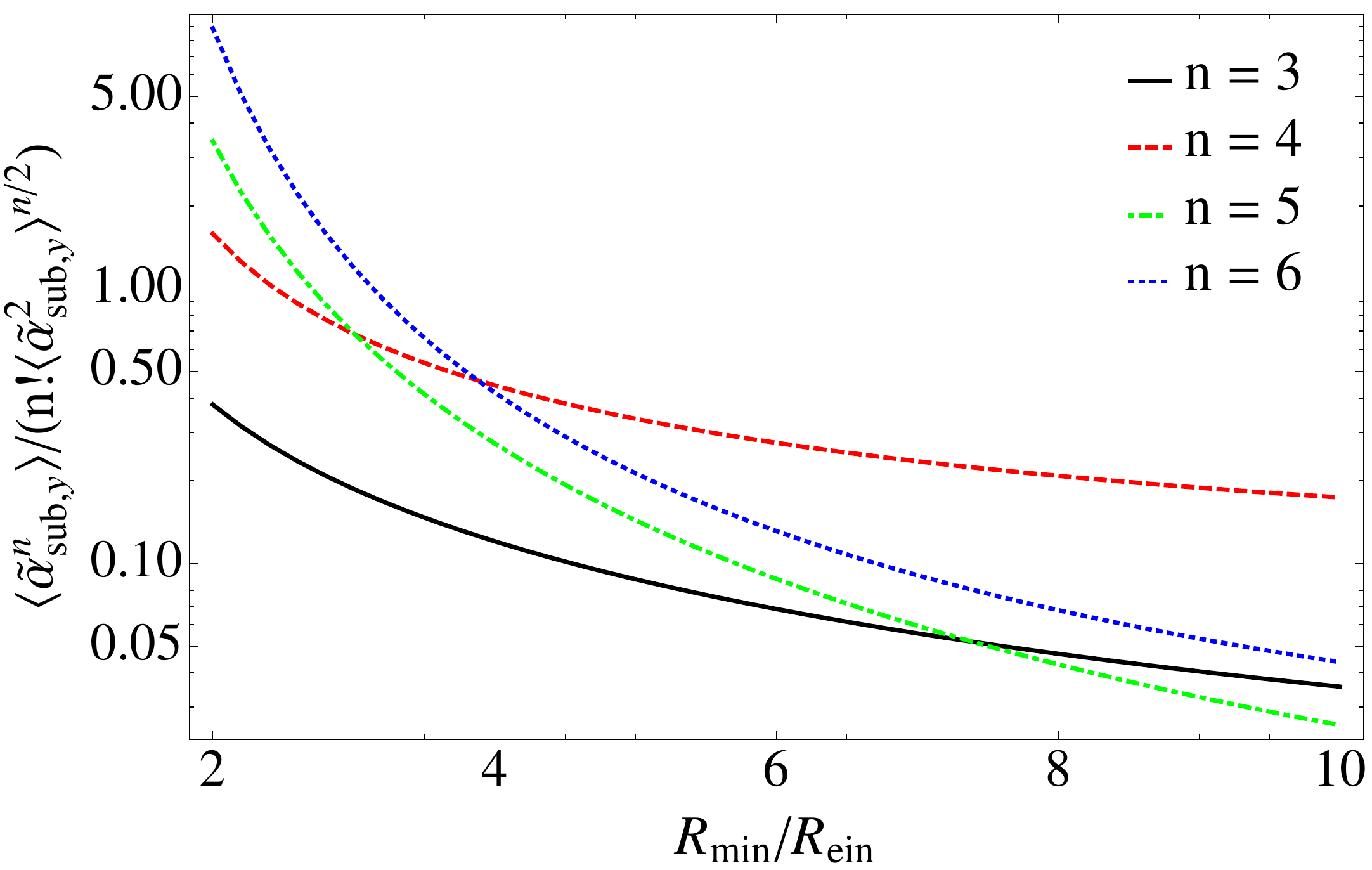}
\caption{Dependence of the non-Gaussian spatial moments of a deflection y-component on the value of $R_{\rm min}$. We take the distributed mass substructures to be spatially distributed  between $R_{\rm min}$ and $R_{\rm max} = 65 R_{\rm ein}$ according to Eq.~(\ref{eq:sub_spatial_dist}) with $r_{\rm c} = 30 R_{\rm ein}$. It is understood here that the mass dependence of the lensing deflection has been divided out, that is, $\tilde{\alpha}_{{\rm sub,} y} \equiv \alpha_{{\rm sub,} y}/m$. We see that the non-Gaussian moments decay as $R_{\rm min}$ is increased.}
\label{fig:Rmin_nonGaussian}
\end{figure}

We note that we can also suppress non-Gaussianities by increasing $R_{\rm min}$. Indeed, the non-Gaussian spatial moments $\langle\mathcal{O}_{\rm L}^n\rangle/\langle\mathcal{O}_{\rm L}^2\rangle^{n/2}$ rapidly decreases as $R_{\rm min}$ is increased as shown in Fig.~\ref{fig:Rmin_nonGaussian}. For definiteness, we illustrate there the ratio of non-Gaussian spatial moments for a single component of a lensing deflection. The results would be very similar for other linear lensing quantities. From a practical point-of-view, we would like $R_{\rm min}$ to be as small as possible in order to encompass as many mass substructures as possible in the distributed analysis. On the other hand, we also need to choose a value of $R_{\rm min}$ large enough for the expansion of Eq.~(\ref{eq:nongaussian_terms}) to rapidly converge. Our tests show that a minimal radius in the range $3R_{\rm ein}\lesssim R_{\rm min} \lesssim 5 R_{\rm ein}$ generally provides a good compromise between these two criteria for the power law mass function considered in this work. Of course, for a different choice of mass function one should adjust $R_{\rm min}$ in order to insure the convergence of the Edgeworth expansion.

The picture that emerges from the considerations above is that for the CDM-relevant case of $\beta\sim-1.9$, $M_{\rm low}/M_{\rm high}\ll 1$, and a realistic normalization of the substructure mass function supported by simulations, the non-Gaussian contributions to Eq.~(\ref{eq:nongaussian_terms}) are subdominant and the joint probability density function of linear lensing quantities will thus be well approximated by a multivariate Gaussian. In this physically relevant region, a useful criterion for the validity of the Gaussian approximation is
\be
a_{\rm 0} \gtrsim \frac{10}{ M_{\rm high}}\frac{(\beta+3)^2}{4! (\beta+5)}\frac{\langle\mathcal{O}_{\rm L}^4\rangle}{\langle\mathcal{O}_{\rm L}^2\rangle^{2}},
\ee
which is valid for $M_{\rm low}/M_{\rm high}\ll 1$, $\beta>-3$, and where $a_{0}$ is the amplitude of the mass function at $M_{0} = M_{\rm high}$. For our choice of spatial distribution given in Eq.~(\ref{eq:sub_spatial_dist}) with $R_{\rm min} = 3R_{\rm ein}$, $R_{\rm max} = 65R_{\rm ein}$, $r_{\rm c} = 30 R_{\rm ein}$, and assuming $M_{\rm high} = 10^{10} M_\odot$, $\beta = -1.9$, and $R_{\rm ein}=1''$, this criterion reads $a_{0}\gtrsim2.3\times10^{-10}M_\odot^{-1}$ when $\mathcal{O}_{\rm L}$ is a lensing deflection. This condition would be slightly relaxed if $\mathcal{O}_{\rm L}$ is taken to be the lensing potential instead. Whenever this condition is satisfied the characteristic function expansion given in Eq.~\eqref{eq:nongaussian_terms} [see also Eq.~\eqref{eq:real_edgeworth_exp}] provides an accurate description of the statistical properties of perturbations to the linear lensing quantities caused by distributed mass substructures.

\subsection{Validity of analytical approach}
We now test the validity of the characteristic function-based approach by comparing its prediction to the distributions of linear lensing quantities obtained by considering a large number of Monte Carlo realizations of distributed substructure populations. As in the analytical calculation of the previous section, we treat the distributed mass substructures as point masses that are spatially distributed according to the cored profile given in Eq.~(\ref{eq:sub_spatial_dist}) with $r_{\rm c} = 30 R_{\rm ein}$, between $R_{\rm min} = 3R_{\rm ein}$ and $R_{\rm max} = 65 R_{\rm ein}$. We consider the distribution of linear lensing quantities at two fiducial image positions located at $\xx_1 = (0,R_{\rm ein})$ and $\xx_2 = (R_{\rm ein},0)$, and take $R_{\rm ein} = 1''$. For concreteness, we assume a lens at redshift $z_{\rm lens} =0.5$ with a source at redshift $z_{\rm src}=1$, which yields a critical density for lensing $\Sigma_{\rm crit} = 1.19\times10^{11} M_\odot/\text{arcsec}^2$. To compute the final probability distribution of linear lensing quantities, we sample the characteristic function given in Eq.~(\ref{eq:real_edgeworth_exp}) on a grid of $\kk_{\rm L}$ and use a fast Fourier transform algorithm to perform the transformation back to configuration space. 

\begin{figure}[t]
\subfigure{\includegraphics[width=0.157\textwidth]{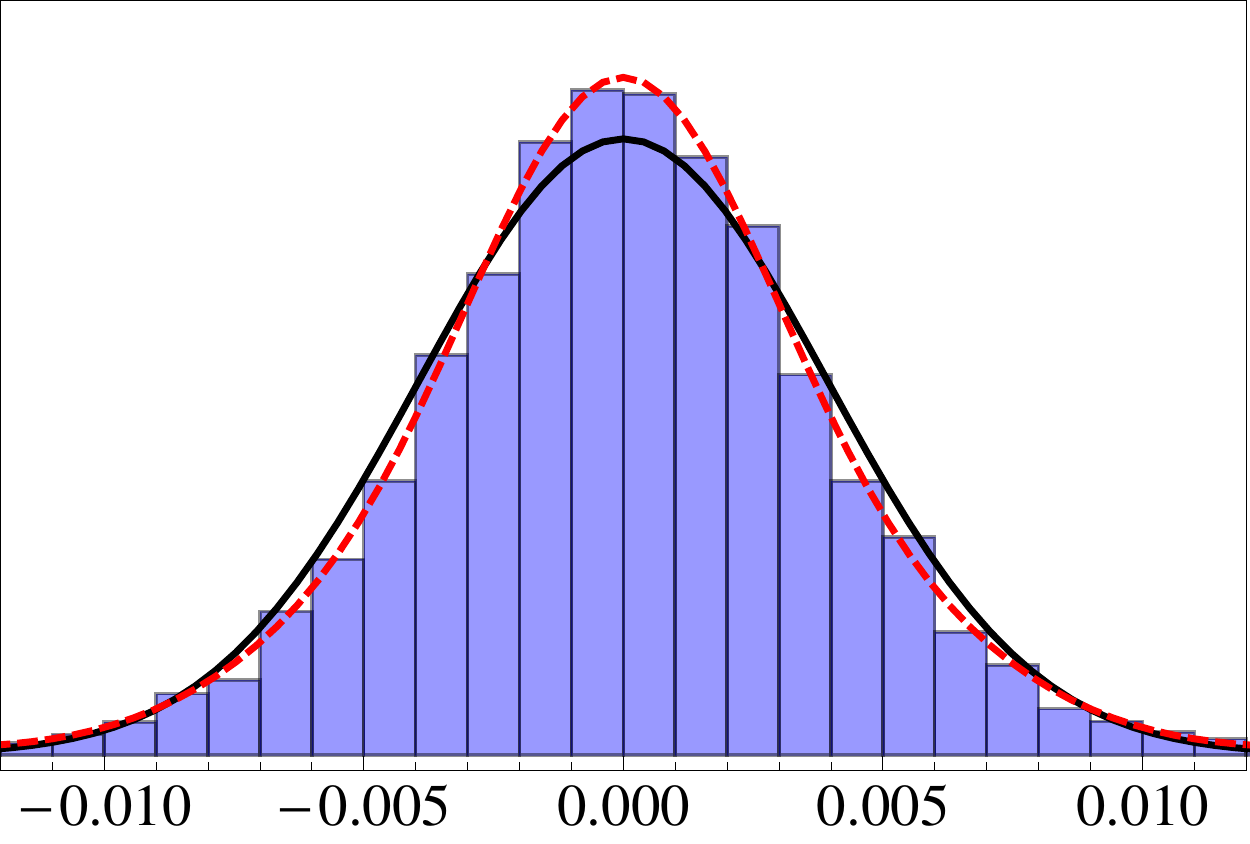}\hspace{0.70\textwidth}}\\
\subfigure{\includegraphics[width=0.22\textwidth]{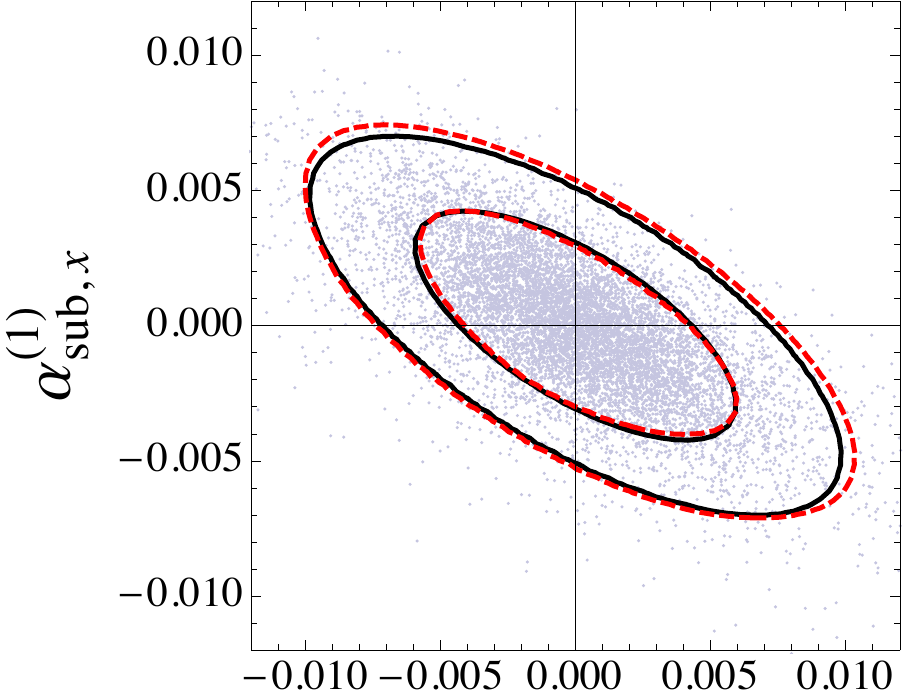}
\hspace{0.022\textwidth}
\vspace{1cm}
\subfigure{\includegraphics[width=0.157\textwidth]{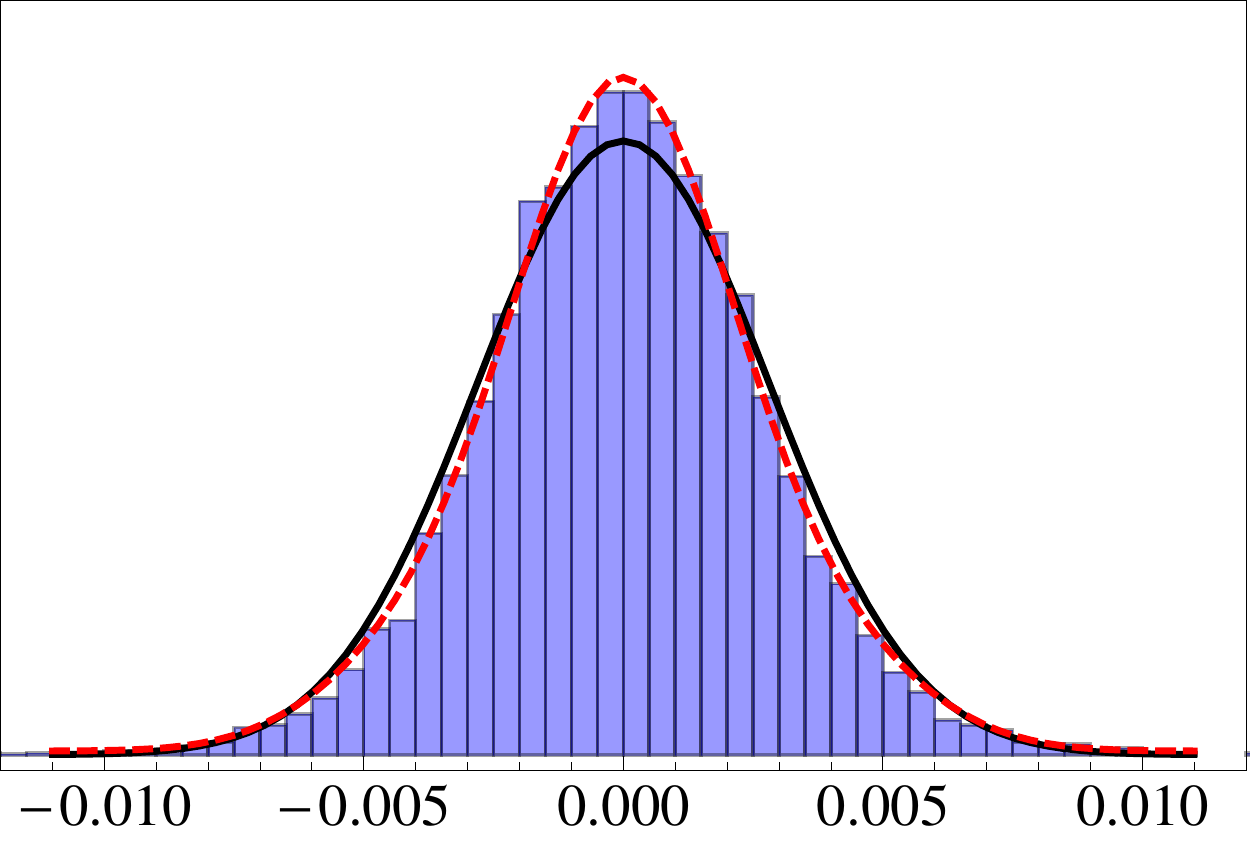}}
\hspace{0.57\textwidth}}\\
\subfigure{\includegraphics[width=.22\textwidth]{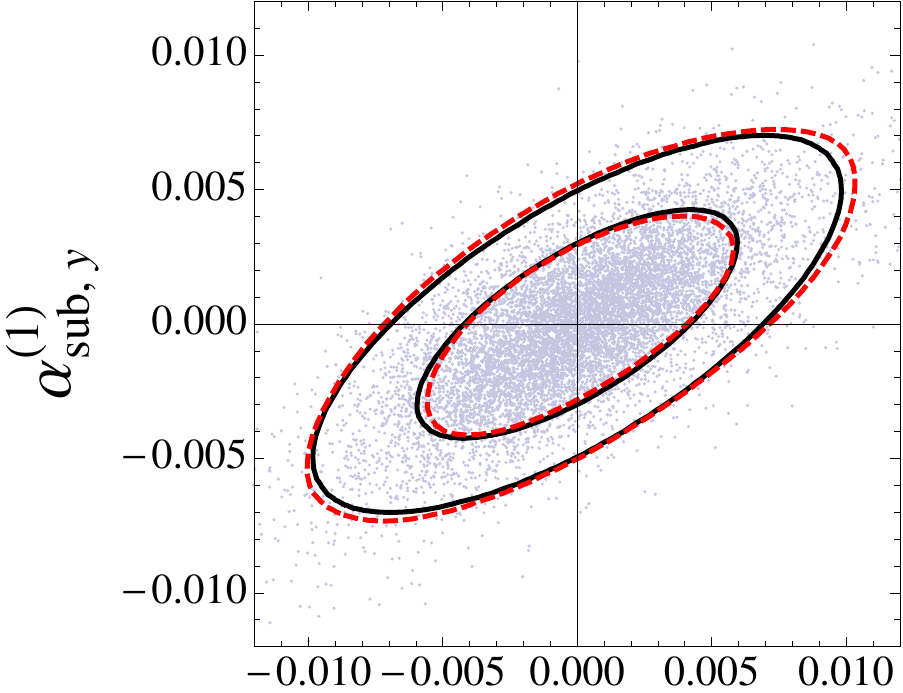}
\includegraphics[width=0.186\textwidth]{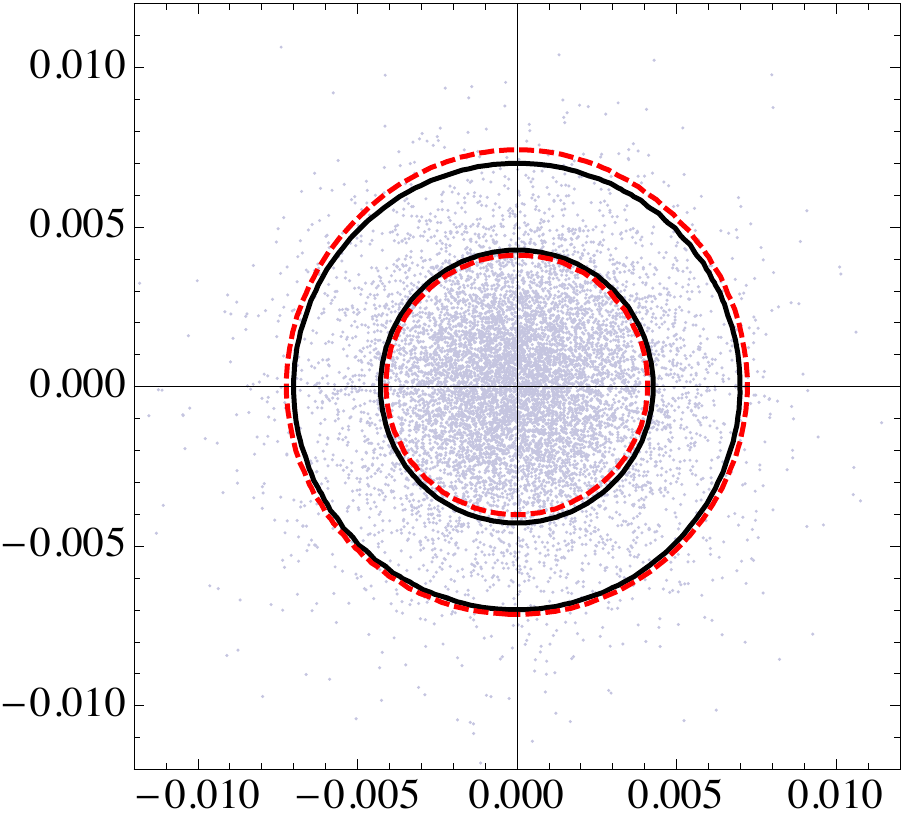}
\hspace{0.022\textwidth}
\vspace{1.1cm}
\includegraphics[width=0.157\textwidth]{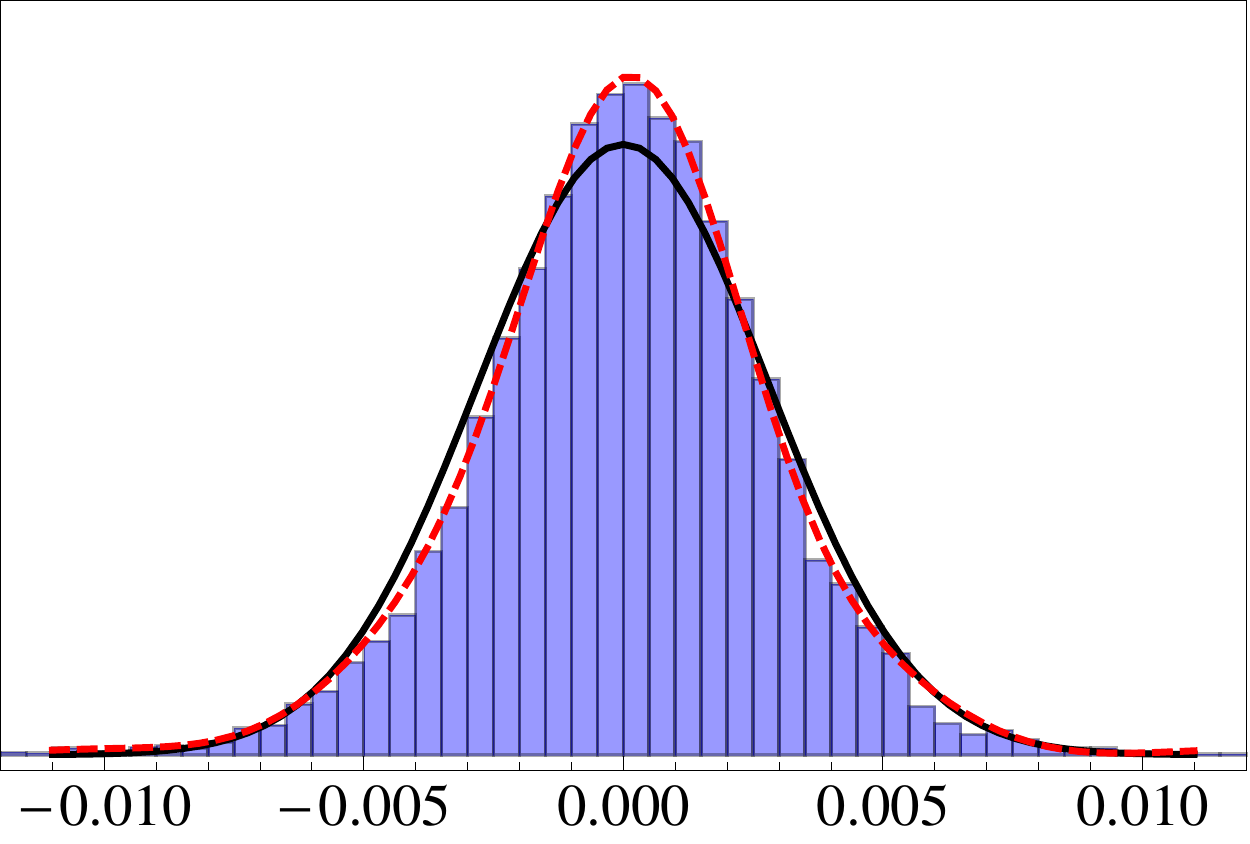}
\hspace{0.38\textwidth}}\\
\subfigure{\includegraphics[width=0.22\textwidth]{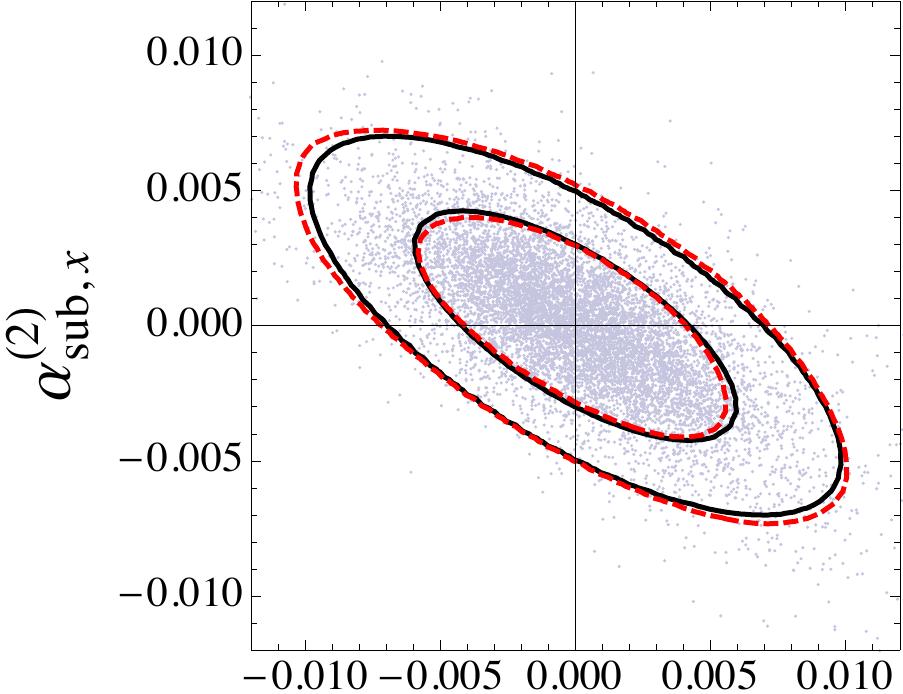}
\includegraphics[width=0.186\textwidth]{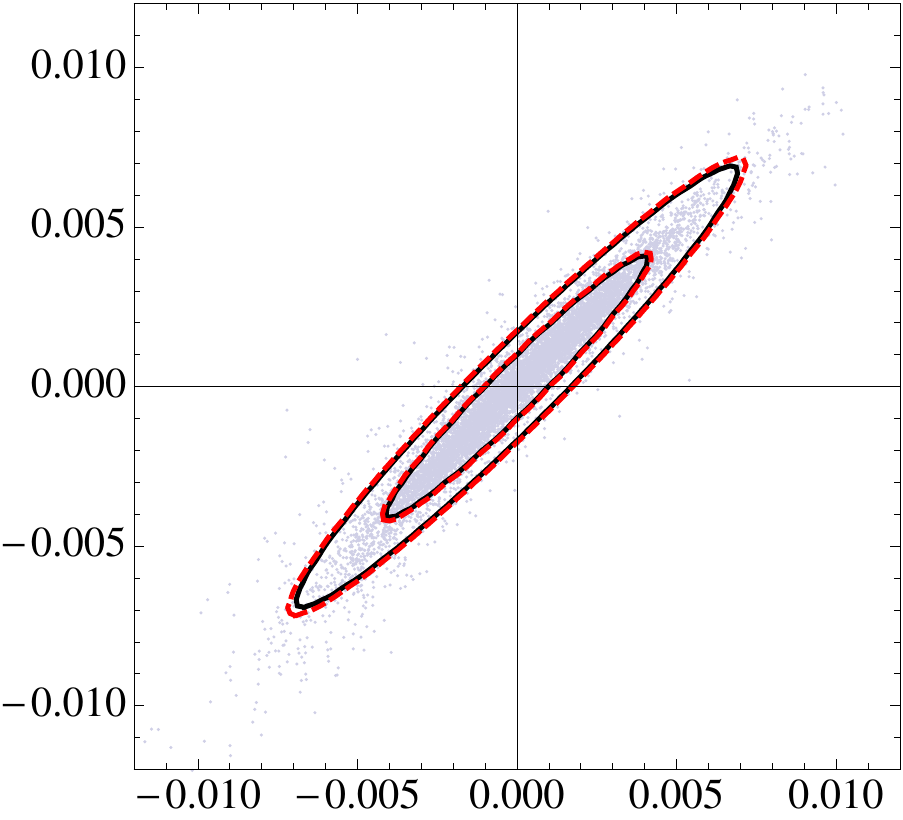}
\includegraphics[width=0.186\textwidth]{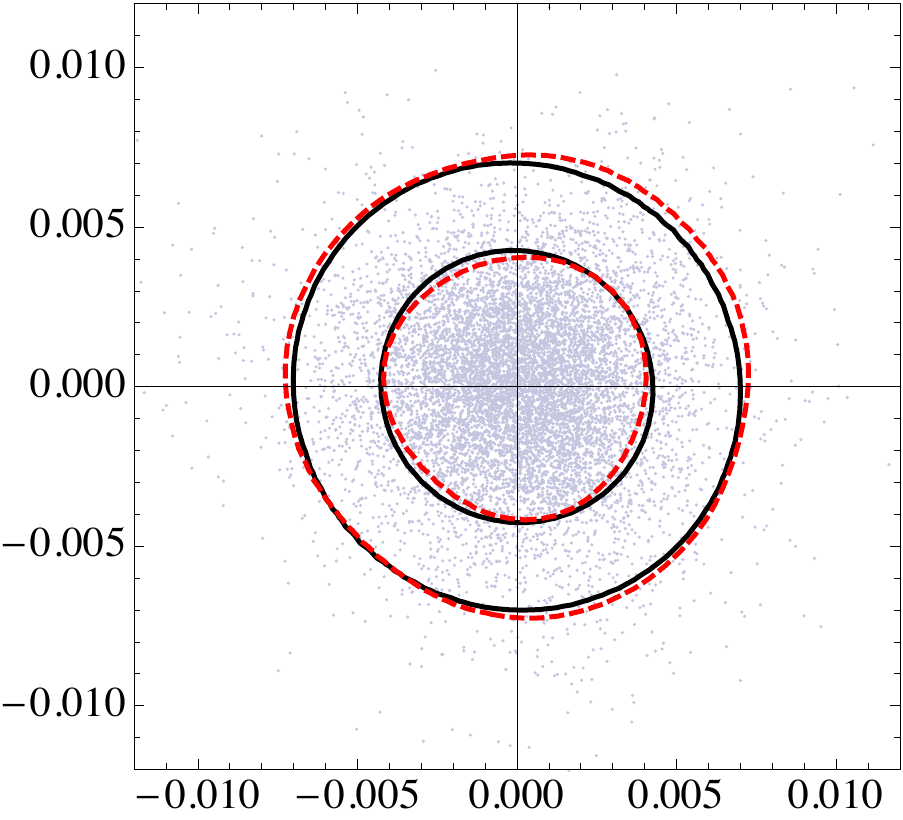}
\hspace{0.025\textwidth}
\vspace{1.1cm}
\includegraphics[width=0.157\textwidth]{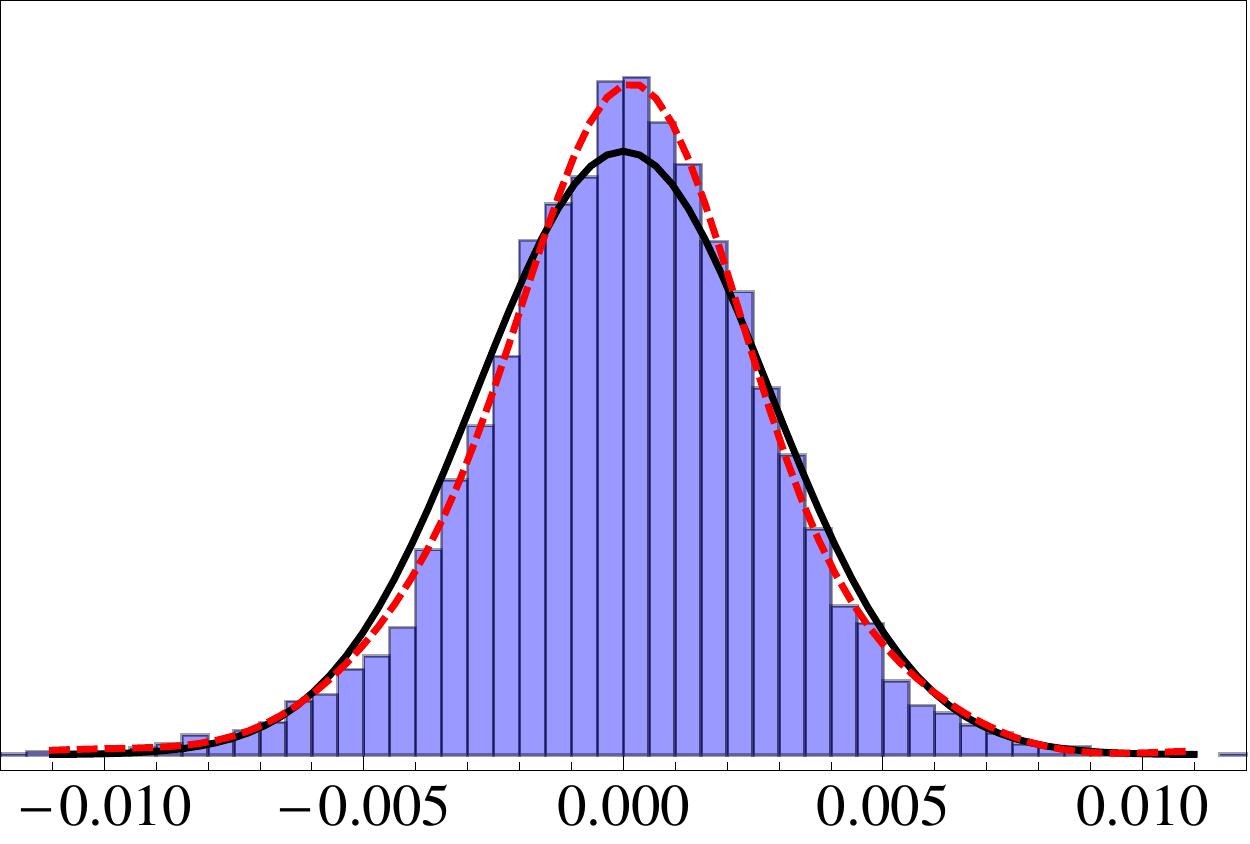}
\hspace{0.185\textwidth}}\\
\subfigure{\includegraphics[width=0.22\textwidth]{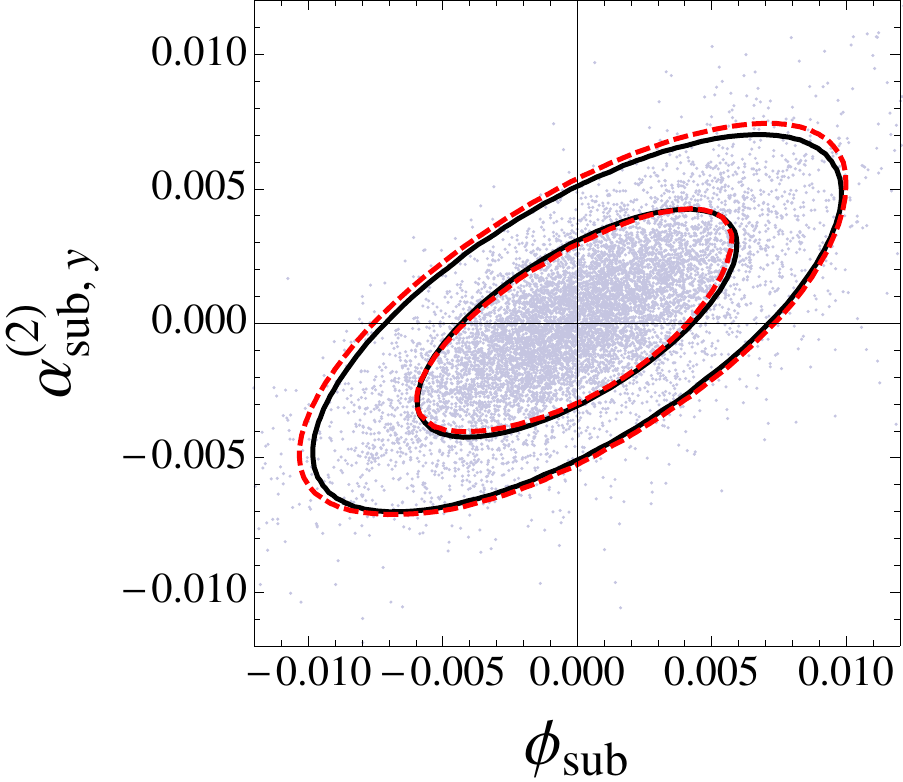}
\includegraphics[width=0.188\textwidth]{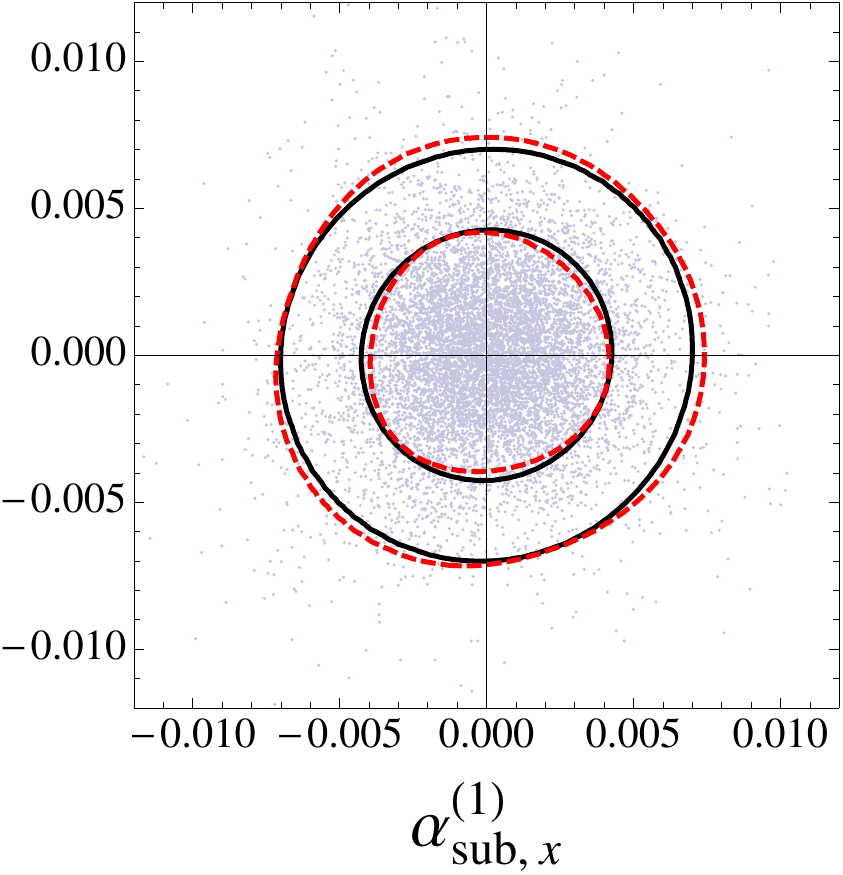}
\includegraphics[width=0.188\textwidth]{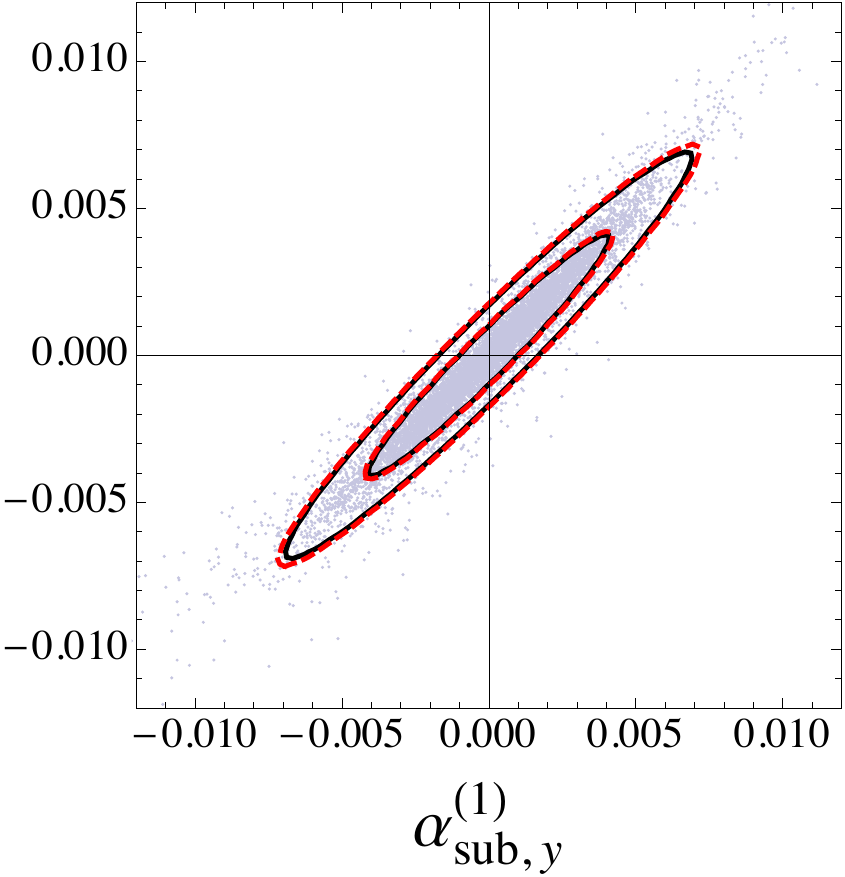}
\includegraphics[width=0.188\textwidth]{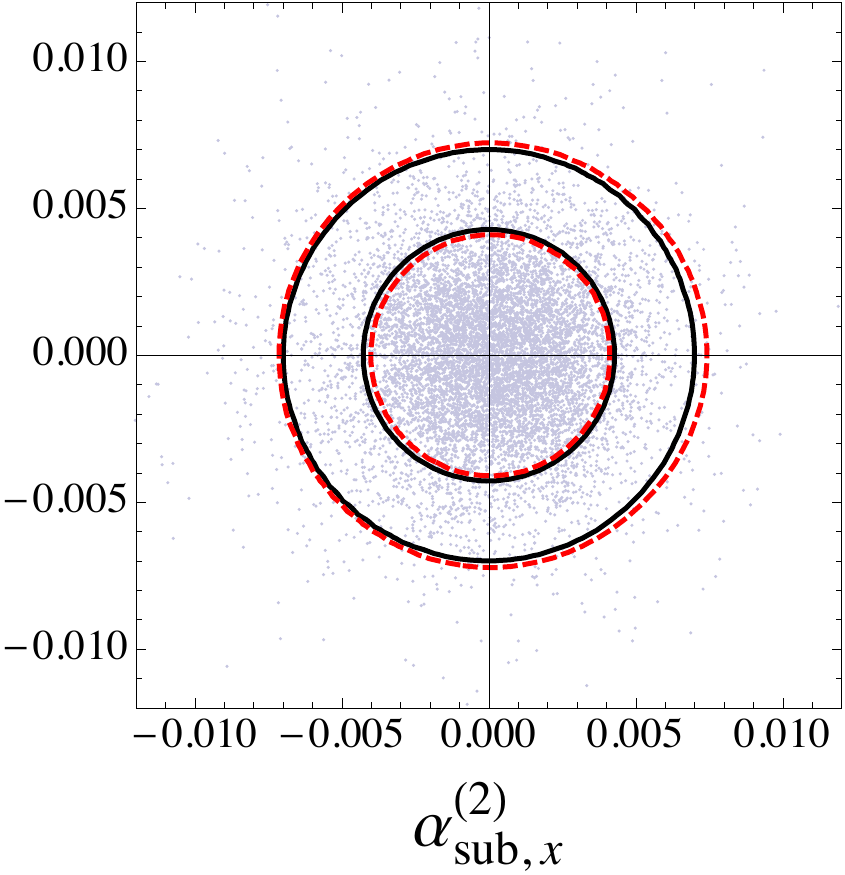}
\includegraphics[width=0.185\textwidth]{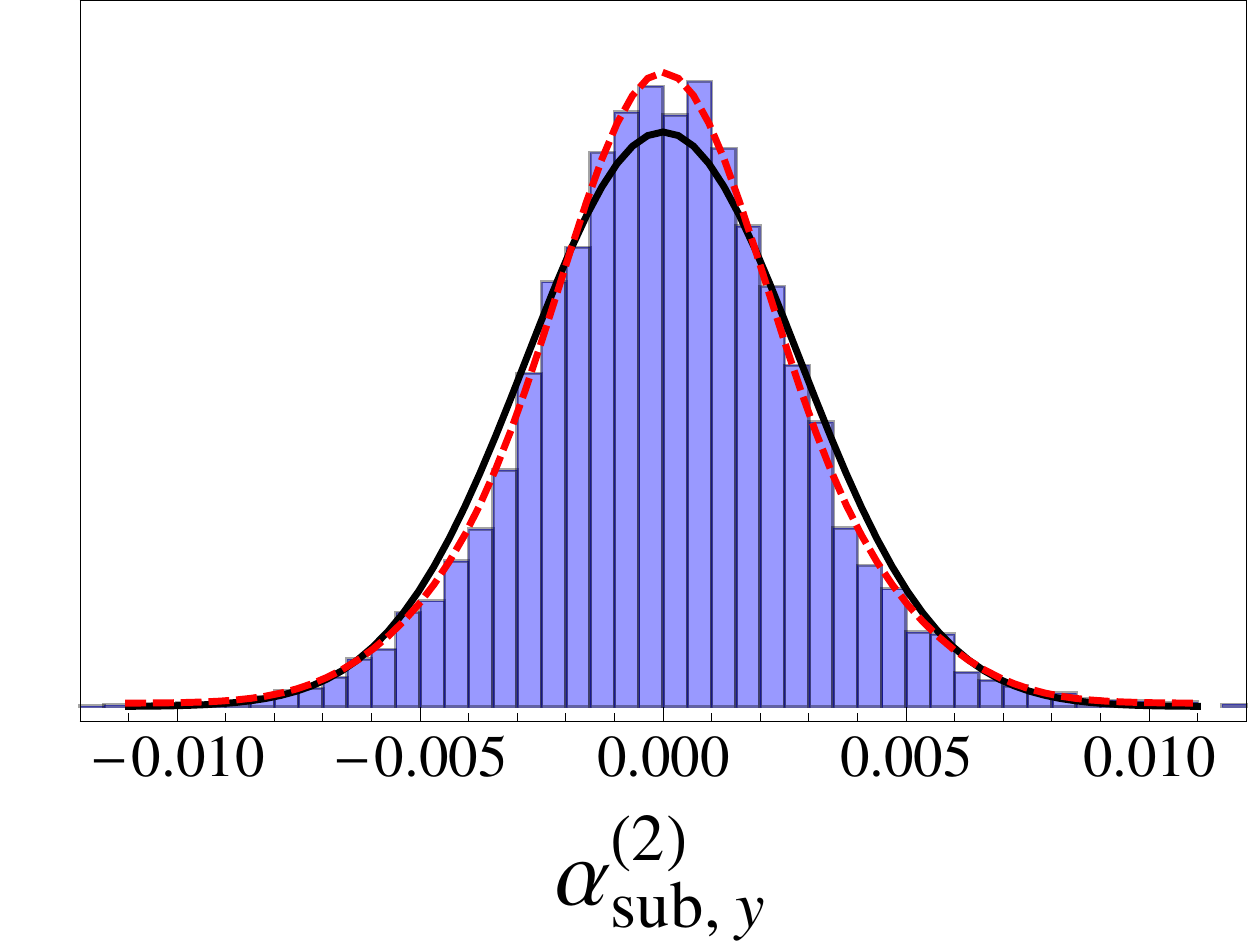}}
\caption{Projections of the probability density function for the linear lensing quantities $\{ \phi_{\rm sub},\alpha_{\rm sub, x}^{(1)},\alpha_{\rm sub, y}^{(1)},\alpha_{\rm sub, x}^{(2)},\alpha_{\rm sub, y}^{(2)}\}$ in the presence of a distributed population of mass substructures. Here, the two images are taken to be $\xx_1 = (0,R_{\rm ein})$ and $\xx_2 =  (R_{\rm ein},0)$, where we take $R_{\rm ein} = 1''$. In the above, $\phi_{\rm sub}$ stands for the projected potential difference between the two images. The gray points in the 2D plots and the blue histograms along the diagonal show the results from $10^4$ Monte Carlo realizations of distributed point mass-like substructure population. The solid black lines display the analytical results from Sec.~\ref{sec:distant_analysis} assuming a purely Gaussian characteristic function, while the dashed red lines show the results obtained by keeping all terms up to order $\langle N_{\rm d}\rangle^{-2}$ in the Edgeworth expansion [see Eq.~(\ref{eq:real_edgeworth_exp})]. In the 2D plots, the inner and outer contours display the $68\%$ and $95\%$ confidence regions, respectively. We assume the mass substructures to be spatially distributed according to Eq.~(\ref{eq:sub_spatial_dist}) with $r_{\rm c} = 30 R_{\rm ein}$. We also take a power law subhalo mass function with slope $\beta =-1.9$ between $M_{\rm low} =10^7 M_\odot$ and $M_{\rm high} = 10^{10} M_\odot$, and take $\langle\kappa_{\rm sub}(R_{\rm ein})\rangle = 0.001$. This yields an expected number of distributed mass substructures $\langle N_{\rm d} \rangle = 3705$.}
\label{fig:Validation_gaussian_good}
\end{figure}
\begin{figure}[t]
\subfigure{\includegraphics[width=0.157\textwidth]{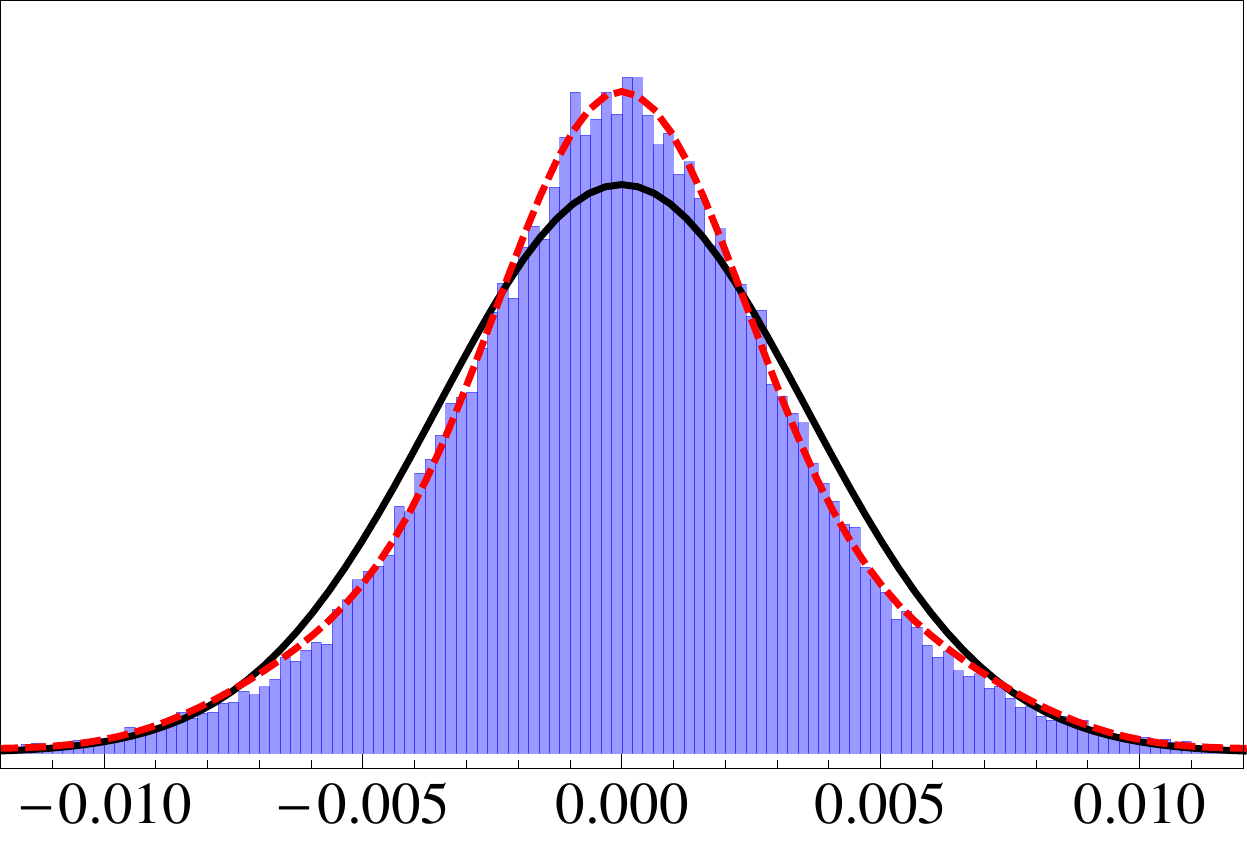}\hspace{0.70\textwidth}}\\
\subfigure{\includegraphics[width=0.22\textwidth]{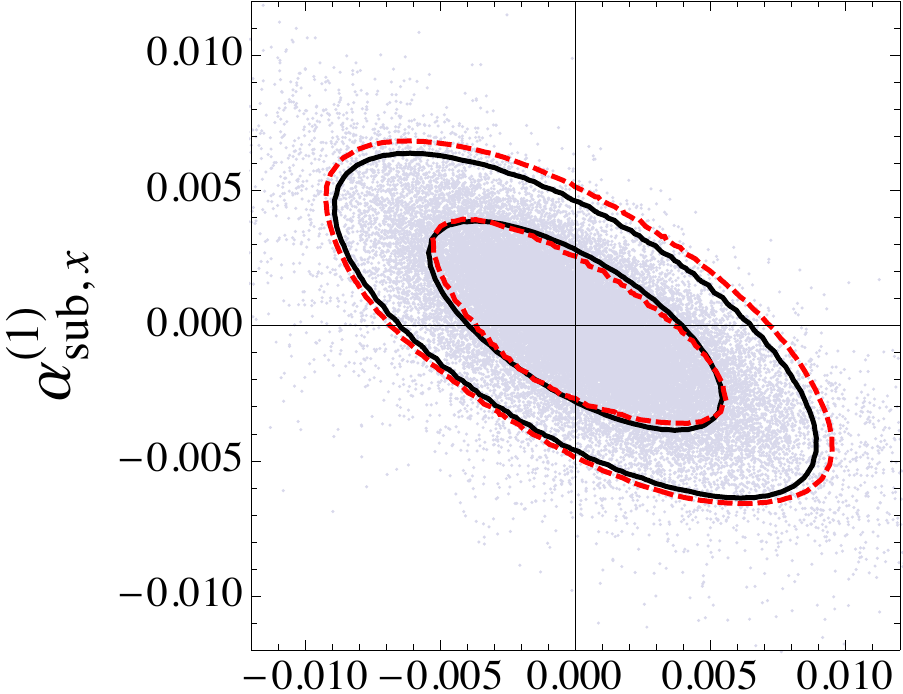}
\hspace{0.022\textwidth}
\vspace{1cm}
\subfigure{\includegraphics[width=0.157\textwidth]{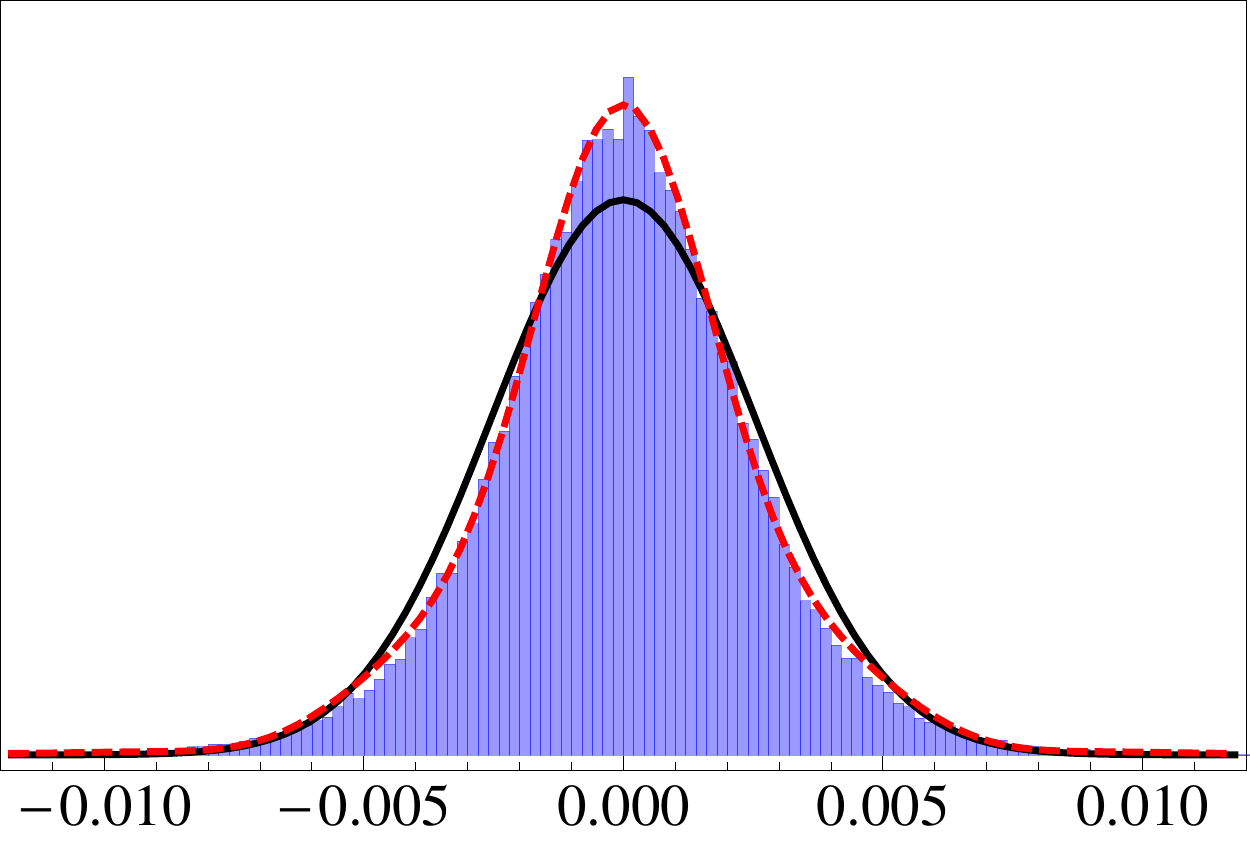}}
\hspace{0.57\textwidth}}\\
\subfigure{\includegraphics[width=.22\textwidth]{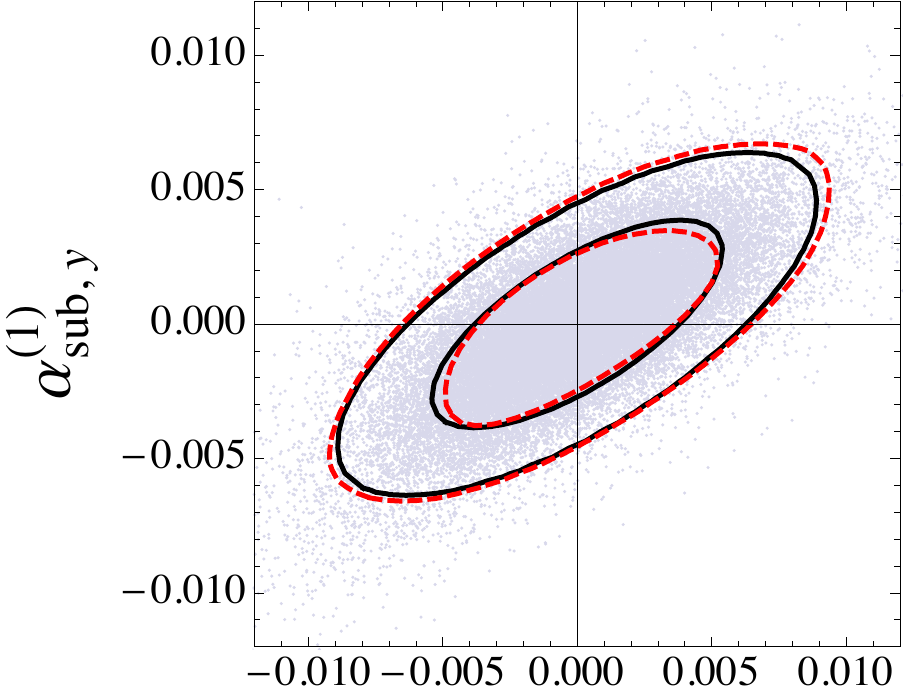}
\includegraphics[width=0.186\textwidth]{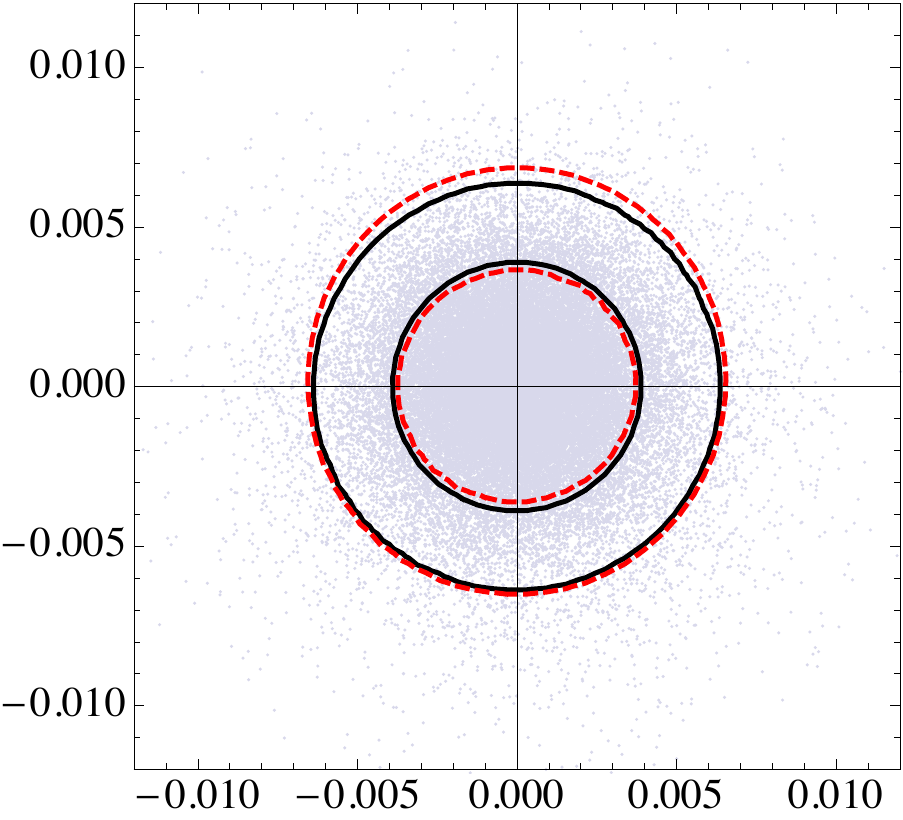}
\hspace{0.022\textwidth}
\vspace{1.1cm}
\includegraphics[width=0.157\textwidth]{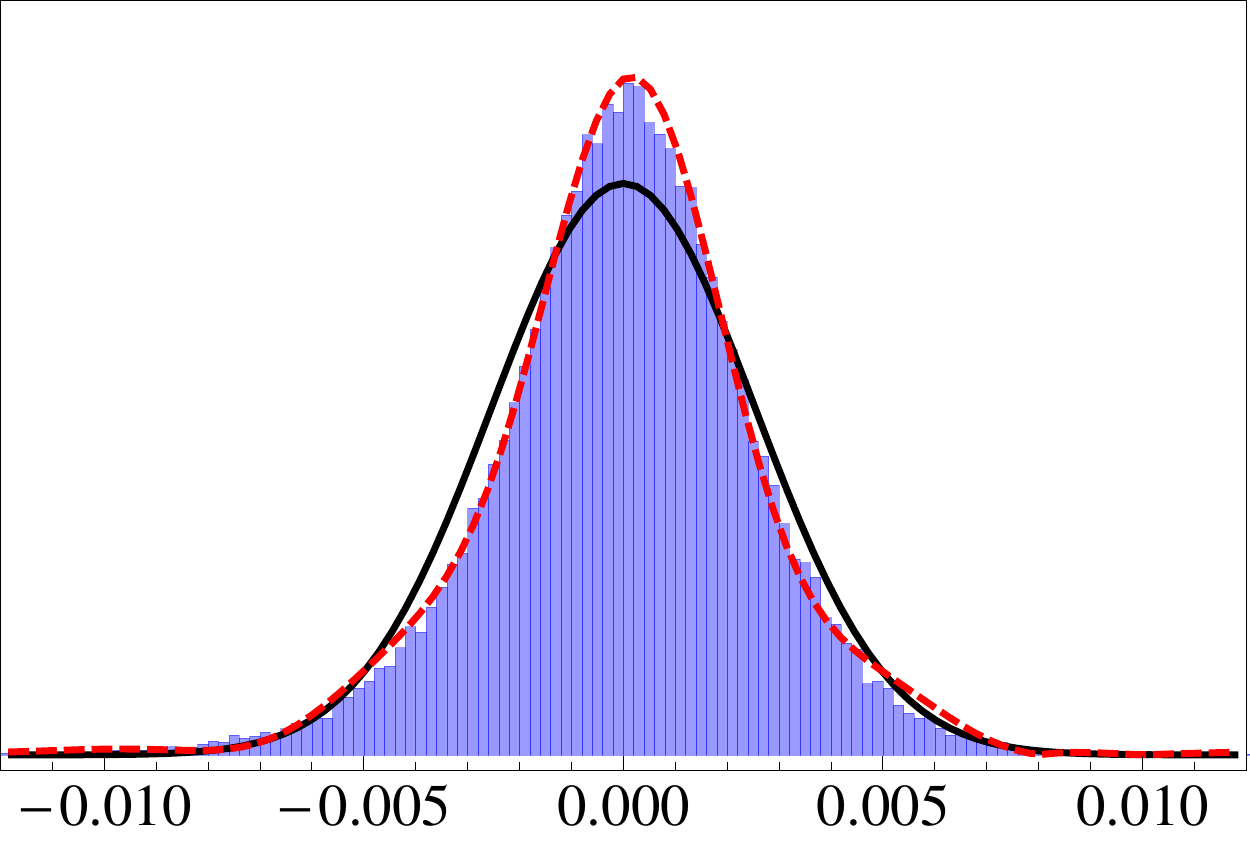}
\hspace{0.38\textwidth}}\\
\subfigure{\includegraphics[width=0.22\textwidth]{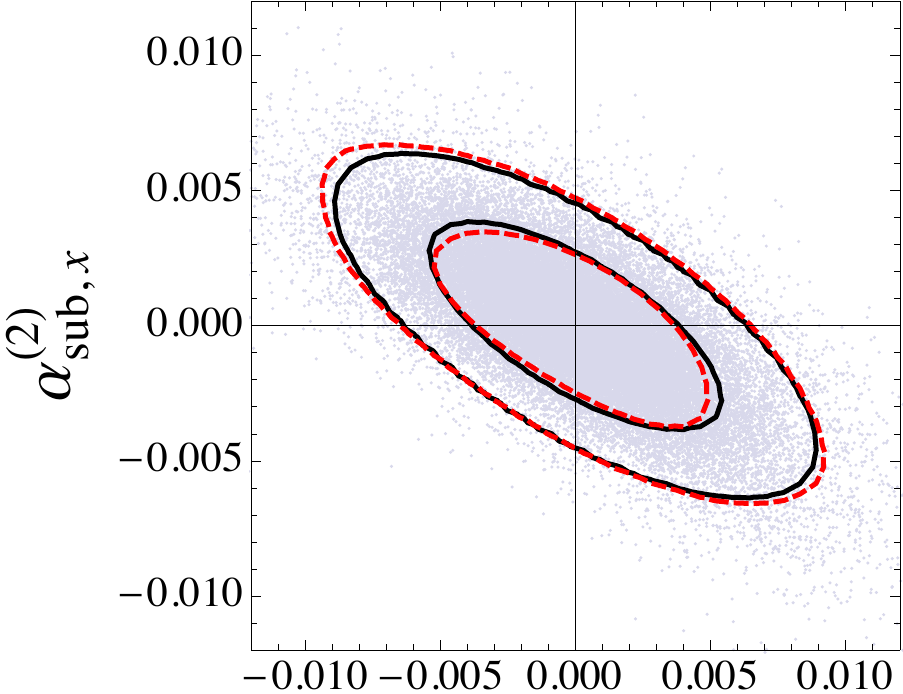}
\includegraphics[width=0.186\textwidth]{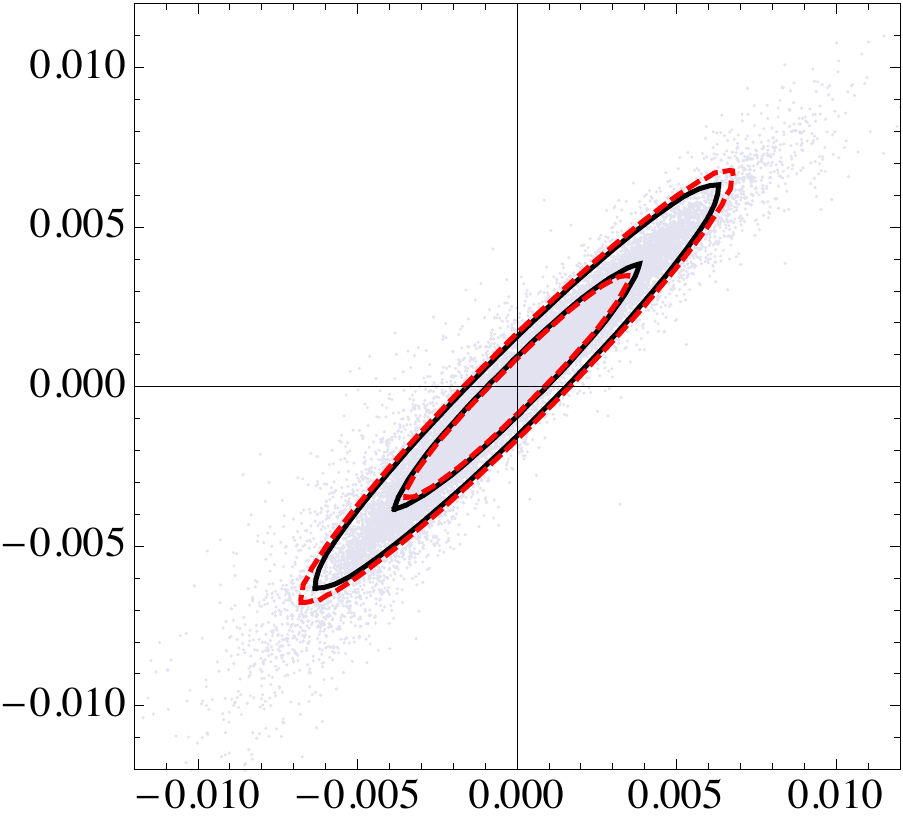}
\includegraphics[width=0.186\textwidth]{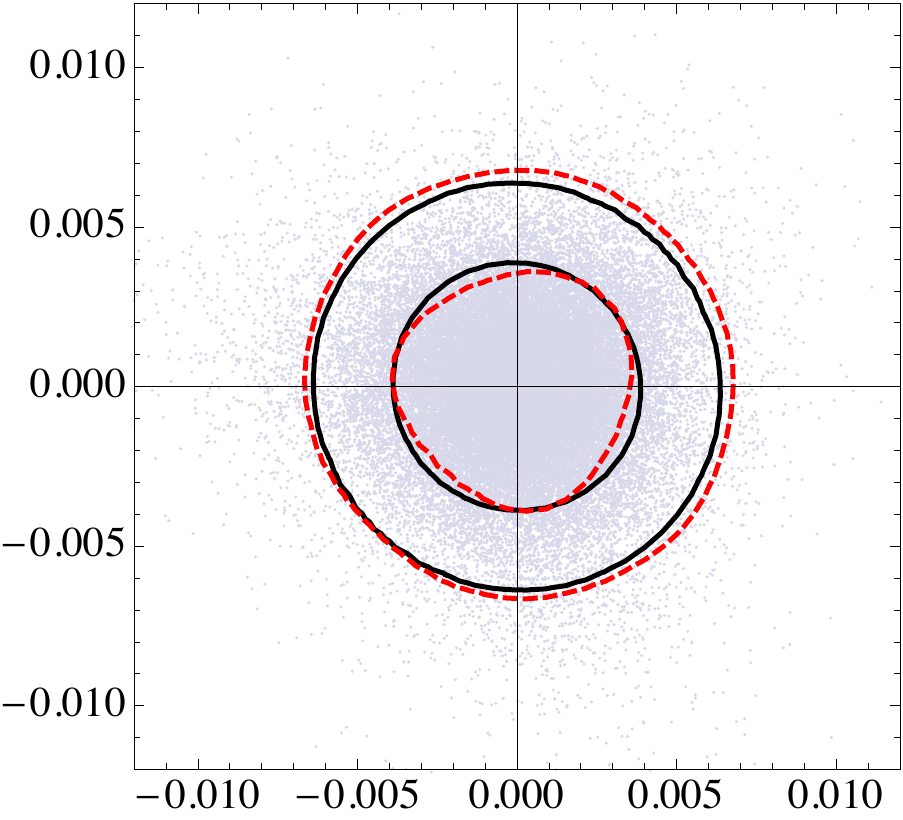}
\hspace{0.025\textwidth}
\vspace{1.1cm}
\includegraphics[width=0.157\textwidth]{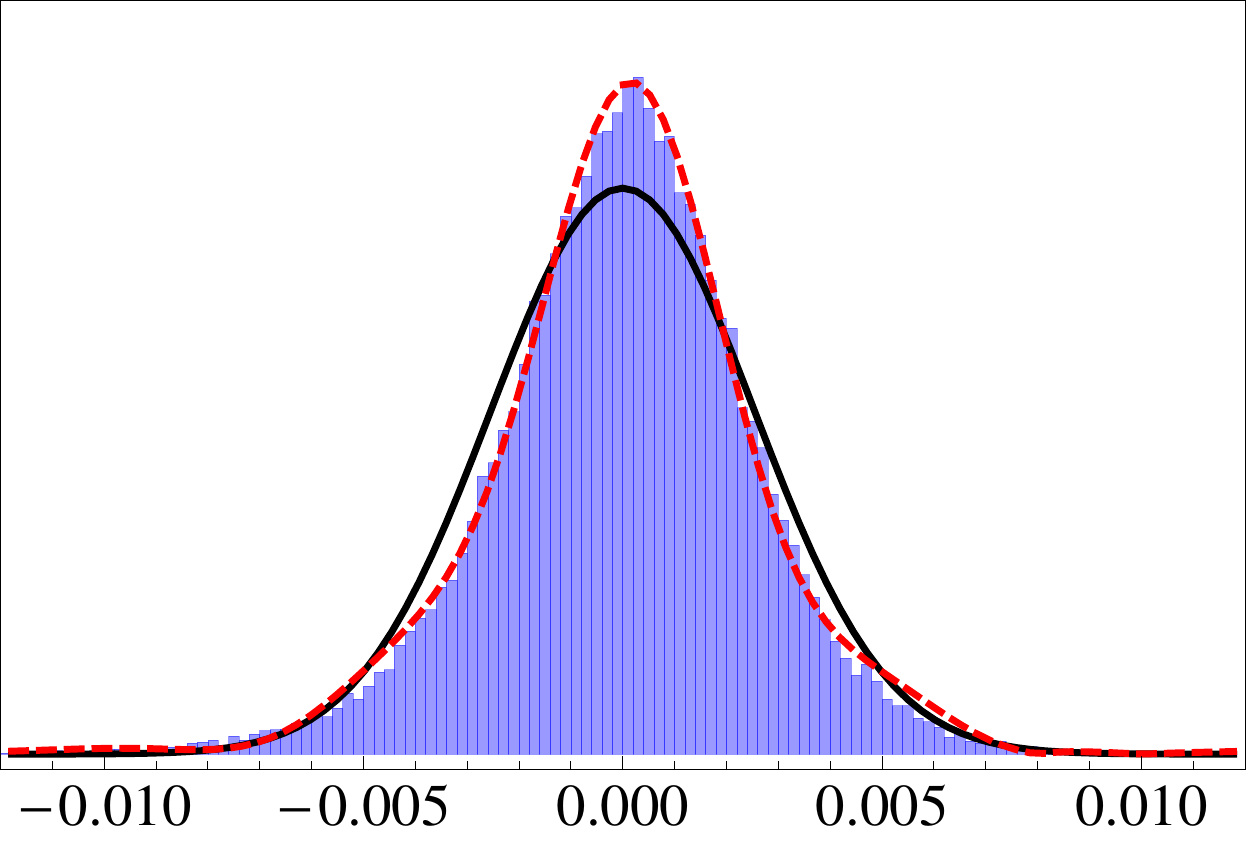}
\hspace{0.185\textwidth}}\\
\subfigure{\includegraphics[width=0.22\textwidth]{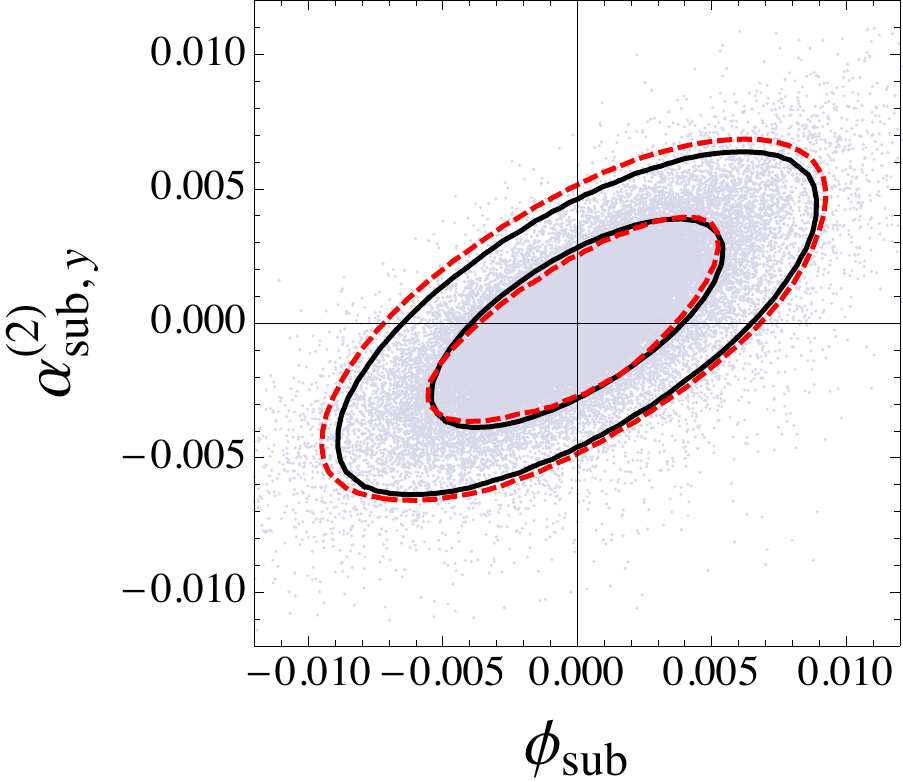}
\includegraphics[width=0.188\textwidth]{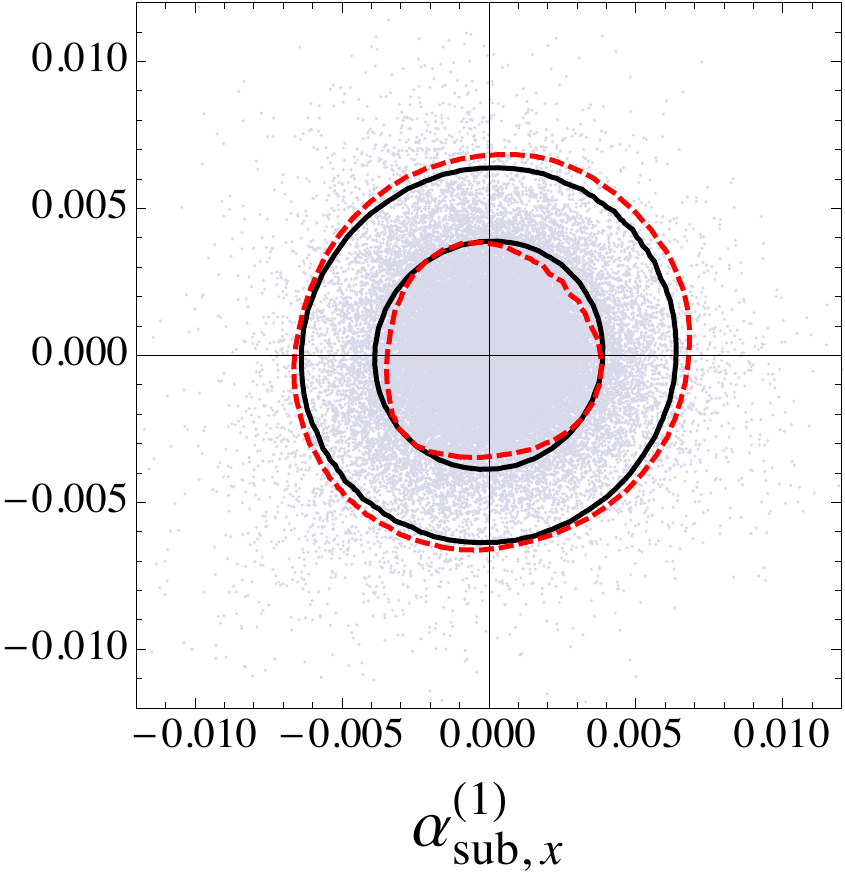}
\includegraphics[width=0.188\textwidth]{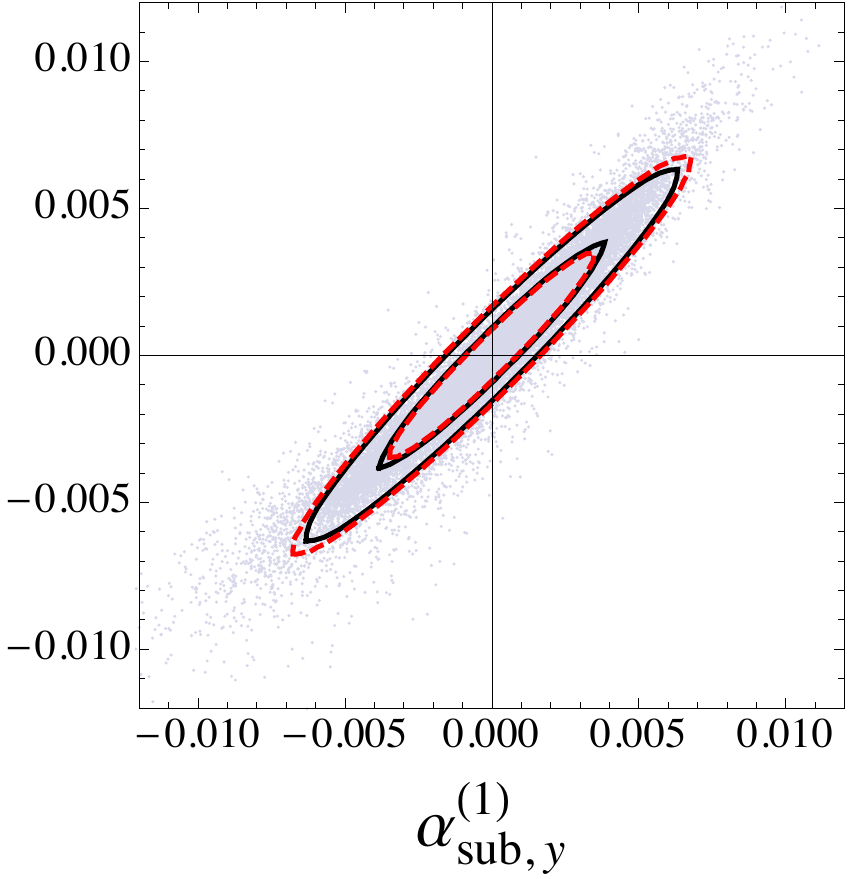}
\includegraphics[width=0.188\textwidth]{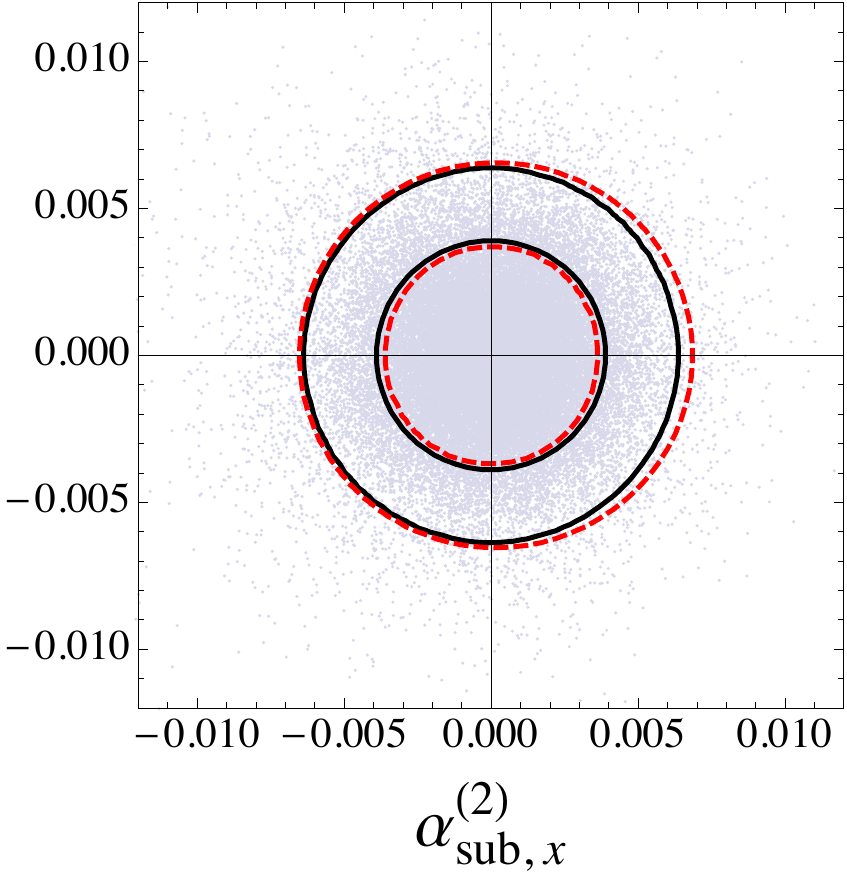}
\includegraphics[width=0.185\textwidth]{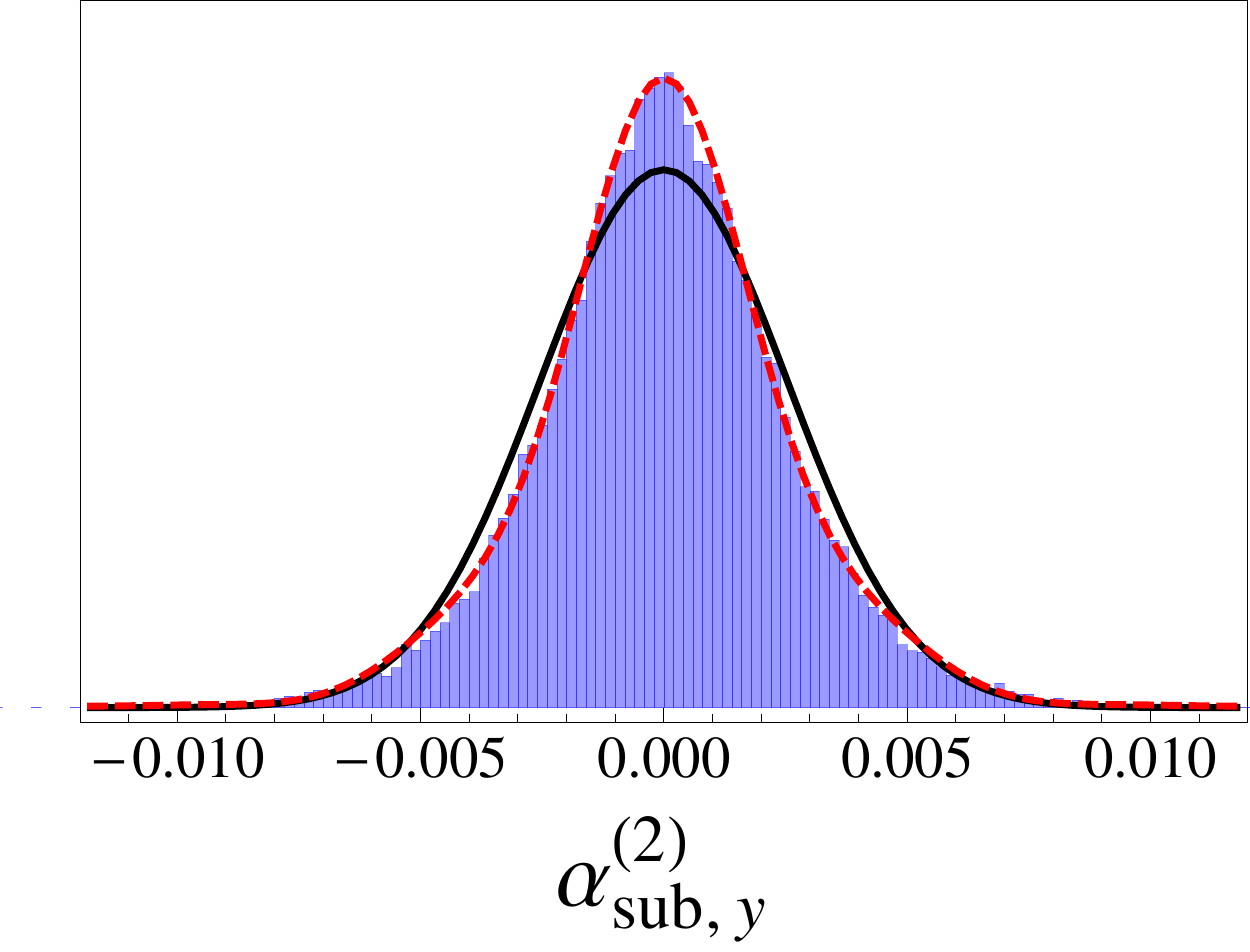}}
\caption{Projections of the probability density function for the linear lensing quantities similar to Fig.~\ref{fig:Validation_gaussian_good} but taking $M_{\rm low} = 2\times10^9 M_\odot$, $M_{\rm high} = 10^{10} M_\odot$, and $\langle\kappa_{\rm sub}(R_{\rm ein})\rangle = 3\times10^{-4}$. The expected number of distributed substructures is $\langle N_{\rm d}\rangle =24$. The gray points in the 2D plots and the blue histograms along the diagonal show the results from $5\times10^4$ Monte Carlo realizations of distributed point mass-like substructure population. }
\label{fig:Validation2_gaussian_good}
\end{figure}

In Fig.~\ref{fig:Validation_gaussian_good}, we compare our analytical predictions to the results from Monte Carlo simulations of distributed substructure populations for a subhalo mass function as given in Eq.~(\ref{eq:subhalo_mass_function}) with $M_{\rm low} = 10^7 M_\odot$, $M_{\rm high} = 10^{10} M_\odot$, and $\beta = -1.9$, with a normalization given by $\langle\kappa_{\rm sub}(R_{\rm ein})\rangle =  0.001$. We display different projections of the joint probability density function for the linear lensing quantities evaluated at the two fiducial image positions.   The gray points in the 2D plots and the blue histograms along the diagonal show the results from the Monte Carlo realizations of distributed substructure population. We show in solid black the results gotten by only keeping the leading Gaussian term in Eq.~(\ref{eq:real_edgeworth_exp}), while the dashed red lines show the results obtained by keeping all terms up to order $O(\langle N_{\rm d}\rangle^{-2})$ in the Edgeworth expansion. Since the mass function parameters listed above predict a relatively large number of mass substructures within the lens halo ($\langle N_{\rm d}\rangle = 3705$), the non-Gaussian contributions in Eq.~(\ref{eq:real_edgeworth_exp}) are suppressed and  the overall behavior of the joint probability density function is very well captured by its leading Gaussian term. Nevertheless, we see that including the higher-order terms in the Edgeworth expansion does improve the concordance of the analytical predictions with the Monte Carlo realizations. This is especially noticeable in the one-parameter probability densities shown along the diagonal where we observe that dashed red lines capture the nonzero excess kurtosis of the Monte Carlo realizations. This indicates that the characteristic function expansion performed in Sec.~\ref{sec:distant_analysis} does lead to the correct probability density function for the linear lensing quantities.

In Fig.~\ref{fig:Validation2_gaussian_good}, we display similar projections of the probability density function of linear lensing quantities, but here we choose a high value of the low mass cutoff $M_{\rm low} = 2\times10^9 M_\odot$, together with $M_{\rm high} = 10^{10} M_\odot$ and $\langle\kappa_{\rm sub}(R_{\rm ein})\rangle = 3\times10^{-4}$. This is an example of a model with very few distributed substructures ($\langle N_{\rm d} \rangle =24$) for which the leading Gaussian approximation still works reasonably well. As clearly shown in the 1D histograms along the diagonal of the plot, this model does have a significant excess kurtosis which is well captured by the Edgeworth expansion. Again, this highlights the usefulness of the expansion given in Eq.~(\ref{eq:real_edgeworth_exp}) to understand the leading departure from Gaussianity. 

Interestingly, we observe in Fig.~\ref{fig:Validation_gaussian_good} that the perturbations to linear lensing quantities from the distributed mass substructures sometimes display strong correlations among themselves. This indicates that the correlation length of perturbations to the linear lensing quantities caused by distributed substructures is larger than the typical image separation in lens systems, consistent with the findings of Ref.~\cite{Chen:2007aa}. More precisely, this implies that the linear lensing quantity perturbations from distributed substructures are dominated by the dipole ($p=1$) term in the multipole expansion of Eq.~(\ref{eq:def_O}). For deflections, this term is independent of image position which explains the very tight correlation between $\alpha_{{\rm sub}, x}^{(1)}$ and $\alpha_{{\rm sub}, x}^{(2)}$, and between $\alpha_{{\rm sub}, y}^{(1)}$ and $\alpha_{{\rm sub}, y}^{(2)}$. The scatter about this almost perfect correlation is due to the contributions from higher multipoles. We note that this scatter tends to increase at large deflection values since these are caused by substructures that are closer to the images and are thus described by higher multipoles. The correlation between deflections and lensing potential perturbations also suggests that  time-delay fluctuations caused by mass substructures are usually accompanied by a corresponding shift to the image position. 

Ultimately, the correlations between linear lensing quantities at different image positions are particularly interesting since it may allow one to distinguish the effects of \emph{local} substructures which mostly affect a single lensed image from those of \emph{distributed} substructures which affect multiple lensed images coherently. However, we generally expect these coherent perturbations to be somewhat degenerate with the smooth lens model. For instance, the dipolar ($p=1$) component of the perturbation can generally be compensated by an appropriate shift to the source position, while the perturbation quadrupole ($p=2$) could be reabsorbed by an appropriate external shear. We leave the study of these potential degeneracies to future work.

\section{Analytical marginalization over distributed mass substructures}\label{marginalization}
In this section, we first describe how we transform from the linear lensing quantities to the actual gravitational lensing observables that can be compared with data. We then explain how we perform the analytical marginalization over the distributed mass substructure population using the characteristic function $Q_{\langle N_{\rm d}\rangle }^{\rm dist}(\kk_{\rm L}|{\bf q},\qsub)$ computed in the previous section. Here, we assume that the effect of the local substructure population is perfectly known, which is equivalent to setting $Q_{\langle N_{\rm l}\rangle }^{\rm local}(\kk_{\rm L}|{\bf q},\qsub)=1$.  We also derive a general expression for the data likelihood in the presence of a distributed population of unresolved mass substructures. We first describe the general calculation, and then specialize to the case where $Q_{\langle N_{\rm d}\rangle }^{\rm dist}(\kk_{\rm L}|{\bf q},\qsub)$ is a multivariate Gaussian. 
\subsection{General case}
The first step in to compute how the probability density function of the stochastic variables $ {\bf t}$ is related to that of the stochastic variables $\Delta\tl$ for which we have computed the characteristic function in the previous section. For definitiveness, we take $\Delta\tl$ to contain all the stochastic deflections $\vec{\alpha}_{\rm sub}^{(i)}$, $i\in N_{\rm img}$, caused by distributed substructures as well as all the stochastic lensing potential shifts $\phi_{\rm sub}^{(i)}$ between the $i$th image and the reference point. In the following, we take the reference point to be the position of the image that is leading the arrival time. We also take the stochastic vector of observables $ {\bf t}$ to contain the image positions $\xx^{(i)}$ and the time delay $\Delta t^{(i)}$ between image $i$ and the reference image caused by distributed substructures. 

The probability density function for the perturbations $\Delta \tl^{\rm dist}$ to the linear lensing quantities caused by distributed substructures is simply
\be
\Phi_{\langle N_{\rm d}\rangle}(\Delta \tl | \qq,\qsub) = \int \frac{d\kk_{\rm L}}{(2\pi)^l} Q_{\langle N_{\rm d}\rangle }^{\rm dist}(\kk_{\rm L}|{\bf q},\qsub)e^{-i \Delta\tl \cdot \kk_{\rm L}},
\ee
where it is understood the $\Delta\tl$ stands for $\Delta\tl^{\rm dist}$ here. We then apply the  transformation given by Eqs.~(\ref{eq:linear_to_actual}) and (\ref{eq:deltatL_2_tL}) to compute the density function for the $ {\bf t}$ stochastic lensing observables
\be\label{eq:Delta_tl_int}
P({\bf t} | \qq,\qsub) = \int d \tl \int \frac{d\kk_{\rm L}}{(2\pi)^l} Q_{\langle N_{\rm d}\rangle }^{\rm dist}(\kk_{\rm L}|{\bf q},\qsub)e^{-i (\tl -\bar{\bf t}_{\rm L}) \cdot \kk_{\rm L}}\de_{\rm D}^{k}\left({\bf t} - {\bf t}(\tl)\right),
\ee
where $k$ is the length of the ${\bf t}$ vector and $\bar{\bf t}_{\rm L}$ is the contribution to the linear lensing quantities from the smooth lens model [Eq.~(\ref{eq:smooth+subs})]. Note that we generally have $l\neq k$ since, for instance, the two shear random variables are mapped to a single magnification perturbation. Nevertheless, as we are only considering potential and deflection perturbations, we have $l=k$ here. In practice, the relation ${\bf t}(\tl)$ is nonlinear but can nonetheless be written down and inverted in a straightforward manner. However, since the distributed substructure population leads to small changes to the lensed image positions and their relative time delays, we can linearize this relation as
\be\label{eq:linear_relation}
{\bf t}(\tl) \approx  \bar{\bf t} + {\bf A}^{-1}(\tl -\bar{\bf t}_{\rm L}) ,
\ee
where $\bar{\bf t}$ are the lensing observables in the absence of substructures, and where ${\bf A}$ is a $l$ by $l$ matrix encoding the transformation between the linear lensing quantities and the actual observables. For instance, in the case where $ {\bf t} = \{\xx^{(i)},\xx^{(j)},\xx^{(k)},\Delta t^{(j)},\Delta t^{(k)}\}$ and $\Delta\tl = \{\vec{\alpha}_{\rm sub}^{(i)},\vec{\alpha}_{\rm sub}^{(j)},\vec{\alpha}_{\rm sub}^{(k)},\phi_{\rm sub}^{(j)},\phi_{\rm sub}^{(k)}\}$, and assuming that image $i$ is the leading image, the inverse of the transformation matrix ${\bf A}$ is given by
\be\label{eq:matrix_A}
{\bf A} =\left(
\begin{array}{ccccc}
{\bf \mu}_{\rm s}(\bar{\xx}_i)^{-1} & 0 &0 &0 &0\\
 0 & {\bf \mu}_{\rm s}(\bar{\xx}_j)^{-1}  & 0 &0 &0\\
 0 & 0 & {\bf \mu}_{\rm s}(\bar{\xx}_k)^{-1}  & 0 &0 \\ 
0& 0& 0 &-t_0^{-1}  & 0 \\
0 & 0 & 0 & 0 &-t_0^{-1} 
\end{array} \right)
\ee
where ${\bf \mu}_{\rm s}(\bar{\xx}_i)$ stands for the $2\times2$ magnification tensor of the smooth lens component evaluated at the unperturbed image position $\bar{\xx}_i$, and $t_0$ is the time constant of the lens which is given by
\be
t_0 = \frac{1+z_{\rm lens}}{c}\frac{D_l D_s}{D_{ls}},
\ee
where $z_{\rm lens}$ is the redshift of the lens, $c$ is the speed of light, and $D_l$, $D_s$, and $D_{ls}$ are the angular diameter distances to the lens, to the source, and from the lens to the source, respectively. As it should be apparent from Eq.~(\ref{eq:matrix_A}), the matrix ${\bf A}\equiv{\bf A}(\qgal,\qenv,\qcos)$ depends only the smooth mass component of the lens, its environment, and the cosmological model. Substituting Eq.~(\ref{eq:linear_relation}) into Eq.~(\ref{eq:Delta_tl_int}), we can then perform the $\tl$ integration

\be
P({\bf t} | \qq,\qsub) =|{\bf A}|\int \frac{d\kk_{\rm L}}{(2\pi)^l} Q_{\langle N_{\rm d}\rangle }^{\rm dist}(\kk_{\rm L}|{\bf q},\qsub)e^{-i  ({\bf t}-\bar{{\bf t}})^{\rm T} {\bf A}^{\rm T} \kk_{\rm L}},
\ee
where $|{\bf A}|$ stands for the determinant of the matrix ${\bf A}$. We can finally substitute the above into Eq.~(\ref{likelihood_qsub}) to compute the data likelihood in the presence of a distributed population of substructures
\ba\label{eq:final_like_general}
\las({\bf d}|{\bf q},\qsub)&=&  \frac{|{\bf A}|}{\sqrt{(2\pi)^l|{\bf C_d}|}}\int d{\bf t}\int \frac{d\kk_{\rm L}}{(2\pi)^l} Q_{\langle N_{\rm d}\rangle }^{\rm dist}(\kk_{\rm L}|{\bf q},\qsub)e^{-i ({\bf t}-\bar{{\bf t}})^{\rm T} {\bf A}^{\rm T}  \kk_{\rm L}} e^{-\frac{1}{2}({\bf t}-{\bf d})^{\rm T}{\bf C_d}^{-1}({\bf t}-{\bf d})}\en
&=&  |{\bf A}|\int \frac{d\kk_{\rm L}}{(2\pi)^l} Q_{\langle N_{\rm d}\rangle }^{\rm dist}(\kk_{\rm L}|{\bf q},\qsub) e^{-\frac{1}{2}\kk_{\rm L}^{\rm T}{\bf A}{\bf C_d} {\bf A}^{\rm T}\kk_{\rm L}} e^{-i ({\bf d}-\bar{{\bf t}})^{\rm T}{\bf A}^{\rm T}  \kk_{\rm L}},
\ea
where we have assumed a Gaussian likelihood, $\las({\bf d}|{\bf t})\propto \exp{[-\frac{1}{2}({\bf t}-{\bf d})^{\rm T}{\bf C_d}^{-1}({\bf t}-{\bf d})]}$, where ${\bf C_d}$ is the data covariance matrix. Thus, the data likelihood is given by the Fourier transform of the product of $Q_{\langle N_{\rm d}\rangle }^{\rm dist}(\kk_{\rm L}|{\bf q},\qsub)$ with the Fourier conjugate distribution of $\las({\bf d}|{\bf t})$ evaluated at the residuals ${\bf d}-\bar{{\bf t}}$ between the data and the predictions from the smooth lens model. This makes sense: The $\kk_{\rm L}$ modes contributing most to the integral are those that can explain the residuals between the actual data and the smooth lens model, and the characteristic function $Q_{\langle N_{\rm d}\rangle }^{\rm dist}$ encodes whether these modes are likely to contribute to the residuals given the input substructure properties.

\subsection{Gaussian case}
We now specialize to the case where $Q_{\langle N_{\rm d}\rangle }^{\rm dist}(\kk_{\rm L}|{\bf q},\qsub)$ is well approximated by a multivariate Gaussian. Starting from Eq.~(\ref{eq:final_like_general}), we have
\ba\label{eq:Gaussian_like_for_code}
\las({\bf d}|{\bf q},\qsub)&=&|{\bf A}|\int \frac{d\kk_{\rm L}}{(2\pi)^l} e^{-\frac{1}{2}\kk_{\rm L}^{\rm T} {\bf C}_{\rm sub} \kk_{\rm L}} e^{-\frac{1}{2}\kk_{\rm L}^{\rm T}{\bf A}{\bf C_d} {\bf A}^{\rm T}\kk_{\rm L}} e^{-i ({\bf d}-\bar{{\bf t}})^{\rm T}{\bf A}^{\rm T}  \kk_{\rm L}}
\en
&=&\frac{|{\bf A}|}{\sqrt{(2\pi)^l |{\bf C}_{\rm sub} +{\bf A}{\bf C_d} {\bf A}^{\rm T}|}}e^{-\frac{1}{2}({\bf d}-\bar{{\bf t}})^{\rm T}{\bf A}^{\rm T}({\bf C}_{\rm sub} +{\bf A}{\bf C_d} {\bf A}^{\rm T})^{-1}{\bf A}({\bf d}-\bar{{\bf t}}) },
\ea
thus leading to a Gaussian data likelihood with total inverse covariance matrix given by
\be
{\bf C}_{\rm tot}^{-1} = {\bf A}^{\rm T}({\bf C}_{\rm sub} +{\bf A}{\bf C_d} {\bf A}^{\rm T})^{-1}{\bf A}.
\ee
This analysis shows that the effect of distributed unresolved mass substructures can be thought of as an additional source of noise that directly contributes to modeling uncertainty. This extra contribution to the net covariance matrix entering the likelihood describes the inherent mass modeling uncertainties due to the lumpiness of massive galaxies acting as strong gravitational lenses. Whether the inherent mass modeling uncertainties caused by mass substructures are relevant or not in the above likelihood depend on the observational precision of a given data set. Conversely, in the event that precise time delay and astrometric observations are available, they can be used with the above likelihood to constrain the physical quantities entering the covariance matrix ${\bf C}_{\rm sub}$.

\section{Discussion}\label{sec:discussion}
%
\begin{figure}
\includegraphics[width=0.69\textwidth]{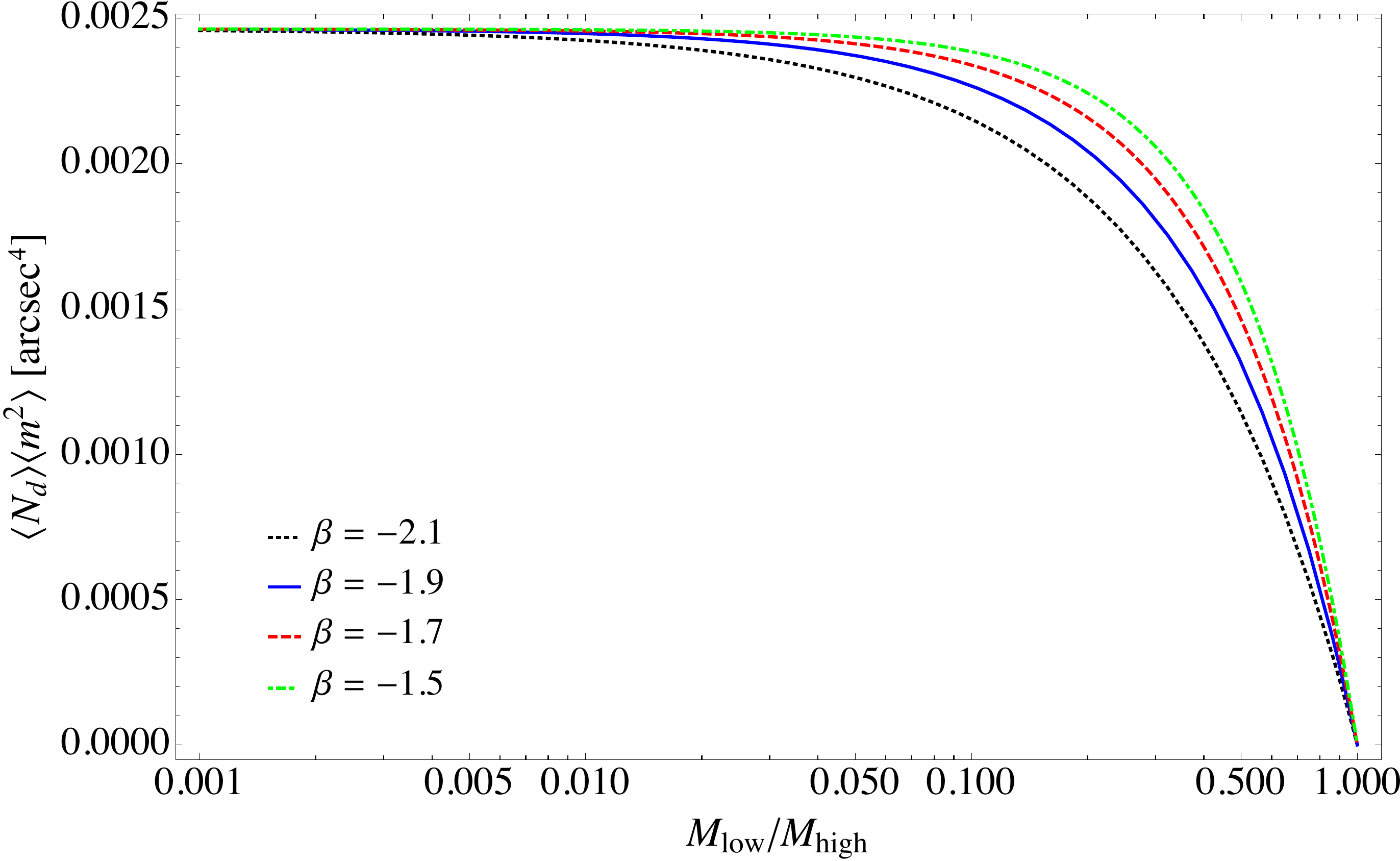}
\caption{Dependence of the product $\langle N_{\rm d}\rangle\langle m^2\rangle$ on the $M_{\rm low}/M_{\rm high}$ for four values of the slope of the mass function. In each case, we choose the normalization of the subhalo mass function $a_0$ such that all curves asymptote to the same value as $M_{\rm low}/M_{\rm high}\rightarrow0$. Our choice of normalization corresponds to $a_0 = 3.8\times10^{-10} M_\odot^{-1}$ at $M_0 = M_{\rm high}=10^{10} M_\odot$ when $\beta=-1.9$.}
\label{fig:m_low_dep}
\end{figure}
The analytical approximation developed in Sec.~\ref{sec:distant_analysis} allows us to not only marginalize over the masses and positions of distributed substructures but also to understand which of their physical properties are most relevant to gravitational lensing observables. For the physically relevant parameter space, the effect of distributed mass substructures on the lensing potential and its first derivative is approximately Gaussian, which implies that most of the relevant physics is encoded in the covariance matrix ${\bf C}_{\rm sub}$. Using Eqs.~(\ref{eq:average_N}) and (\ref{eq:covmat_sub_decomp}), this covariance matrix can be decomposed as follow
\be\label{eq:gen_cov_matrix_dis}
{\bf C}_{\rm sub}^{ij} = \langle N_d\rangle \langle m^2\rangle\langle \mathcal{O}_{\rm L}^i\mathcal{O}_{\rm L}^j\rangle =  \langle \kappa_{\rm sub}(r_{\rm ref})\rangle\frac{\langle m^2\rangle}{\langle m \rangle} \frac{ \langle \mathcal{O}_{\rm L}^i\mathcal{O}_{\rm L}^j\rangle}{ \mathcal{P}_r(r_{\rm ref})},
\ee
where $r_{\rm ref}$ is a reference radius where the amplitude of the convergence in mass substructures is set (taken to be the Einstein radius of the main lens in earlier sections of this paper), and where 
\be\label{eq:spatial_2-pt_correlator}
\langle \mathcal{O}_{\rm L}^i\mathcal{O}_{\rm L}^j\rangle \equiv \int d^2r \mathcal{P}_r(r,\theta) \mathcal{O}_{\rm L}^i\mathcal{O}_{\rm L}^j.
\ee
In going from the first to the second equality in Eq.~\eqref{eq:gen_cov_matrix_dis}, we use Eqs.~\eqref{eq:how_to_compute_N} and \eqref{eq:kappa_sub} to express $\langle N_{\rm d}\rangle$ in terms of $\ksub$. We give in Appendix \ref{app:cov_mat} useful expressions for the different entries of $\bf{C}_{\rm sub}$. The expected number of distributed substructures is given by Eq.~(\ref{eq:how_to_compute_N}), which for $\beta< -1$ and $M_{\rm low} \ll M_{\rm high}$ is approximately given by
\be
\langle N_{\rm d}\rangle \sim -\frac{a_0M_{\rm low}^{\beta+1}}{M_{0}^\beta(\beta+1)}(1-\mathcal{P}_r(<R_{\rm min})).
\ee
The second moment of the mass function $\langle m^2\rangle$ is easily computed from Eq.~(\ref{eq:subhalo_mass_function}), 
\be
\langle m^2\rangle = \frac{1}{\pi^2\Sigma_{\rm crit}^2} \frac{\beta+1}{\beta+3} \frac{M_{\rm high}^{\beta+3}-M_{\rm low}^{\beta+3}}{M_{\rm high}^{\beta+1}-M_{\rm low}^{\beta+1}} \sim -  \frac{1}{\pi^2\Sigma_{\rm crit}^2} \frac{\beta+1}{\beta+3}\frac{M_{\rm high}^{\beta+3}}{M_{\rm low}^{\beta+1}}\quad\text{for}\quad M_{\rm low} \ll M_{\rm high}, 
\ee
where the last approximation is valid for when $-3<\beta<-1$. The second moment of the mass function thus has a rather strong dependence on the minimal subhalo mass. Now, if we look at the product $ \langle N_{\rm d}\rangle\langle m^2\rangle$, we immediately see that the leading dependence on $M_{\rm low}$ cancels out for the physically relevant case $-3<\beta<-1$ and $M_{\rm low} \ll M_{\rm high}$
\be\label{eq:exact_dep_cov}
 \langle N_{\rm d}\rangle\langle m^2\rangle =  \frac{1}{\pi^2\Sigma_{\rm crit}^2}\frac{a_0}{\beta+3}\frac{M_{\rm high}^{\beta+3}-M_{\rm low}^{\beta+3}}{M_0^\beta}(1-\mathcal{P}_r(<R_{\rm min}))\sim \frac{1}{\pi^2\Sigma_{\rm crit}^2}\frac{a_0}{\beta+3}\frac{M_{\rm high}^{\beta+3}}{M_0^\beta}(1-\mathcal{P}_r(<R_{\rm min})). 
\ee
This shows that the scaling of the substructure covariance matrix depends mostly on the normalization of the mass function $a_0$ and on the largest subhalo mass $M_{\rm high}$.  We illustrate this behavior in Fig.~\ref{fig:m_low_dep} for different values of the mass function slope $\beta$. In each case, we choose the mass function normalization such that all curves asymptote to the same value as $M_{\rm low}/M_{\rm high}\rightarrow0$. In the regime where Gaussianity holds ($M_{\rm low}/M_{\rm high} \lesssim0.1$), we observe that $\langle N_{\rm d}\rangle\langle m^2\rangle$ is roughly constant as $M_{\rm low}/M_{\rm high}$ is varied. Measurement of this constant could provide a consistency test for standard cold dark matter theory. Since there are strong degeneracies between the different parameters in Eq.~(\ref{eq:exact_dep_cov}), the extraction of individual mass function parameters would require strong priors from either simulations or complementary observations. 

In addition to its dependence on the subhalo mass function, the covariance matrix ${\bf C}_{\rm sub}$ also encodes important information about the spatial distribution of distributed mass substructures. Since each entry of the covariance matrix depends on the spatial distribution through a different kernel [Eq.~(\ref{eq:spatial_2-pt_correlator})], it is reasonable to believe that lensing observables will provide good sensitivity to the spatial distribution of substructures. A detailed analysis of the sensitivity of different lensing observations to unresolved substructures will be carried in an upcoming work. Looking ahead, we expect that combining magnification information (mostly sensitive to local substructures), with astrometric fluctuations (sensitive to both local and distributed substructures) and time delay perturbations (sensitive to distributed substructures with some local sensitivity) will lead to a comprehensive picture of the satellite populations of distant lens galaxies. 

\section{Conclusion}\label{sec:conclusion}
In this paper, we have computed from first principles the probability distribution of lensing potentials and deflections in the presence of an unresolved population of mass substructures that are located beyond the strong lensing region. We have determined that for a realistic substructure population, the distribution of lensing potential and deflection perturbations is close to a multivariate Gaussian. We have computed the leading order deviations from Gaussianity and used them to determine when the probability distribution ceases to be well approximated by Gaussian statistics. We have shown in Sec.~\ref{marginalization} how our technique can be used to efficiently marginalize over the properties of distributed mass substructures without having to perform costly numerical simulations of mass substructure populations. 

For simplicity, we have treated distributed substructures as independent point masses, which we believe is an excellent approximation for subhalos far away from lensed images. We note that our approach can straightforwardly be generalized to clustered and extended substructures if we compute the theory covariance matrix as
\be
{\bf C}_{\rm sub}^{ij} \propto \int d^2r \int d^2r' K_i({\bf r}) K_j({\bf r'}) \langle \kappa_{\rm sub}({\bf r})\kappa_{\rm sub}({\bf r}')\rangle,
\ee
where the kernels $K_i$ and $K_j$ depend on which linear lensing quantities are being used. We note that in the point-mass limit the above expression reduces to Eq.~(\ref{eq:gen_cov_matrix_dis}). Therefore, we can see that, in the general case, we are really probing the ensemble-averaged two-point correlation function of the distributed substructure convergence field. This two-point function, which is in general neither homogeneous nor isotropic, can be directly measured in $N$-body or semi-analytic simulations, hence providing a way to assess the importance of subhalo clustering (the two-halo term) and to test the accuracy of the point-mass approximation. We leave such tests to future work.

In the present manuscript, we have focused our attention on time delay and astrometry perturbations since these are the lensing observables that are most sensitive to distributed mass substructures. Expanding our analysis to include mass substructures near lensed images would allow the incorporation of magnification information into our framework. Together, the relative flux measurements, positions, and time delays between lensed images have power to constrain both the local and distributed substructure populations of a lens galaxy, given appropriate levels of measurement precision. Quantifying these precision levels in detail will vary from system to system, and will be the subject of future work. As an example, in the case of time delays, the fluctuations caused by distributed substructures are demonstrated to be $\sqrt{\Delta t^2}<1$~day  \cite{Keeton:2009ab}, suggesting time delay precision levels on the order of hours.

One of the main advantages of having an analytical framework to handle mass substructures is that it allows efficient exploration of degeneracies between substructure effects ($\qsub$), on the one hand, and the macrolens ($\qgal$), its environment ($\qenv$) and possible line-of-sight structures ($\qlos$) on the other. Exploring and marginalizing over these degeneracies is important in assessing the sensitivity of current and future data to the detection of a population of nonluminous mass substructures in the outskirts of distant galaxies. Such a detection would confirm a key prediction, or offer a quantitative challenge, of our current paradigm for structure formation based on the CDM scenario. The synthesis of all lensing observables which are sensitive to different combinations of local and distant substructures, measured with sufficient precision, have the potential to produce a complete picture of the substructure mass function. The stochastic millilensing framework developed here is a necessary step toward this goal.

\acknowledgments
We thank Geoffrey Bryden, James Bullock, Curt Cutler, Olivier Dor\'e, David Hogg, Jeffrey Jewel, James Taylor, and Michele Vallisneri for useful conversations. The work of F.-Y. C.-R. was performed in part at the California Institute of Technology for the Keck Institute for Space Studies, which is funded by the W. M. Keck Foundation. F.-Y. C.-R. thanks the Aspen Center for Physics, where some of this work was performed. The Aspen Center for Physics is supported by the National Science Foundation under Grant No. 1066293.  Part of the research described in this paper was carried out at the Jet Propulsion Laboratory, California Institute of Technology, under a contract with the National Aeronautics and Space Administration (NASA). L. A. M.  gratefully acknowledges support by the NASA ATFP program through Award No. 399131.02.02.02.98. C. R. K. acknowledges support from the National Science Foundation under Grant No. AST-0747311. The research of K. S. is supported in part by a Natural Science and Engineering Research Council (NSERC) of Canada Discovery Grant. D. A. G. acknowledges the support of the NASA Undergraduate Internship and Student Internship programs. 

\appendix
%
\section{Subhalo Scale and Truncation Radii}\label{app:scale_and_trunc}
\subsection{Scale radius}
In this appendix, we derive a relation between the total mass $M_{\rm sub}$ of a subhalo and its scale radius $r_{\rm s}$. Our starting point is the relation between the maximum circular velocity inside a subhalo $v_{\rm max}$ and the radius $r_{\rm max}$ at which this velocity occurs \cite{Garrison-Kimmel:2014vqa}
\be\label{eq:velocity_radius_rel}
\left(\frac{r_{\rm max}}{1 \,{\rm kpc}}\right) = (0.72 \pm 0.25) \left(\frac{v_{\rm max}}{10\,{\rm km\,s^{-1}}}\right)^{1.47}.
\ee
By the virial theorem, we also have 
\be\label{eq:virial}
\frac{ G\, M_{\rm sub}(< r_{\rm max})}{r_{\rm max}}=  v_{\rm max}^2,
\ee
where $G$ is the gravitational constant and $M_{\rm sub}(< r_{\rm max})$ is the subhalo mass enclosed within $r_{\rm max}$, which for a smoothly truncated NFW density profile given in Eq.~(\ref{eq:truncated_NFW}) is
\be\label{eq:mass_within_Rmax}
M_{\rm sub}(<r_{\rm max}) = M_{\rm sub}\frac{4(1+x_{\rm max})\tau \arctan{(x_{\rm max}/{\tau)}} -2 x_{\rm max} (1+\tau^2) +(1+x_{\rm max})(\tau^2-1)\ln{\left[\frac{\tau^2(1+x_{\rm max})^2}{(x_{\rm max}^2+\tau^2)}\right]}}{2(1+x_{\rm max})\left((\tau^2-1)\ln{\tau}+\pi\tau-(\tau^2+1)\right)},
\ee
where $M_{\rm sub}$ is the total mass of the subhalo given in Eq.~(\ref{eq:total_tNFW_mass}), $\tau \equiv r_{\rm t}/r_{\rm s}$ (where $r_{\rm t}$ is the tidal radius), and $x_{\rm max} \equiv r_{\rm max}/r_{\rm s}$. We substitute the above expression into Eq.~(\ref{eq:virial}) and maximize the left-hand side to find the radius at which the maximum circular velocity occurs. The resulting equation is not solvable analytically, but for realistic values of $\tau$, we obtain 
\be\label{eq:xmax_relation}
x_{\rm max} \simeq 2.1626 \left(1-\frac{1}{1+\tau^2}\right)^2\quad\text{for}\quad \tau \gtrsim 7.
\ee
We then substitute the above into Eqs.~(\ref{eq:mass_within_Rmax}) and (\ref{eq:virial}) and use Eq.~(\ref{eq:velocity_radius_rel}) to eliminate $v_{\rm max}$ from Eq.~(\ref{eq:virial}) in order to obtain a relation between the scale radius $r_{\rm s}$ and the total mass of the subhalo $M_{\rm sub}$
\be
\left(\frac{r_{\rm s}}{1\,{\rm kpc}}\right) = 1.0\pm0.3\left(\frac{M_{\rm sub}}{10^9\,M_\odot}\right)^{0.735},
\ee
where we have neglected a weak dependence on $\tau$ since it leads to changes that are smaller than the scatter about the mean.
\subsection{Tidal truncation radius}
We use the result of Refs.~\cite{Binney2008,Tormen:1997ik} for the tidal truncation radius $r_{\rm t}$ of a subhalo of mass $M_{\rm sub}$ located at a distance $r_{3{\rm D}}$ from the main halo center, 
\be
r_{\rm t} =  \left(\frac{M_{\rm sub}}{[2-{\rm d}\ln{M_{\rm main}}/{\rm d}\ln{r_{3{\rm D}}}]M_{\rm main}(<r_{3{\rm D}})}\right)^{1/3}r_{3{\rm D}},
\ee
where $M_{\rm main}$ is the mass of the main lens halo and where both $r_{\rm t}$ and $r_{3{\rm D}}$ are radii in three-dimensional space (not projected). Since the truncation radius is most relevant to the stochastic lensing signal for subhalos lying close or within the Einstein radius of the lens, it is sufficient to specify the mass distribution in the vicinity of the lensed images. It this work, we focus on power-law mass models which have been shown to provide good fits to many gravitational lenses. The projected mass density divided by the critical density for lensing for these models is given by \cite{Keeton:ab}
\be
\kappa_{\rm main}(r) = \frac{1}{2}\left(\frac{b}{r}\right)^{2-\alpha_{\rm main}},\qquad (\alpha_{\rm main}\neq2)
\ee
where $b$ is a length scale closely related to the Einstein radius of the main lens and $\alpha_{\rm main}$ is the power-law index of the density profile. We can deproject this relation to obtain the 3-d mass profile of the main lens
\be
\rho_{\rm main}(r_{3{\rm D}} ) = \frac{1}{2\sqrt{\pi} b}\Sigma_{\rm crit} \frac{\Gamma\left(\frac{3-\alpha_{\rm main}}{2}\right)}{\Gamma\left(\frac{2-\alpha_{\rm main}}{2}\right)}\left(\frac{b}{r_{3{\rm D}} }\right)^{3-\alpha_{\rm main}},
\ee
where $\Gamma(x)$ is the gamma function. This relation is easily integrated to obtain $M_{\rm main}(<r_{3{\rm D}} )$, which leads a tidal truncation radius given by
\be
r_{\rm t} =  \left(\frac{\alpha_{\rm main}}{2-\alpha_{\rm main} }\frac{ \Gamma\left(\frac{2-\alpha_{\rm main}}{2}\right)}{ \Gamma\left(\frac{3-\alpha_{\rm main}}{2}\right)} \frac{ M_{\rm sub}}{2\sqrt{\pi}\Sigma_{\rm crit}b^2}\right)^{1/3}\left(\frac{b}{r_{3{\rm D}} }\right)^\frac{\alpha_{\rm main}}{3}r_{3{\rm D}} , \qquad (\alpha_{\rm main}\neq2).
\ee
We are generally interested in quasi-isothermal inner density profiles ($\alpha_{\rm main}\sim 1$) for the main lens halo, which implies that we generally have
\be
 r_{\rm t}\propto M_{\rm sub}^{1/3} r_{3{\rm D}} ^{2/3}
 \ee
for a subhalo of mass $M_{\rm sub}$ located at a distance $r_{3{\rm D}}$ from the center of the main lens halo.

\section{Spatial moments of the substructure distribution}\label{app:spatial_momts}
In this appendix, we simplify the structure of the spatial integral appearing in Eq.~\eqref{eq:moment_exp} using a multinomial expansion. Keeping only the first $N_{\rm mult}$ multipoles in Eq.~(\ref{eq:def_O}), the spatial integral takes the form
\ba
\int_{\cH_{\rm d}} d^2r \mathcal{P}_r(r,\theta;\qsub) ( \kk_{\rm L}\cdot \vec{\mathcal{O}}_{\rm L} )^n&=& \int_{\cH_{\rm d}} d^2r \mathcal{P}_r(r,\theta;\qsub)  \Bigg(- \sum_{p=1}^{N_{\rm mult}} \frac{1}{r^p}\left[\kk_{\rm L}\cdot\vec{A}_p\cos{(p\,\theta)} + \kk_{\rm L}\cdot\vec{B}_p \sin{(p\,\theta)}\right]\Bigg)^n\en
&=&(-1)^n\int_{\cH_{\rm d}} d^2r \mathcal{P}_r(r,\theta;\qsub) \Bigg(\sum_{\sum p_i =n} \binom{n}{p_1,p_2,\ldots,p_{2N_{\rm mult}}}\en
&&\qquad \times \prod_{t=1}^{N_{\rm mult}} \left[\frac{1}{r^{t}}\kk_{\rm L}\cdot\vec{A}_t\cos{(t\,\theta)}\right]^{p_t}\prod_{s=1}^{N_{\rm mult}}  \left[\frac{1}{r^{s}} \kk_{\rm L}\cdot\vec{B}_s \sin{(s\,\theta)}\right]^{p_{N_{\rm mult}+s}}\Bigg)\en
&=&(-1)^n\sum_{\Vert \pp{}\Vert =n} \binom{n}{\pp{}} K_{ \pp{}} \prod_{t=1}^{N_{\rm mult}} (\kk_{\rm L}\cdot\vec{A}_t)^{p_t}\prod_{s=1}^{N_{\rm mult}}(\kk_{\rm L}\cdot\vec{B}_s)^{p_{N_{\rm mult}+s}},
\ea
where $\pp{} = \{p_1,p_2,\ldots,p_{2 N_{\rm mult}}\}$ is a multi-index with $\Vert \pp{}\Vert = \sum_{j=1}^{2N_{\rm mult}} p_j$, and where the kernel $K_{\pp{}}$ is given by
\be\label{app:eq:spatial_kernel}
K_{\pp{}} = \int_{\cH_{\rm d}} d^2r\, \mathcal{P}_r(r,\theta;\qsub) \left(\frac{1}{r}\right)^{\sum_{j=1}^{ N_{\rm mult}} j\, (p_j+p_{N_{\rm mult}+j})}\,\,\prod_{t=1}^{N_{\rm mult}} \cos{(t\,\theta)}^{p_t} \prod_{s=1}^{N_{\rm mult}}\sin{(s\,\theta)}^{p_{N_{\rm mult}+s}},
\ee
and where
\be
\binom{n}{\pp{}} = \frac{n!}{p_1!p_2!\ldots p_{2 N_{\rm mult}}!}
\ee
is the multinomial coefficient.

\section{Edgeworth expansion of the characteristic function}\label{app:edgeworth_exp}
In this appendix, we perform the Edgeworth expansion of the characteristic function $Q_{\langle N_{\rm d}\rangle }^{\rm dist}$ in powers of $ \langle N_{\rm d}\rangle^{-1/2}$. We start by writing the Cholesky decomposition of the covariance matrix ${\bf C}_{\rm sub}$ as
\be\label{eq:covmat_sub_decomp}
{\bf C}_{\rm sub} = \langle N_{\rm d}\rangle\langle m^2\rangle \Lambda \Lambda^{\rm T},
\ee
which is always possible since ${\bf C}_{\rm sub}$ is a symmetric positive-definite matrix, and where we have pulled out the overall scaling with the average number of mass substructures and the second moment of the mass function. We can then define a new normalized Fourier conjugate variable $\tilde{\kk}_{\rm L}$ as
\be
\tilde{\kk}_{\rm L} = \sqrt{  \langle N_{\rm d}\rangle\langle m^2\rangle} \Lambda^{\rm T} \kk_{\rm L}
\ee
and express the characteristic function as a function of it. Since the characteristic function $\tilde{Q}_{\langle N_{\rm d}\rangle }^{\rm dist}(\tilde{\kk}_{\rm L})$ of the normalized variable  $\tilde{\kk}_{\rm L}$ is related to that given in Eq.~(\ref{eq:nongaussian_terms}) by
\be
 \tilde{Q}_{\langle N_{\rm d}\rangle }^{\rm dist}(\tilde{\kk}_{\rm L})  = \frac{1}{ \sqrt{\langle N_{\rm d}\rangle\langle m^2\rangle}|\Lambda|}Q_{\langle N_{\rm d}\rangle }^{\rm dist}\left(\frac{(\Lambda^{\rm T})^{-1}\tilde{\kk}_{\rm L}}{\sqrt{\langle N_{\rm d}\rangle\langle m^2\rangle}} \right),
\ee
we obtain 
\ba\label{eq:tilde_Q_before_exp}
\tilde{Q}_{\langle N_{\rm d}\rangle }^{\rm dist}(\tilde{\kk}_{\rm L}) &= &\frac{e^{-\frac{1}{2}\tilde{\kk}_{\rm L}\cdot  \tilde{\kk}_{\rm L}}}{\sqrt{\langle N_{\rm d}\rangle\langle m^2\rangle}|\Lambda|} \exp{\Bigg[\sum_{n=3}^\infty\frac{(-1)^n\langle m^n\rangle}{n!\langle N_{\rm d}\rangle^{n/2-1}\langle m^2\rangle^{n/2}}}\en
&&\qquad\qquad\times \left(\sum_{\Vert p\Vert =n} \binom{n}{\pp{}} K_{ \pp{}} \prod_{t=1}^{N_{\rm mult}} (i\tilde{\kk}_{\rm L}^{\rm T}\Lambda^{-1}\vec{A}_t)^{p_t}\prod_{s=1}^{N_{\rm mult}}(i\tilde{\kk}_{\rm L}^{\rm T}\Lambda^{-1} \vec{B}_s)^{p_{N_{\rm mult}+s}}\right)\Bigg],
\ea
where we have taken the mean vector ${\bf u}\equiv \langle \Delta\tl\rangle$ to vanish, but the above result can straightforwardly be generalized to a nonzero mean values of linear lensing quantities. To make the notation more compact, we define 
\be
\langle (i\tilde{\kk}_{\rm L}\cdot \vec{\mathcal{O}}_{\rm L})^n\rangle \equiv\sum_{\Vert p\Vert =n} \binom{n}{\pp{}} K_{ \pp{}} \prod_{t=1}^{N_{\rm mult}} (i\tilde{\kk}_{\rm L}^{\rm T}\Lambda^{-1}\vec{A}_t)^{p_t}\prod_{s=1}^{N_{\rm mult}}(i\tilde{\kk}_{\rm L}^{\rm T}\Lambda^{-1} \vec{B}_s)^{p_{N_{\rm mult}+s}}.
\ee
We then expand the exponential in Eq.~(\ref{eq:tilde_Q_before_exp}) to obtain the proper Edgeworth expansion of $\tilde{Q}_{\langle N_{\rm d}\rangle }^{\rm distant}(\tilde{\kk}_{\rm L})$
\ba\label{eq:real_edgeworth_exp}
\tilde{Q}_{\langle N_{\rm d}\rangle }^{\rm distant}(\tilde{\kk}_{\rm L}) &=& \frac{e^{-\frac{1}{2}\tilde{\kk}_{\rm L}\cdot  \tilde{\kk}_{\rm L}}}{\sqrt{\langle N_{\rm d}\rangle\langle m^2\rangle}|\Lambda|}\Bigg(1-\frac{1}{ \langle N_{\rm d}\rangle^{1/2}}\left[ \frac{ \langle m^3\rangle\langle (i\tilde{\kk}_{\rm L}\cdot \vec{\mathcal{O}}_{\rm L})^3\rangle}{3!\langle m^2\rangle^{3/2}}\right]+\frac{1}{ \langle N_{\rm d}\rangle}\left[ \frac{ \langle m^4\rangle\langle (i\tilde{\kk}_{\rm L}\cdot \vec{\mathcal{O}}_{\rm L})^4\rangle}{4!\langle m^2\rangle^2} + \frac{ \left( \langle m^3\rangle\langle (i\tilde{\kk}_{\rm L}\cdot \vec{\mathcal{O}}_{\rm L})^3\rangle\right)^2}{72\langle m^2\rangle^3}\right]\en
&& -\frac{1}{ \langle N_{\rm d}\rangle^{3/2}}\left[ \frac{ \langle m^5\rangle\langle (i\tilde{\kk}_{\rm L}\cdot \vec{\mathcal{O}}_{\rm L})^5\rangle}{5!\langle m^2\rangle^{5/2}} + \frac{ \langle m^3\rangle\langle (i\tilde{\kk}_{\rm L}\cdot \vec{\mathcal{O}}_{\rm L})^3\rangle\langle m^4\rangle\langle (i\tilde{\kk}_{\rm L}\cdot \vec{\mathcal{O}}_{\rm L})^4\rangle}{144\langle m^2\rangle^{7/2}}+\frac{\left(\langle m^3\rangle\langle (i\tilde{\kk}_{\rm L}\cdot \vec{\mathcal{O}}_{\rm L})^3\rangle\right)^3}{1296\langle m^2\rangle^{9/2}}\right]\en
&&+\,O\left(\langle N_{\rm d}\rangle^{-2}\right)\Bigg).
\ea
%

\section{Covariance Matrix for Linear Lensing Quantities}\label{app:cov_mat}
%
\subsection{General expressions in the presence of circular symmetry}
In general, the covariance matrix for linear lensing quantities is given by
\be
{\bf C}_{\rm sub}^{ij} = \langle N_{\rm d}\rangle\langle m^2\rangle \int d^2r \mathcal{P}_{r}(r,\theta) \left[ \sum_{p=1}^\infty \frac{1}{r^p}\left[A^{(i)}_p\cos{(p\,\theta)} + B^{(i)}_p \sin{(p\,\theta)}\right] \right]\left[ \sum_{t=1}^\infty \frac{1}{r^t}\left[A^{(j)}_t\cos{(t\,\theta)} + B^{(j)}_t \sin{(t\,\theta)}\right] \right],
\ee
where $A^{(i)}_p$ denotes the $i$th component of the vector $\vec{A}_p$. In the case of a circular halo with $\mathcal{P}_r(r,\theta)\equiv \mathcal{P}_r(r)$, the above expression dramatically simplifies. In the following, we provide convenient expressions for the different entries of the covariance matrix for the linear lensing quantities. We take the position of image $i$ to be $\xx_i = (r_i \cos{\theta_i}, r_i\sin{\theta_i})$, that of image $j$ to be $\xx_j = (r_j \cos{\theta_j}, r_j\sin{\theta_j})$, and the reference point for the projected lensing potential is $\xx_{\rm ref} = (r_{\rm ref}\cos{\theta_{\rm ref}}, r_{\rm ref}\sin{\theta_{\rm ref}})$. We use the notation $\phi_{\rm sub}^{(i)} $ to denote the difference in projected potential between image $i$ and the reference point, that is, $\phi_{\rm sub}^{(i)} \equiv \phi_{\rm sub}(\xx_i) - \phi_{\rm sub}(\xx_{\rm ref})$. The covariances for the deflections are given by
\be
\langle \alpha_{{\rm sub}, x}^{(i)} \alpha_{{\rm sub}, x}^{(j)} \rangle = \langle \alpha_{{\rm sub}, y}^{(i)} \alpha_{{\rm sub}, y}^{(j)} \rangle=\langle N_{\rm d}\rangle\langle m^2\rangle\sum_{p=0}^\infty r_i^{p} r_j^{p}\mathcal{K}[2(p+1)]\cos{\{p(\theta_i-\theta_j)\}},
\ee
\be
\langle \alpha_{{\rm sub}, x}^{(i)} \alpha_{{\rm sub}, y}^{(j)} \rangle = \langle N_{\rm d}\rangle\langle m^2\rangle\sum_{p=1}^\infty r_i^{p} r_j^{p}\mathcal{K}[2(p+1)]\sin{\{p(\theta_i-\theta_j)\}},
\ee
where the kernel $\mathcal{K}[n]$ is given by
\be\label{eq:actual_spatial_kernel}
\mathcal{K}[n] \equiv \pi \int_{R_{\rm min}}^{R_{\rm max}} dr\, r\, \mathcal{P}_r(r) \frac{1}{r^n}
\ee
where the leading factor of $\pi$ comes from the angular integration over $\theta$. We give explicit expressions for $\mathcal{K}[n]$ in the next subsection for two choices of spatial distributions. The cross terms between deflections and lensing potential are
\be
\langle \alpha_{{\rm sub}, x}^{(i)} \phi_{\rm sub}^{(j)} \rangle =\langle N_{\rm d}\rangle\langle m^2\rangle\sum_{p=1}^\infty \frac{r_i^{p-1}\mathcal{K}[2p]}{p} \left( r_j^{p}\cos{[(p-1)\theta_i-p\theta_j]}-r_{\rm ref}^p\cos{[(p-1)\theta_i-p\theta_{\rm ref}]} \right),
\ee
\be
\langle \alpha_{{\rm sub}, y}^{(i)} \phi_{\rm sub}^{(j)} \rangle =\langle N_{\rm d}\rangle\langle m^2\rangle\sum_{p=1}^\infty \frac{r_i^{p-1}\mathcal{K}[2p]}{p} \left( -r_j^{p}\sin{[(p-1)\theta_i-p\theta_j]}+r_{\rm ref}^p\sin{[(p-1)\theta_i-p\theta_{\rm ref}]} \right).
\ee
Finally, the covariance between projected lensing potential is given by
\be
\langle \phi_{\rm sub}^{(i)} \phi_{\rm sub}^{(j)} \rangle =\langle N_{\rm d}\rangle\langle m^2\rangle\sum_{p=1}^\infty \frac{\mathcal{K}[2p]}{p^2} \left( r_i^{p} r_j^{p}\cos{[p(\theta_i-\theta_j)]}+r_{\rm ref}^p\left( r_{\rm ref}^p-r_i^p\cos{[p(\theta_i-\theta_{\rm ref})]}-r_j^p\cos{[p(\theta_j-\theta_{\rm ref})]} \right)\right).
\ee
\subsection{$\mathcal{K}$ kernel for different choices of spatial distributions}
We now provide explicit expressions for the spatial kernel given in Eq.~(\ref{eq:actual_spatial_kernel}) above.
\subsubsection{Power-law spatial distribution}
We consider the following power-law spatial distribution
\be
\mathcal{P}_r(r) = \frac{\eta r^{\eta-2}}{2\pi(R_{\rm max}^{\eta}-R_{\rm min}^{\eta})},\quad (0<\eta\leq2),
\ee
where the case $\eta=1$ corresponds to an isothermal profile, while $\eta = 2$ reduces to the case of a uniform spatial distribution. For this distribution, the kernel takes the form
\be
\mathcal{K}[n]=
\begin{cases}
\frac{\eta(R_{\rm min}^{\eta-n} - R_{\rm max}^{\eta-n})}{2(n-\eta)(R_{\rm max}^\eta-R_{\rm min}^\eta)} & \text{if } n \neq \eta,\\
\frac{\eta}{2(R_{\rm max}^\eta-R_{\rm min}^\eta)}\ln{\left(\frac{R_{\rm max}}{R_{\rm min}}\right)} & \text{if } n=\eta.
\end{cases}
\ee
\subsubsection{Cored spatial distribution}
In this case, the spatial distribution of substructure is given by
\be
 \mathcal{P}_r(r)  =\left( \frac{1}{2\pi r_{\rm c}^2}\frac{1}{W(R_{\rm max}/r_{\rm c})-W(R_{\rm min}/r_{\rm c})}\right)\frac{1}{(1+ (r/r_{\rm c}))^2},\quad\text{where}\quad W(x) = \frac{1}{1+x}+\ln{(1+x)},
 \ee
 and where $r_{\rm c}$ is the core radius. The kernel is then given by
 \ba
 \mathcal{K}[n]&=&\frac{1}{2}\frac{1}{W(R_{\rm max}/r_{\rm c})-W(R_{\rm min}/r_{\rm c})}\Bigg[\frac{R_{\rm min}^{1-n}}{r_{\rm c}+R_{\rm min}} -\frac{R_{\rm max}^{1-n}}{r_{\rm c}+R_{\rm max}}\en
&&\qquad\qquad +\frac{(n-1)}{n}\left[\frac{1}{R_{\rm max}^n} {}_2F_1\left(1,n;1+n;-\frac{r_{\rm c}}{R_{\rm max}}\right)-\frac{1}{R_{\rm min}^n} {}_2F_1\left(1,n;1+n;-\frac{r_{\rm c}}{R_{\rm min}}\right)\right]\Bigg],
 \ea
 where ${}_2F_1(a,b;c;x)$ is the ordinary (Gaussian) hypergeometric function.
 
 \newpage
 \section{Tables of used symbols}
\begin{table}[h!]
\begin{tabular}{|c|c|}
\hline
{\bf Symbol} & {\bf Description}  \\
\hline
$\vec{\alpha}$ & Lensing deflection vector \\
\hline
$\alpha_{\rm main}$ &  Slope of the main lens galaxy projected density profile \\
\hline
$\asub$ & Deflection vector caused by substructures \\
\hline
$\alpha_{{\rm sub},x}^{(i)}$ & x-component of the net deflection caused by substructures at the $i$th image\\
\hline
$\beta$ & Slope of the subhalo mass function \\ 
\hline
$\gamma_{\rm c}, \gamma_{\rm s}$ & Lensing shear components \\
\hline
$\gamma_{\rm sub}$ & Magnitude of shear caused by substructures \\
\hline
$\de_{\rm D}^k$ & $k$-dimensional Dirac delta function\\
\hline
$\kappa$ & Lensing convergence \\
\hline
$\ksub$ & Lensing convergence in mass substructures \\
\hline
$\Lambda$ & Upper triangular matrix from the Cholesky decomposition of the covariance matrix \\
\hline
$\mu_{\rm s}(\xx)$ & Magnification tensor for the smooth lens component, evaluated at position $\xx$ \\
\hline
$\Pi(\qq)$ & Prior probability distribution of $\qq$ \\
\hline
$\rho({\bf r})$ & Three-dimensional density profile \\
\hline
$\Sigma_{\rm crit}$ & Critical density for gravitational lensing \\
\hline
$\tau$ & $\equiv r_{\rm t}/r_{\rm s}$ \\
\hline
$\phi$ & Projected gravitational potential \\
\hline
$\psub$ & Substructure contribution to $\phi$ \\
\hline
$\psub^{(i)}$ & Projected gravitational potential difference between the $i$th image and the reference point \\
\hline
$\Phi_1(\xx|\qq)$ & PDF of $\xx$ given $\qq$, where $\xx$ is a single independent random variable\\
\hline 
$\Phi_N(\xx|\qq)$ & PDF of $\xx$ given $\qq$, where $\xx$ is the sum of $N$ independent random variables\\
\hline
\end{tabular}
\caption{Summary Greek symbols used throughout the manuscript. }\label{param_sum_table1}
\end{table}
\begin{table}[b!]
\begin{tabular}{|c|c|}
\hline
{\bf Symbol} & {\bf Description}  \\
\hline
$a_0$ & Normalization of the subhalo mass function \\
\hline
${\bf A}$ & Transformation matrix between the linear lensing quantities and the lensing observables \\
\hline
$\vec{A}_p$, $\vec{B}_p$ & Vectors of $p$th-order multipole coefficients \\
\hline
$b$ & Approximate Einstein radius of main lens galaxy\\
\hline
${\bf c}_{\rm sub}$ & Vector containing the individual substructure parameters \\
\hline 
${\bf c}_{\rm sub}^{(1)}$ & Substructure parameters for a single mass clump \\
\hline
${\bf C}_{\rm sub}$ & Covariance matrix for the linear lensing quantities \\
\hline
${\bf C}_{\bf d}$ & Data covariance matrix \\
\hline 
${\bf d}$ &  Data vector \\
\hline
$D_l$, $D_s$, $D_{ls}$ & Angular diameter distances to the lens, to the source, and from the lens to the source \\
\hline
$h$ & Projection of $r_{\rm 3D}$ along the line of sight \\
\hline
$\mathcal{H}_{\rm d}$ & Area occupied by the distributed substructures \\
\hline
$\kk_{\rm L}$ & Fourier variable conjugate to $\Delta\tl$ \\
\hline
$K_{\bf p}$ & Spatial kernel for multipole expansion \\
\hline
$\mathcal{K}[n]$ & $n$th-order spatial kernel for the multipole expansion of the substructure covariance matrix \\
\hline
$\las(\xx|{\bf d})$ & Likelihood of theory vector $\xx$ given data vector ${\bf d}$\\
\hline
$m$ & Normalized substructure mass $\equiv M_{\rm sub}/(\pi \Sigma_{\rm crit})$ \\
\hline
$M_0$ & Reference mass for subhalo mass function  \\
\hline
$M_{\rm main}$ & Mass of main lens galaxy \\
\hline
\end{tabular}
\caption{Summary of roman and scripted symbols used throughout the manuscript.}\label{param_sum_table2}
\end{table}
\begin{table}[b!]
\begin{tabular}{|c|c|}
\hline
{\bf Symbol} & {\bf Description}  \\
\hline
$M_{\rm min}$ & Low-mass bound of the subhalo mass function \\
\hline
$M_{\rm max}$ & High-mass bound of the subhalo mass function \\
\hline 
$M_{\rm main}$ & Mass of main lens halo \\
\hline
$M_{\rm NFW}$ & Mass normalization of the NFW profile \\
\hline
$M_{\rm sub}$ & Substructure mass \\
\hline
$N$, $\langle N \rangle$ & Number of mass substructure, average number of mass substructures\\
\hline
$N_{\rm l}$, $N_{\rm d}$ & Number of local and distributed substructures \\
\hline
$N_{\rm img}$ & Number of lensed images  \\
\hline
$N_{\rm mult}$ & Maximum number of multipole included in the expansion \\
\hline
$\vec{\mathcal{O}}_{\rm L}$ & Vector of stochastic random variables ($\equiv \Delta\tl/m$)  \\
\hline
${\bf p}$ & Multi-index (vector of multiple indices) \\
\hline
$P(\xx|\qq)$ & PDF of vector $\xx$ given parameter vector $\qq$\\
\hline
$P_{\rm sub}$ & PDF for the mass and position of substructures \\
\hline
$\mathcal{P}_M(M_{\rm sub})$ & PDF for the substructure mass \\
\hline
$\mathcal{P}_r(r)$ & PDF for the spatial distribution of substructures \\
\hline
$\qcos,\qdm$ & Cosmological parameters, Dark matter parameters \\
\hline
$\qenv,\qlos$ & Lens environment parameters, Line-of-sight structure parameters   \\
\hline
$\qgal$ & Macrolens parameters  \\
\hline
$\qsub$ & Substructure population parameters \\
\hline
$\qq$ & $\equiv\{\qgal,\qenv,\qlos \}$  \\
\hline
$q_1(\kk|\qq)$ & Characteristic function of $\Phi_1(\xx|\qq)$  \\
\hline
$q_1^{\rm local}$, $q_1^{\rm dist}$ & Local and distributed contribution to $q_1$ \\
\hline
$Q_N(\kk|\qq)$ & Characteristic function of $\Phi_N(\xx|\qq)$ \\
\hline
$Q_N^{\rm local}$,  $Q_N^{\rm dist}$ & Local and distributed contribution to $Q_N$\\
\hline
$r,\theta$ & Two-dimensional polar coordinates \\
\hline 
$r_{\rm 3D}$ & Three-dimensional position of subhalo within the lens galaxy \\
\hline
$r_{\rm c}$ & Core radius of the substructures' spatial distribution \\
\hline
$r_{\rm max}$ & Radius where $v_{\rm max}$ occurs \\
\hline
$r_{\rm ref}$ & Reference radius where zero of projected potential is defined \\
\hline
$r_{\rm s}$ & Scale radius of subhalo \\
\hline
$r_{\rm sub}$ & Three-dimensional distance from center of subhalo \\
\hline
$r_{\rm t}$ & Tidal radius of subhalo\\
\hline
$R_{\rm ein}$ & Einstein radius of the lens \\
\hline 
$R_{\rm min}$ &  Minimum radius of the distributed substructure population \\
\hline
$R_{\rm max}$ & Maximum radius of the distributed substructure population \\
\hline
$R_{\rm vir}$ & Virial radius of the main lens galaxy \\
\hline
${\bf t}$ & Vector of theory observables \\
\hline
$\bar{\bf t}$ & Contribution to ${\bf t}$ from smooth mass component and environment\\
\hline
$\tl$ & Vector of linear lensing quantities \\
\hline
$\bar{\bf t}_{\rm L}$ & Contribution to $\tl$ from smooth mass component and environment\\
\hline
$\de\tl^{(i)}$  &  Contribution to $\tl$ from the $i$th mass substructure \\
\hline
$\Delta\tl$ & $\equiv \sum_i \de\tl^{(i)}$ \\
\hline
$\Delta\tl^{\rm local}, \Delta\tl^{\rm dist}$ & Local and distributed contributions to $\Delta\tl$ \\
\hline
$\Delta t^{(i)}$ & Arrival time delay between image $i$ and the leading image \\ 
\hline
${\bf u}$ & $\equiv \langle \Delta \tl \rangle$ \\
\hline
$v_{\rm max}$ & Maximum circular velocity of a dark matter halo \\
\hline
$\bar{\bf x}_i$ & Unperturbed (from smooth model only) position of the $i^{\rm th}$ image \\
\hline
${\bf x}^{(i)}$ & Actual position of $i^{\rm th}$ lensed image \\
\hline 
$x_{\rm max}$ & $\equiv r_{\rm max}/r_{\rm s}$ \\
\hline
$z_{\rm lens}$ & Redshift of the main lens \\
\hline 
\end{tabular}
\caption{Summary of roman and scripted symbols used throughout the manuscript (continued).}\label{param_sum_table3} 
\end{table}

 \newpage
 
\bibliography{Stochastic_lensing}

\end{document}